\documentclass[12pt]{article}
\usepackage{jheppub}
\pdfoutput=1
\usepackage{epsfig}
\usepackage{amssymb}
\usepackage{hyperref}
\usepackage{amsmath}
\usepackage{mathrsfs}
\usepackage{multirow}
\usepackage{scalerel}
\usepackage{mathtools}
\usepackage{textcomp}
\usepackage{color}
\usepackage{rotating}
\usepackage{subcaption}
\usepackage[all]{xy}
\usepackage{arydshln}

\usepackage{tikz}
\usetikzlibrary{tikzmark}
\usetikzlibrary{decorations.markings}

\usetikzlibrary{calc}

\DeclareMathOperator{\Gr}{Gr}
\DeclareMathOperator{\Li}{Li}
\DeclareMathOperator{\sgn}{sgn}

\def\ket#1{\langle #1 \rangle}
\def\nl{\nonumber\\}
\def\nn{\nonumber}
\def\x{\mathcal{X}}
\def\xcoord{$\mathcal{X}$-coordinate}
\def\xcoords{$\mathcal{X}$-coordinates}
\def\a{\mathcal{A}}
\def\acoord{$\mathcal{A}$-coordinate}
\def\acoords{$\mathcal{A}$-coordinates}

\def\pdfeq#1{\texorpdfstring{$#1$}{a}}

\def\fd5{f_{D_5}}
\def\fa3{f_{A_3}}

\def\drawLabeledPentagon{
\coordinate (P1) at (90:1);
\coordinate (P2) at (18:1);
\coordinate (P3) at (306:1);
\coordinate (P4) at (234:1);
\coordinate (P5) at (162:1);
\draw[white,fill=white] (0,0) circle (1cm);
\draw (P1) -- (P2) -- (P3) -- (P4) -- (P5) -- cycle;
\draw (0,1.2) node {1};
\draw (.9,.3) node[anchor=west] {2};
\draw (.5,-.9) node[anchor=west] {3};
\draw (-.5,-.9) node[anchor=east] {4};
\draw (-.9,.3) node[anchor=east] {5};
}

\def\drawHexagon{
\coordinate (P1) at (90:1);
\coordinate (P2) at (30:1);
\coordinate (P3) at (330:1);
\coordinate (P4) at (270:1);
\coordinate (P5) at (210:1);
\coordinate (P6) at (150:1);
\draw[white,fill=white] (0,0) circle (1.1cm);
\draw (P1) -- (P2) -- (P3) -- (P4) -- (P5) -- (P6) -- cycle;
}

\def\drawLabeledHexagon{
\coordinate (P1) at (90:1);
\coordinate (P2) at (30:1);
\coordinate (P3) at (330:1);
\coordinate (P4) at (270:1);
\coordinate (P5) at (210:1);
\coordinate (P6) at (150:1);
\draw (P1) -- (P2) -- (P3) -- (P4) -- (P5) -- (P6) -- cycle;
\draw (0,1.2) node {1};
\draw (30:1.05) node[anchor=west] {2};
\draw (330:1.05) node[anchor=west] {3};
\draw (270:1.2) node {4};
\draw (210:1.05) node[anchor=east] {5};
\draw (150:1.05) node[anchor=east] {6};
}

\title{Cluster Algebras and the Subalgebra Constructibility of the Seven-Particle Remainder Function} 

\author{John~Golden$^{1,2}$}
\author{and Andrew~J.~McLeod$^{2,3,4}$}

\affiliation{\footnotesize $^1$ Leinweber  Center for Theoretical Physics and
Randall Laboratory of Physics, Department of Physics,
University of Michigan
Ann Arbor, MI 48109, USA}

\affiliation{\footnotesize $^2$ Kavli Institute for Theoretical Physics, 
UC Santa Barbara, Santa Barbara, CA 93106, USA}

\affiliation{\footnotesize $^3$ SLAC National Accelerator Laboratory,
Stanford University, Stanford, CA 94309, USA}

\affiliation{\footnotesize $^4$ Niels Bohr International Academy, Blegdamsvej 17, 2100 Copenhagen, Denmark}

\abstract{We review various aspects of cluster algebras and the ways in which they appear in the study of loop-level amplitudes in planar ${\cal N} = 4$ supersymmetric Yang-Mills theory. In particular, we highlight the different forms of cluster-algebraic structure that appear in this theory's two-loop MHV amplitudes---considered as functions, symbols, and at the level of their Lie cobracket---and recount how the `nonclassical' part of these amplitudes can be decomposed into specific functions evaluated on the $A_2$ or $A_3$ subalgebras of $\Gr(4,n)$. We then extend this line of inquiry by searching for other subalgebras over which these amplitudes can be decomposed. We focus on the case of seven-particle kinematics, where we show that the nonclassical part of the two-loop MHV amplitude is also constructible out of functions evaluated on the $D_5$ and $A_5$ subalgebras of $\Gr(4,7)$, and that these decompositions are themselves decomposable in terms of the same $A_4$ function. These nested decompositions take an especially canonical form, which is dictated in each case by constraints arising from the automorphism group of the parent algebra.}

\emailAdd{amcleod@nbi.ku.dk}
\emailAdd{johngolden@gmail.com}

\begin{document}
\hypersetup{pageanchor=false}

\begin{flushright} SLAC--PUB--17354
\end{flushright}

\maketitle
\hypersetup{pageanchor=true}

\section{Introduction}

Multi-loop scattering amplitudes in the planar limit of ${\cal N} = 4$ supersymmetric Yang-Mills (sYM) theory exhibit a great deal of intriguing mathematical structure. Much of this structure, at least at low loops and particle multiplicity, seems to be intimately tied to the cluster algebras associated with the Grassmannian Gr(4,$n$)~\cite{ArkaniHamed:2012nw,Golden:2013xva,Bourjaily:2012gy}. This is especially true in the maximally-helicity-violating (MHV) sector, where amplitudes have been computed at high loop orders in six- and seven-particle kinematics~\cite{Dixon:2013eka,Dixon:2014voa,Drummond:2014ffa,Caron-Huot:2016owq,Dixon:2016nkn}, and algorithms exist for calculating two loop amplitudes for any number of particles~\cite{CaronHuot:2011ky,Golden:2014xqf}. Remarkably, each of the branch cuts in these amplitudes ends at the vanishing locus of some cluster coordinate on Gr(4,$n$)~\cite{Golden:2013xva,Golden:2013lha,Golden:2014xqa,Golden:2014pua}, and---even more strikingly---their iterated discontinuities vanish unless sequentially taken in coordinates that appear together in a cluster of Gr(4,$n$)~\cite{Drummond:2017ssj}. All six- and seven-particle next-to-MHV (NMHV) amplitudes that have currently been computed in this theory share these remarkable properties~\cite{CaronHuot:2011kk,Dixon:2014iba,Drummond:2014ffa,Dixon:2015iva,Caron-Huot:2016owq,Dixon:2016nkn,Dixon:2016apl,Drummond:2018dfd,Caron-Huot:2019vjl,Caron-Huot:2019bsq}, as do certain classes of Feynman integrals~\cite{Drummond:2010cz,Drummond:2017ssj,Bourjaily:2018aeq,Henn:2018cdp}, some of which have been computed to all loop orders~\cite{Caron-Huot:2018dsv}. While this collection of amplitudes and integrals represents the simplest this theory has to offer, it remains suggestive that cluster algebras combinatorially realize these salient aspects of their analytic structure, thereby encoding locality in a non-obvious way.

The fact that cluster algebras appear in this context is not totally surprising, given that the plabic graphs that describe the integrands of this theory to all loop orders are themselves dual to cluster algebras. In particular, the boundaries of the positive Grassmannian, where it is known that these integrands can develop physical singularities, all lie on the vanishing loci of cluster coordinates on Gr(4,$n$)~\cite{ArkaniHamed:2012nw}. Despite this, it's far from obvious that the location of \emph{all} physical singularities will be picked out by cluster coordinates in this way---and indeed, even at one loop, N$^2$MHV amplitudes have singularities at points involving square roots when expressed in terms of cluster coordinates~\cite{Prlina:2017azl}. This obfuscates the general connection between cluster algebras and the amplitudes of this theory, as does the eventual appearance of functions beyond polylogarithms~\cite{Paulos:2012nu,CaronHuot:2012ab,Bourjaily:2015jna,Bourjaily:2017wjl,Bourjaily:2017bsb,Bourjaily:2018ycu,Bourjaily:2018yfy}. Both complications point to the need for more general objects than cluster algebras to describe the analytic structure of this theory at higher loops and particle multiplicities.

There is reason, however, to be optimistic that an analogously simple characterization of this (more complicated) analytic structure might be found. This optimism stems from the observation that the infinite class of amplitudes we currently have access to---the two-loop MHV amplitudes---have properties beyond branch cuts that seem to be indelibly tied to cluster algebras. In particular, the `nonclassical' part of each of these amplitudes (that is, the part that cannot be expressed in terms of classical polylogarithms) is uniquely determined by a small set of physical and cluster-algebraic properties~\cite{Golden:2014pua}.  Seemingly unrelated, a pair of functions can be associated with the simplest cluster algebras, related to the Dynkin diagrams $A_2$ and $A_3$, in terms of which these nonclassical components can be decomposed into a sum over the $A_2$ or $A_3$ subalgebras of Gr(4,$n$)~\cite{Golden:2014xqa}. Furthermore, the remaining `classical' part of these amplitudes can always be written as products of classical polylogarithms involving only negative cluster coordinates as arguments~\cite{Golden:2014xqf}. It can be hoped that the pervasiveness of such cluster-algebraic structure points to the existence of a deeper and more general combinatorial structure that extends to all particle multiplicities and helicity configurations. If so, better understanding the many ways in which cluster algebras appear in these amplitudes can help us identify the features this structure must have.

The motivation for this work is to review the connections already mentioned between cluster algebras, as well as to further mine the two-loop MHV amplitudes for additional cluster-algebraic structure. The review portion of this paper, comprising most of sections \ref{sec:brief_intro} and \ref{sec:cluster_polylog_MHV_review}, aims to be a self-contained pedagogical introduction to cluster algebras and their appearance in loop-level amplitudes in $\mathcal{N}=4$ sYM. These sections in particular focus on the subalgebra structure of cluster algebras, as these subalgebras will be used in the second half of the paper to build progressively more complicated polylogarithmic functions. A central ingredient in this story is the set of automorphisms respected by a given cluster algebra, as these automorphisms constrain the space of polylogarithms that can be defined on them.

We will in particular be interested in the space of `cluster polylogarithms'---poly- logarithms defined on (both finite and infinite) cluster algebras that exhibit interesting forms of cluster-algebraic structure at both the level of their symbol and their Lie cobracket. Such functions naturally arise in the study of the two-loop MHV amplitudes in planar ${\cal N} =4$ sYM theory~\cite{Golden:2014xqa}, but also constitute an interesting class of functions in their own right. In the case of finite cluster algebras these functions can be studied systematically, and in the second half of this paper we describe an efficient method for constructing all nonclassical cluster polylogarithms on a given finite cluster algebra. This procedure is based on what we call the `cluster subalgebra constructibility' of these functions, and in particular on the conjecture that all nonclassical cluster polylogarithms (on both finite and infinite cluster algebras) can be decomposed into sum over their $A_2$ subalgebras~\cite{Golden:2014xqa}. We illustrate this method by constructing all (nonclassical) cluster polylogarithms on $\Gr(4,7)$ and its subalgebras (this space was also explored in~\cite{Harrington:2015bdt}, using a slightly different approach).

While the (nonclassical parts of the) two-loop MHV amplitudes in planar ${\cal N} =4$ sYM~\cite{Golden:2014xqa} are already known to be $A_2$- and $A_3$- constructible, it is interesting to ask whether they can be constructed in terms of cluster polylogarithms on larger algebras. In particular, using the construction described above, we can ask this question of the two-loop seven-point MHV amplitude---itself a cluster polylogarithm associated with the Grassmannian $\Gr(4,7)$. We show that such decompositions not only exist over the $D_5$, $A_5$, and $A_4$ subalgebras of $\Gr(4,7)$, but that these decompositions take an especially canonical form, uniquely dictated (up to an overall scale) by the interplay of the subalgebra and automorphism structures of $\Gr(4,7)$. Specifically, we will find that the seven-particle remainder function can be decomposed as
\begin{align}
	R^{(2)}_7 &=  {\color{white} - } \frac{1}{20} \sum_{D_5\subset \Gr(4,7)} \sum_{A_4\subset D_5} f_{A_4}^{+-}(x_1\to x_2 \to x_3 \to x_4) + \dots \nonumber \\
	& = - \frac{1}{20}  \sum_{A_5\subset \Gr(4,7)} \sum_{A_4\subset A_5} f_{A_4}^{+-}(x_1\to x_2 \to x_3 \to x_4) + \dots, \nonumber
\end{align}
in terms of a single function $f_{A_4}^{+-}$ defined in section~\ref{sec:r27-sub-constructibility}, where the trailing dots indicate a contribution consisting of purely classical polylogarithms. 

In a forthcoming companion paper, we also use this technology to construct the full eight-point remainder function in terms of cluster polylogarithms. That is, we describe methods for searching for decompositions of the nonclassical part of $R_n^{(2)}$ even when $\Gr(4,n)$ is infinite, and then---using such a decomposition---systematically construct the classical and beyond-the-symbol components of $R_8^{(2)}$ using the techniques described in~\cite{Golden:2014xqf}. We also show that there exists a `generalized BDS-like ansatz' that preserves all Steinmann and cluster adjacency relations in eight-particle kinematics (and for all multiplicities that are a mutliple of 4), despite the nonexistence of the standard BDS-like ansatz for these numbers of particles (this normalization is also discussed in section~\ref{sec:cluster-algebra-R2n} of the present work).

This paper is organized as follows. In section~\ref{sec:brief_intro} we provide a self-contained introduction to cluster algebras and why they appear in ${\cal N} =4$ sYM theory. While this section will largely constitute review for those familiar with recent developments at the intersection of these topics, the discussion of automorphisms in section~\ref{sec:automorphisms} will likely be new to even those familiar with the physics literature. In section~\ref{sec:cluster_polylog_MHV_review} we discuss some of the tools relevant for working with polylogarithms, particularly their associated coaction and Lie cobracket. We then recap the ways in which the coproduct and cobracket of the two-loop MHV amplitudes in planar ${\cal N} =4$ sYM theory have been found to exhibit curious cluster-algebraic structure. In section~\ref{sec:sub-constructibility} we turn to the more general space of cluster polylogarithms, and initiate a systematic exploration of the nonclassical functions in this space, showing (in the case of finite cluster algebras) that there are surprisingly few such functions. Finally, as an application of this technology, we explore the ways in which these functions can be used to express the two-loop seven-point MHV amplitude in section~\ref{sec:r27-sub-constructibility}. In particular, we show that this amplitude admits a (nested) decomposition in terms of cluster polylogarithms defined on cluster algebras of every rank smaller than that of $\Gr(4,7)$, and thereby identy cluster polylogarithms of physical interest on the $D_5$, $A_5$, and $A_4$ cluster algebras, supplementing the two that are already known to be of interest on $A_2$ and $A_3$. 

We also include two appendices. Appendix~\ref{appendix:subalgebras} tabulates all the (nested) subalgebras of $\Gr(4,7)$, while appendix~\ref{appendix:cobrackets} tabulates the number of independent functions that have a nonzero Lie cobracket on each of these finite cluster algebras.

\section{A Brief Introduction to Cluster Algebras} \label{sec:brief_intro}

Cluster algebras were first introduced by Fomin and Zelevinsky \cite{1021.16017} as a tool for identifying which algebraic varieties come equipped with a natural notion of positivity, and what quantities determine this positivity. As a consequence, they naturally appear in the study of the positive Grassmannian $\Gr_{+}(k,n)$, i.e.~the space of $k\times n$ matrices modulo the action of $\text{GL}(k)$ where all ordered $k\times k$ minors are positive. They correspondingly also appear in the study of planar ${\cal N} = 4$ sYM theory, since the integrands in this theory are encoded to all orders by $\Gr_+(4,n)$~\cite{ArkaniHamed:2012nw}.

A simple example of the type of questions cluster algebras help address is: if one were to check just the positivity of a set of matrix minors (not their actual value), what are the minimal sets of minors that suffice to determine whether a point is in $\Gr_+(k,n)$? These minors are not all independent; they satisfy identities known as Pl\"ucker relations, for example
\begin{equation}
  \label{eq:plucker-rel}
  \ket{abI} \ket{cdI} = \ket{acI} \ket{bdI} - \ket{adI}\ket{bcI},
\end{equation}
where each Pl\"ucker coordinate $\ket{i_1,\ldots,i_k}$ corresponds to the minor of columns $i_1, \ldots,i_k$, and $I$ is a multi-index object with $k-2$ entries.

To gain some intuition for this problem, let us explore the case of $\Gr_+(2,5)$. The five cyclically adjacent minors $\ket{12}$, $\ket{23}$, $\ket{34}$, $\ket{45}$, and $\ket{15}$ cannot be eliminated in terms of each other by Pl\"ucker relations, so each gives rise to an independent positivity constraint. However, making use of the Pl\"ucker relations 
\begin{equation} \label{eq:plucker}
\begin{split}
	\ket{24} &= (\ket{12}\ket{34} + \ket{23}\ket{14})/\ket{13},\\
	\ket{25} &= (\ket{12}\ket{45} + \ket{24}\ket{15})/\ket{14},\\
	\ket{35} &= (\ket{25}\ket{34} + \ket{23}\ket{45})/\ket{24},
\end{split}	 	
\end{equation} 
we can eliminate three of the nonadjacent minors---for instance, $\ket{24}$, $\ket{25}$, and $\ket{35}$---in terms of the remaining ${{5}\choose{2}} - 3 = 7$ adjacent and nonadjacent ones. It can be checked that all further Pl\"ucker relations are implied by those in~\eqref{eq:plucker}, telling us that seven minors must be computed to determine if a matrix is in $\Gr_+(2,5)$. However, as should be clear, it is not sufficient to check the positivity of {\it any} seven (ordered) minors; only certain triples of minors can be eliminated. It would therefore be advantageous to have a method for generating all sets of minors that are sufficient to answer this question.

To motivate how cluster algebras address this problem, consider the following triangulation of the pentagon:
\begin{equation} \label{eq:pent_triangulation}
\begin{gathered}
\begin{tikzpicture}
  \drawLabeledPentagon
  \draw[color=red] (P1) -- (P3);
  \draw[color=red] (P1) -- (P4);
\end{tikzpicture} 
\end{gathered} .
\end{equation}
We can associate the line connecting points $i$ and $j$ with the Pl\"ucker coordinate $\ket{ij}$; if we further assign these lines length $\ket{ij}$, the resulting pentagon is cyclic (in the sense that all its vertices reside on a common circle) due to Ptolemy's theorem. Conversely, all cyclic $n$-gons represent a point in $\Gr_+(2,n)$~\cite{Arkani-Hamed:2014bca}. Note that the length of the three diagonals that are not present in this triangulation are determined by the length of the seven lines that are present (including edges); the problem has been reduced to geometry. This makes clear why these three diagonals---the three eliminated above---are redundant for the purpose of determining whether a matrix is in $\Gr_+(2,5)$. 

\begin{figure}[t]   \centering
  \begin{tikzpicture}[scale=0.8, every node/.style={scale=0.8}]
	\coordinate (P1) at (90:4);
	\coordinate (P2) at (18:4);
	\coordinate (P3) at (306:4);
	\coordinate (P4) at (234:4);
	\coordinate (P5) at (162:4);
	\draw (P1) -- (P2) -- (P3) -- (P4) -- (P5) -- cycle;
	\begin{scope}[shift = {(90:3.8)}]
	\drawLabeledPentagon
	\draw[color=red] (P1) -- (P3);
	\draw[color=black] (P1) -- (P4);
	\end{scope}
	\begin{scope}[shift = {(18:3.8)}]
	\drawLabeledPentagon
	\draw[color=black] (P2) -- (P4);
	\draw[color=red] (P1) -- (P4);
	\end{scope}
	\begin{scope}[shift = {(306:3.8)}]
	\drawLabeledPentagon
	\draw[color=red] (P2) -- (P4);
	\draw[color=black] (P2) -- (P5);
	\end{scope}
	\begin{scope}[shift = {(234:3.8)}]
	\drawLabeledPentagon
	\draw[color=black] (P3) -- (P5);
	\draw[color=red] (P2) -- (P5);
	\end{scope}
	\begin{scope}[shift = {(162:3.8)}]
	\drawLabeledPentagon
	\draw[color=black] (P1) -- (P3);
	\draw[color=red] (P3) -- (P5);
	\end{scope}
\end{tikzpicture}
  \caption{All possible triangulations of the pentagon. Mutating on the red chord moves you clockwise around the figure.}
\label{fig:pent_triangulations}
\end{figure}
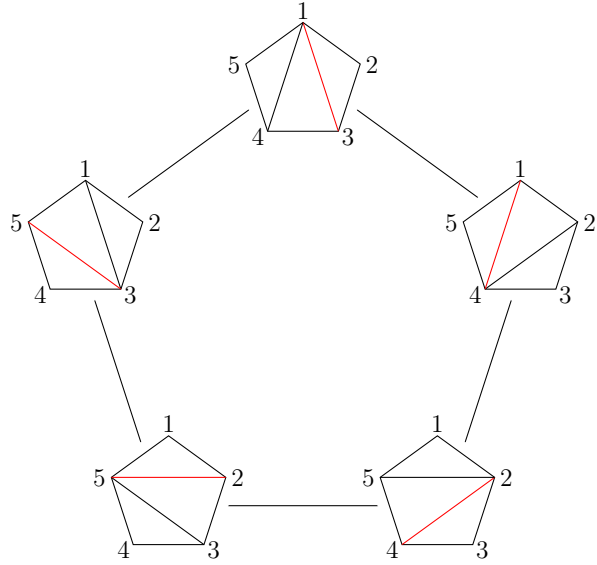

From the first relation in~\eqref{eq:plucker} we see that we could have instead chosen to check the positivity of $\ket{24}$ rather than $\ket{13}$. This corresponds to choosing a different triangulation, which we get by trading the latter diagonal for the former:
\begin{equation}
\begin{gathered}
\begin{tikzpicture}
  \drawLabeledPentagon
  \draw[color=blue, thick] (P1) -- (P2) -- (P3) -- (P4) -- cycle;
  \draw[color=red] (P1) -- (P3);
  \draw (2,0) node {\scalebox{1.4}{$\Rightarrow$}};
\begin{scope}[xshift = 4cm]
  \drawLabeledPentagon
  \draw[color=blue, thick] (P1) -- (P2) -- (P3) -- (P4) -- cycle;
  \draw[color=red] (P2) -- (P4); 
\end{scope}
\end{tikzpicture}
\end{gathered}.  
\end{equation}
We have highlighted in blue the fact that both diagonals are framed by the same quadrilateral face. More generally, we can pick any quadrilateral face and flip the diagonal it contains to generate a different triangulation. Repeatedly performing these flips generates all possible triangulations of the pentagon, as can be seen in figure~\ref{fig:pent_triangulations}. Each triangulation provides a set of edges/minors whose positivity ensures that a matrix is in $\Gr_+(2,5)$. 

In an analogous way, cluster algebras answer questions about positivity for a larger class of algebraic varieties (and in particular for all $\Gr_+(k,n)$) by considering `clusters' that can all be generated by an operation called `mutation' just as all triangulations of the pentagon are generated by flipping the diagonals of quadrilateral faces. We now turn to the definition of these objects, considering first how the example of $\Gr_+(2,5)$ can be rephrased in this language.

\subsection{Clusters, Mutations, and Cluster ${\cal A}$-coordinates}

Clusters can be defined to be quiver diagrams---namely, oriented graphs equipped with arrows connecting different nodes---in which each node is assigned a cluster coordinate.\footnote{Here and throughout, we are implicitly restricting ourselves to skew-symmetric cluster algebras of geometric type.} We can form one of the clusters of $\Gr(2,5)$ out of our original triangulation~\eqref{eq:pent_triangulation} by assigning an orientation to the pentagon and all subtriangles such as
\begin{equation}\label{eq:oriented-pentagon}
\begin{gathered}
\begin{tikzpicture}
  \drawLabeledPentagon
  \draw[color=red] (P1) -- (P3);
  \draw[color=red] (P1) -- (P4);
  \draw (18:.6) node[rotate=144] {$\circlearrowleft$};
  \draw (0:0) node[rotate=144] {$\circlearrowleft$};
  \draw (162:.6) node[rotate=144] {$\circlearrowleft$};
\end{tikzpicture} 
\end{gathered} \ .
\end{equation}
The nodes of our cluster are then given by the lines of this triangulation (making the minors $\ket{ab}$ our cluster coordinates), where an arrow is assigned from $\ket{ab}$ to $\ket{cd}$ if the triangle orientations in~\eqref{eq:oriented-pentagon} have segment $(ab)$ flowing into segment $(cd)$ and these segments border a common triangle. This gives us the quiver
\begin{equation}\label{eq:gr25-seed}
\begin{gathered}
\begin{xy} 0;<1pt,0pt>:<0pt,-1pt>::
	(25,30) *+{\langle 13\rangle} ="0",
	(75,30) *+{\langle 14\rangle} ="1",
	(125,30) *+{\framebox[5ex]{$\langle 15\rangle$}} ="2",
	(125,70) *+{\framebox[5ex]{$\langle 45\rangle$}} ="3",
	(75,70) *+{\framebox[5ex]{$\langle 34\rangle$}} ="4",
	(25,70) *+{\framebox[5ex]{$\langle 23\rangle$}} ="5",
	(0,0) *+{\framebox[5ex]{$\langle 12\rangle$}} ="6",
	(145,75) *+{},
	"0", {\ar"1"},
	"4", {\ar"0"},
	"0", {\ar"5"},
	"6", {\ar"0"},
	"1", {\ar"2"},
	"3", {\ar"1"},
	"1", {\ar"4"},
\end{xy}
\end{gathered} ,
\end{equation}
where the boxes around the $\ket{ii+1}$ indicate that they are \emph{frozen}---they can never change because they aren't in the interior of a quadrilateral face. One can also draw arrows (with partial weight) connecting these frozen nodes (see, for instance,~\cite{ArkaniHamed:2012nw}), but we will not keep track of them in the present work.

We have now drawn our first cluster (also sometimes called a seed). The Pl\"ucker coordinates in this cluster are referred to as cluster $\a$-coordinates, and they come in two flavors: mutable (for example, $\ket{13}$ and $\ket{14}$ above) and frozen ($\ket{ii+1}$ above). In any quiver, the information encoded by the arrows can also be represented in terms of a skew-symmetric matrix
\begin{equation}
b_{i j} = (\# \text{ of arrows}\; i \to j) - (\# \text{ of arrows}\; j \to i)
\label{eq:bijdef}
\end{equation}
called the exchange matrix (or the signed adjacency matrix).

The process of mutation, which we have described geometrically in terms of flipping diagonals, has a simple interpretation at the level of quivers. Given a quiver such as~\eqref{eq:gr25-seed}, we can choose any mutable node $k$ to mutate on (this is equivalent to picking which diagonal to flip). Mutation gives us back a new quiver in which the $\a$-coordinate $a_{k}$ has been sent to $a_{k}'$, where 
\begin{equation}
  \label{eq:a-coord-mutation}
  a_{k} a_{k}' = \prod_{i \vert b_{i k} > 0} a_{i}^{b_{i k}} + \prod_{i \vert b_{i k} < 0} a_{i}^{-b_{i k}},
\end{equation} (with the understanding that an empty product is set to one), while all other cluster $\a$-coordinates remain unchanged. The arrows connecting the nodes in this new quiver are also modified by carrying out the following sequence of operations:
\begin{itemize}
	\item for each path $i\to k \to j$, add an arrow $i\to j$,
	\item reverse all arrows on the edges incident with $k$,
	\item remove any two-cycles (oppositely-oriented arrows) that may have formed.
\end{itemize}
This creates a new adjacency matrix $b_{ij}'$ via 
\begin{equation}
  \label{eq:b-mutation}
  b'_{i j} =
  \begin{cases}
    -b_{i j}, &\quad \text{if $k \in \lbrace i, j\rbrace$,}\\
    b_{i j}, &\quad \text{if $b_{i k} b_{k j} \leq 0$,}\\
    b_{i j} + b_{i k} b_{k j}, &\quad \text{if $b_{i k}, b_{k j} > 0$,}\\
    b_{i j} - b_{i k} b_{k j}, &\quad \text{if $b_{i k}, b_{k j} < 0$.}
  \end{cases}
\end{equation}
Mutation is an involution, so mutating on $a_k'$ will take you back to the original cluster (just as flipping the same diagonal twice will take you back to where you started). 

In terms of these ingredients, a cluster algebra can be defined to be a set of clusters that is closed under mutation. Thus, mutating on any non-frozen node of any cluster will generate a different cluster in the same cluster algebra. In practice, one therefore constructs cluster algebras by starting from a seed such as~\eqref{eq:gr25-seed}, and iteratively mutating on all available nodes until the set of clusters closes (or it becomes clear the cluster algebra is infinite). 

It is common to refer to certain cluster algebras by particularly nice representative clusters, when the mutable notes of the corresponding quiver form an oriented Dynkin diagram. For instance, the Gr(2,5) cluster algebra is often referred to as  $A_2$, since the mutable part of the seed~\eqref{eq:gr25-seed} is given by $\ket{13}\to\ket{14}$. Thus, we will often speak interchangeably of the cluster algebras for $\Gr(2,5)$ and $A_2$. This is a slight abuse of notation, as the $\Gr(2,5)$ cluster algebra corresponds specifically to the cluster algebra generated by the collection of frozen and mutable nodes in eq.~(\ref{eq:gr25-seed}), whereas an $A_2$ cluster algebra can in principle be dressed with any number of frozen nodes. We will see why this language is useful in the next section.

\subsection{Cluster \texorpdfstring{$\mathcal{X}$}{X}-coordinates}

Cluster algebras can also be formulated in terms of a different set of cluster coordinates, called Fock-Goncharov coordinates or $\x$-coordinates~\cite{FG03b}. As we will see in future sections, cluster \xcoords\ play a crucial role in connecting cluster algebras to polylogarithms and scattering amplitudes. While it is always possible to phrase results involving cluster algebras directly in terms of \acoords, \xcoords\ often allow for cluster-algebraic structure to be made more manifest. 

Clusters formed out of \xcoords\ can be directly constructed out of clusters involving \acoords. Given a quiver equipped with $\a$-coordinates and described by the exchange matrix $b_{ij}$, we can compute an $\x$-coordinate to assign to each mutable node by
\begin{equation} \label{eq:x_from_a_coordinates}
	x_i = \prod_j a_j^{b_{ji}}. 	
\end{equation} 
For example, the $\x$-coordinate cluster associated with~(\ref{eq:gr25-seed}) is formed by associating cluster \xcoords\ with all its mutable nodes, where these \xcoords\ are constructed by putting all Pl\"ucker coordinates that point to that node in the numerator, and all Pl\"ucker coordinates that are pointed to by that node in the denominator. That is, we get
\begin{equation} \label{eq:a2_x_seed}
	\frac{\ket{12}\ket{34}}{\ket{14}\ket{23}} \to \frac{\ket{13}\ket{45}}{\ket{15}\ket{34}},
\end{equation}
which again takes the form of the generic $A_2$ quiver $x_1 \to x_2$. In the pentagon-triangulation picture, these $\x$-coordinates describe overlapping quadrilaterals, for instance
\begin{align}\begin{gathered}
\begin{tikzpicture}
  \drawLabeledPentagon
  \draw[color=blue, thick] (P1) -- (P2) -- (P3) -- (P4) -- cycle;
\end{tikzpicture} 
\begin{aligned}
\sim \frac{\ket{12}\ket{34}}{\ket{14}\ket{23}},  \\[1.6cm] \
\end{aligned} \qquad 
\begin{tikzpicture}
  \drawLabeledPentagon
  \draw[color=blue, thick] (P1) -- (P3) -- (P4) -- (P5) -- cycle;
\end{tikzpicture}  
\begin{aligned}
\sim \frac{\ket{13}\ket{45}}{\ket{15}\ket{34}}, \\[1.6cm] \
\end{aligned}
\vspace{-1.4cm}
\end{gathered}
\end{align}
which come in one-to-one correspondence with the diagonals in a (single) triangulation.

Mutation rules for \xcoords\ are different than for \acoords, and are given by
\begin{equation}
  \label{eq:x-coord-mutation}
  x_{i}' =
  \begin{cases}
    x_{k}^{-1}, &\quad i=k,\\
    x_{i} (1+x_{k}^{\sgn b_{i k}})^{b_{i k}}, &\quad i \neq k
  \end{cases}\ ,
\end{equation}
where mutation has been carried out on node $k$. Mutation still changes the arrows in the quiver diagram as it did in the case of \acoords.  Given just a quiver diagram, it can sometimes be unclear whether a given quiver should be mutated using the \acoord\ or \xcoord\ rules~\eqref{eq:a-coord-mutation} or~\eqref{eq:x-coord-mutation}. We adopt the convention in this work that if a quiver is given with no frozen nodes, it should be thought of as equipped with \xcoords.

Just as with $\a$-coordinate clusters, we can generate all $\x$-coordinate clusters by mutation. Mutating on alternating nodes of our $A_2$ cluster (starting with $x_2$), we generate the following sequence of clusters:
\begin{align}
  x_1 &\to x_2 \nl
  x_1(1+x_2) &\leftarrow \frac{1}{x_2} \nl
  \frac{1}{x_1(1+x_2)} &\to \frac{1+x_1+x_1 x_2}{x_2} \\
  \frac{1+x_1}{x_1 x_2} &\leftarrow \frac{x_2}{1+x_1+x_1 x_2} \nl
  \frac{x_1 x_2}{1+x_1} &\to \frac{1}{x_1} \nl
  x_2 &\leftarrow x_1 \nl
  &\ \ \vdots \nn
\end{align}
The series then repeats, with all arrows reversed. Note that (specifically in the case of $A_2$), if we label these $\x$-coordinates by
\begin{gather}\label{def:a2-xcoords}
  \x_1 = 1/x_1, \qquad \qquad \x_2 = x_2, \qquad \qquad \x_3 = x_1(1+x_2), \\ 
  \x_4 = \frac{1+x_1+x_1 x_2}{x_2}, \qquad \qquad \x_5 = \frac{1+x_1}{x_1 x_2}, \nonumber
\end{gather}
the mutation rule in~\eqref{eq:x-coord-mutation} takes the simple form
\begin{equation}\label{eq:exchange-relation}
  1+\x_i = \x_{i-1}\x_{i+1},
\end{equation}
while all the clusters take the form $1/\x_i \to \x_{i+1}$. Eq. (\ref{eq:exchange-relation}) is commonly referred to as the $A_2$ exchange relation. 
Putting this all together, we will generically refer to an $A_2$ cluster algebra as any set of clusters $1/\x_{i-1} \to \x_i$ for $i=1\ldots5$ where the $\x_i$ satisfy eq. (\ref{eq:exchange-relation}). We believe it is useful at this point to emphasize that one can take as input any $x_1$ and $x_2$, and generate an associated $A_2$. 

A very useful feature of cluster \xcoords\ is that they come equipped with a natural Poisson bracket structure, making the Grassmannian $\Gr(k,n)$ a cluster Poisson variety~\cite{GSV}. Namely, when two \xcoords\ appear together in a cluster of $\Gr(k,n)$, there exists a Poisson bracket that evaluates to
\begin{equation}
\{x_i, x_j \} = b_{ij} x_i x_j .
\end{equation}
This structure respects mutation, implying that the entry $b_{ij}$ (which counts the number of arrows from $x_i$ to $x_j$ in a given cluster's quiver) will be the same in all clusters containing both $x_i$ and $x_j$. The Poisson bracket (and associated Sklyanin bracket) will play a larger role in forthcoming work~\cite{cluster_subalgebras_ii}, so we defer further discussion of this structure (for existing discussions in the literature, see~\cite{PoissonVarieties,Vergu:2015svm}).

\subsection{Subalgebras and Cluster Polytopes}\label{sec:subalgebras_cluster_polytopes}

Cluster algebras contain a rich and intricate subalgebra structure, which will play a central role in our analysis. It is simple to illustrate how these subalgebras arise by considering $\Gr(2,6)$, which triangulates the hexagon. In figure~\ref{eq:gr26-seed} we give the seed cluster for $\Gr(2,6)$ in the triangulation, \acoord, and \xcoord\ representations. Since the mutable nodes take the form of an $A_3$ Dynkin diagram, we often speak of $\Gr(2,6)$ and $A_3$ interchangeably, just as we did with $\Gr(2,5) \simeq A_2$. 

\begin{figure}
\begin{equation*}
\begin{gathered}
\begin{tikzpicture}
	\drawLabeledHexagon
  	\draw[color=red] (P1) -- (P3);
  	\draw[color=red] (P1) -- (P4);
  	\draw[color=red] (P1) -- (P5);
\end{tikzpicture}
\\ \scalebox{1.4}{$\Updownarrow$}
\\ \ \vspace{-1.1cm} \ \\
\begin{xy} 0;<1pt,0pt>:<0pt,-1pt>::
	(25,30) *+{\langle 13\rangle} ="0",
	(75,30) *+{\langle 14\rangle} ="1",
	(125,30) *+{\langle 15\rangle} ="2",
	(0,0) *+{\framebox[5ex]{$\langle 12\rangle$}} ="3",
	(25,70) *+{\framebox[5ex]{$\langle 23\rangle$}} ="4",
	(75,70) *+{\framebox[5ex]{$\langle 34\rangle$}} ="5",
	(125,70) *+{\framebox[5ex]{$\langle 45\rangle$}} ="6",
	(175,70) *+{\framebox[5ex]{$\langle 56\rangle$}} ="7",
	(175,30) *+{\framebox[5ex]{$\langle 16\rangle$}} ="8",
	"0", {\ar"1"},
	"1", {\ar"2"},
	"3", {\ar"0"},
	"0", {\ar"4"},
	"5", {\ar"0"},
	"1", {\ar"5"},
	"6", {\ar"1"},
	"2", {\ar"6"},
	"7", {\ar"2"},
	"2", {\ar"8"},
\end{xy} \\ 
\ \vspace{-.4cm} \ \\ 
\scalebox{1.4}{$\Updownarrow$} \\ 
\ \vspace{-.4cm} \ \\ 
\begin{xy} 0;<1pt,0pt>:<0pt,-1pt>::
	(0,0) *+\txt{\fontsize{16pt}{16pt} $\frac{\ket{12}\ket{34}}{\ket{14}\ket{23}}$} ="0",
	(75,0) *+\txt{\fontsize{16pt}{16pt} $\frac{\ket{13}\ket{45}}{\ket{15}\ket{34}}$} ="1",
	(150,0) *+\txt{\fontsize{16pt}{16pt} $\frac{\ket{14}\ket{56}}{\ket{16}\ket{45}}$} ="2",
	"0", {\ar"1"},
	"1", {\ar"2"},
\end{xy}
\end{gathered} 
\end{equation*}
\caption{A triangulation of the hexagon along with its assocated \acoord\ and \xcoord\ seed quivers.}
\label{eq:gr26-seed}
\end{figure}

The $\Gr(2,6)$ cluster algebra features 14 clusters, which can be grouped into multiple (overlapping) subalgebras. A simple example is the collection of all triangulations which involve the chord $\ket{15}$. This set contains 5 clusters and is itself a cluster algebra, which can be generated by treating $\ket{15}$ as a frozen node (or in \xcoords, freezing the node $\frac{\ket{14}\ket{56}}{\ket{16}\ket{45}}$). This of course is the cluster algebra corresponding to the triangulations of the pentagon formed by points $1,\ldots,5$, outlined here in blue:
\begin{equation}
\begin{gathered}
\begin{tikzpicture}
	\drawHexagon
  	\draw[color=red] (P1) -- (P3);
  	\draw[color=red] (P1) -- (P4);
	\draw[color=blue] (P1) -- (P2) -- (P3) -- (P4) -- (P5) -- cycle;
\end{tikzpicture} 
\end{gathered}
{\ ,\ }
\begin{gathered}
\begin{tikzpicture}
	\begin{scope}[xshift=2cm]
	\drawHexagon
  	\draw[color=red] (P2) -- (P4);
  	\draw[color=red] (P1) -- (P4);
	\draw[color=blue] (P1) -- (P2) -- (P3) -- (P4) -- (P5) -- cycle;
	\end{scope}
\end{tikzpicture} 
\end{gathered}
{\ ,\ }
\begin{gathered}
\begin{tikzpicture}
	\begin{scope}[xshift=4cm]
	\drawHexagon
  	\draw[color=red] (P2) -- (P4);
  	\draw[color=red] (P2) -- (P5);
	\draw[color=blue] (P1) -- (P2) -- (P3) -- (P4) -- (P5) -- cycle;
	\end{scope}
\end{tikzpicture} 
\end{gathered}
{\ ,\ }
\begin{gathered}
\begin{tikzpicture}
	\begin{scope}[xshift=6cm]
	\drawHexagon
  	\draw[color=red] (P2) -- (P5);
  	\draw[color=red] (P3) -- (P5);
	\draw[color=blue] (P1) -- (P2) -- (P3) -- (P4) -- (P5) -- cycle;
	\end{scope}
\end{tikzpicture} 
\end{gathered}
{\ ,\ }
\begin{gathered}
\begin{tikzpicture}
	\begin{scope}[xshift=8cm]
	\drawHexagon
  	\draw[color=red] (P1) -- (P3);
  	\draw[color=red] (P3) -- (P5);
	\draw[color=blue] (P1) -- (P2) -- (P3) -- (P4) -- (P5) -- cycle;
	\end{scope}
\end{tikzpicture}
\end{gathered}\ .
\end{equation}
We therefore refer to the collection of clusters that involve this pentagon as an $A_2$ subalgebra of $\Gr(2,6)$. 

It should be clear, upon referring back to the mutation rules~\eqref{eq:a-coord-mutation} and~\eqref{eq:x-coord-mutation}, that this $A_2$ subaglebra is truly identical to what we have been calling $\Gr(2,5)$. That is, neither the $\a$- or $\x$-coordinate mutation rule (or the rule for constructing the $\x$-coordinate cluster out of the $\a$-coordinate one) depends on nodes further than a single arrow away from the node on which one is mutating. Correspondingly, this subalgebra doesn't know about the existence of nodes involving point/column 6. (In the \xcoord\ case, the coordinate associated with the newly frozen node will change when one mutates on the node it is connected to, but the presence of this frozen node does not effect the coordinates appearing in the $A_2$ subalgebra itself.) We consider two subalgebras to be identical when the clusters they appear in only differ by nodes that have no effect on the mutable nodes of the subalgebra.  

What if we instead disallow mutation on the chord $\ket{13}$ (and the corresponding $\x$-coordinate node $\frac{\ket{12}\ket{34}}{\ket{14}\ket{23}}$)? Dropping the nodes that play no role in any of the mutations that remain gives rise to the effective quiver
\begin{equation}\label{eq:gr26-subalgebra-seed}
\begin{gathered}
\begin{xy} 0;<1pt,0pt>:<0pt,-1pt>::
	(25,0) *+{\langle 14\rangle} ="0",
	(75,0) *+{\langle 15\rangle} ="1",
	(125,0) *+{\framebox[5ex]{$\langle 16\rangle$}} ="2",
	(125,40) *+{\framebox[5ex]{$\langle 56\rangle$}} ="3",
	(75,40) *+{\framebox[5ex]{$\langle 45\rangle$}} ="4",
	(25,40) *+{\framebox[5ex]{$\langle 34\rangle$}} ="5",
	(-25,0) *+{\framebox[5ex]{$\langle 13\rangle$}} ="6",
	"0", {\ar"1"},
	"4", {\ar"0"},
	"0", {\ar"5"},
	"6", {\ar"0"},
	"1", {\ar"2"},
	"3", {\ar"1"},
	"1", {\ar"4"},
\end{xy}
\end{gathered} \ ,
\end{equation}
where we have put a box around $\ket{13}$ to make clear we are now treating it as frozen. We have also dropped the arrow from $\ket{34}$ to $\ket{13}$ since we are ignoring arrows between frozen nodes. The comparison to~\eqref{eq:gr25-seed} should be clear; this just represents a re-labeled version of $\Gr(2,5)$. 

Similarly, if we disallow mutation on the chord $\ket{14}$ (and $\frac{\ket{13}\ket{45}}{\ket{15}\ket{34}}$), we generate an $A_1 \times A_1$ subalgebra, since the chord $\ket{14}$ divides the hexagon in to two non-overlapping squares, each of which are triangulated by $A_1$ (or really $\Gr(2,4)$):
\begin{equation}
\begin{gathered}
\begin{tikzpicture}
	\drawHexagon
  	\draw[color=blue] (P1) -- (P2) -- (P3) -- (P4) -- cycle;
  	\draw[color=blue] (P1) -- (P4) -- (P5) -- (P6) -- cycle;
  	\draw[color=red] (P1) -- (P5);
  	\draw[color=red] (P1) -- (P3); 
\end{tikzpicture} 
\end{gathered}
{\ ,\ }
\begin{gathered}
\begin{tikzpicture}
  	\begin{scope}[xshift=2cm]
	\drawHexagon
  	\draw[color=blue] (P1) -- (P2) -- (P3) -- (P4) -- cycle;
  	\draw[color=blue] (P1) -- (P4) -- (P5) -- (P6) -- cycle;
  	\draw[color=red] (P1) -- (P5);
  	\draw[color=red] (P4) -- (P2);
	\end{scope}
\end{tikzpicture} 
\end{gathered}
{\ ,\ }
\begin{gathered}
\begin{tikzpicture}
	\begin{scope}[xshift=4cm]
	\drawHexagon
  	\draw[color=blue] (P1) -- (P2) -- (P3) -- (P4) -- cycle;
  	\draw[color=blue] (P1) -- (P4) -- (P5) -- (P6) -- cycle;
  	\draw[color=red] (P6) -- (P4);
  	\draw[color=red] (P4) -- (P2);
	\end{scope}
\end{tikzpicture} 
\end{gathered}
{\ ,\ }
\begin{gathered}
\begin{tikzpicture}
	\begin{scope}[xshift=6cm]
	\drawHexagon
  	\draw[color=blue] (P1) -- (P2) -- (P3) -- (P4) -- cycle;
  	\draw[color=blue] (P1) -- (P4) -- (P5) -- (P6) -- cycle;
  	\draw[color=red] (P6) -- (P4);
  	\draw[color=red] (P1) -- (P3);
	\end{scope}
\end{tikzpicture}
\end{gathered} \ .
\end{equation}
In appendix~\ref{appendix:subalgebras} we have tabulated the number of such $A_2$ and $A_1 \times A_1$ subalgebras in $A_3$, as well as the subalgebras of other cluster algebras that appear in $\Gr(4,7)$. There it will be found that there are in fact six $A_2$ subalgebras and three $A_1 \times A_1$ subalgebras of $A_3$.

This subalgebra structure can be nicely visualized by constructing an object known as the cluster polytope of a given cluster algebra. The vertices of this polytope each represent a cluster, while its edges represent the mutations that map these clusters into each other. For instance, figure \ref{fig:pent_triangulations} corresponds to the $\Gr(2,5) \simeq A_2$ cluster polytope, which also coincidentally takes the form of a pentagon. Note that every node has valency two since each cluster has two mutable vertices.

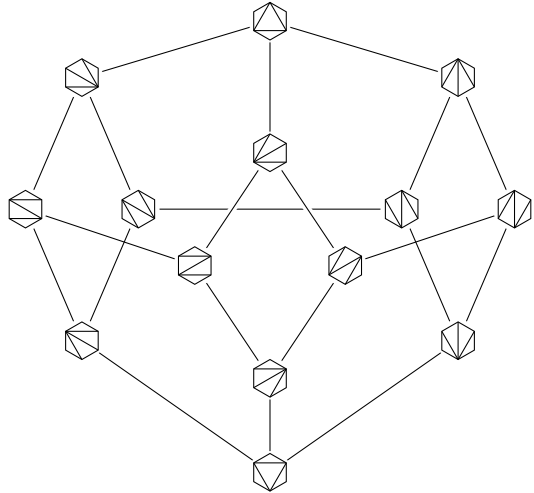
\begin{figure}[t]
  \centering
  \begin{tikzpicture}[scale=2.5]
\coordinate (p1) at (0,1);
\coordinate (p2) at (1,.7);
\coordinate (p3) at (1.3,0);
\coordinate (p4) at (1,-.7);
\coordinate (p5) at (0,-1.4);
\coordinate (p6) at (-1,-.7);
\coordinate (p7) at (-1.3,0);
\coordinate (p8) at (-1,.7);
\coordinate (p9) at (-.4,-.3);
\coordinate (p10) at (0,.3);
\coordinate (p11) at (.4,-.3);
\coordinate (p12) at (0,-.9);
\coordinate (p13) at (-.7,0);
\coordinate (p14) at (.7,0);
\draw (p13) -- (p14);
\draw[color=white, line width=1mm] (p9) -- (p10);
\draw[color=white, line width=1mm] (p10) -- (p11);
\draw (p1) -- (p2) -- (p3) -- (p4) -- (p5) -- (p6) -- (p7) -- (p8) -- cycle;
\draw (p9) -- (p10) -- (p11) -- (p12) -- cycle;
\draw (p8) -- (p13) -- (p6);
\draw (p2) -- (p14) -- (p4);
\draw (p1) -- (p10);
\draw (p7) -- (p9);
\draw (p11) -- (p3);
\draw (p12) -- (p5);
\begin{scope}[shift = {(p1)}, scale=0.1, every node/.style={scale=0.5}]
	\drawHexagon
	\draw (P1) -- (P3);
	\draw (P3) -- (P5);
	\draw (P1) -- (P5);
\end{scope}
\begin{scope}[shift = {(p2)}, scale=0.1, every node/.style={scale=0.5}]
	\drawHexagon
	\draw (P1) -- (P3);
	\draw (P1) -- (P4);
	\draw (P1) -- (P5);
\end{scope}
\begin{scope}[shift = {(p3)}, scale=0.1, every node/.style={scale=0.5}]
	\drawHexagon
	\draw (P2) -- (P4);
	\draw (P1) -- (P4);
	\draw (P1) -- (P5);
\end{scope}
\begin{scope}[shift = {(p4)}, scale=0.1, every node/.style={scale=0.5}]
	\drawHexagon
	\draw (P2) -- (P4);
	\draw (P1) -- (P4);
	\draw (P6) -- (P4);
\end{scope}
\begin{scope}[shift = {(p5)}, scale=0.1, every node/.style={scale=0.5}]
	\drawHexagon
	\draw (P2) -- (P4);
	\draw (P4) -- (P6);
	\draw (P6) -- (P2);
\end{scope}
\begin{scope}[shift = {(p6)}, scale=0.1, every node/.style={scale=0.5}]
	\drawHexagon
	\draw (P2) -- (P6);
	\draw (P6) -- (P3);
	\draw (P4) -- (P6);
\end{scope}
\begin{scope}[shift = {(p7)}, scale=0.1, every node/.style={scale=0.5}]
	\drawHexagon
	\draw (P2) -- (P6);
	\draw (P3) -- (P6);
	\draw (P3) -- (P5);
\end{scope}
\begin{scope}[shift = {(p8)}, scale=0.1, every node/.style={scale=0.5}]
	\drawHexagon
	\draw (P1) -- (P3);
	\draw (P3) -- (P6);
	\draw (P3) -- (P5);
\end{scope}
\begin{scope}[shift = {(p9)}, scale=0.1, every node/.style={scale=0.5}]
	\drawHexagon
	\draw (P2) -- (P6);
	\draw (P2) -- (P5);
	\draw (P3) -- (P5);
\end{scope}
\begin{scope}[shift = {(p10)}, scale=0.1, every node/.style={scale=0.5}]
	\drawHexagon
	\draw (P1) -- (P5);
	\draw (P2) -- (P5);
	\draw (P3) -- (P5);
\end{scope}
\begin{scope}[shift = {(p11)}, scale=0.1, every node/.style={scale=0.5}]
	\drawHexagon
	\draw (P1) -- (P5);
	\draw (P2) -- (P5);
	\draw (P2) -- (P4);
\end{scope}
\begin{scope}[shift = {(p12)}, scale=0.1, every node/.style={scale=0.5}]
	\drawHexagon
	\draw (P2) -- (P6);
	\draw (P2) -- (P5);
	\draw (P2) -- (P4);
\end{scope}
\begin{scope}[shift = {(p13)}, scale=0.1, every node/.style={scale=0.5}]
	\drawHexagon
	\draw (P1) -- (P3);
	\draw (P3) -- (P6);
	\draw (P6) -- (P4);
\end{scope}
\begin{scope}[shift = {(p14)}, scale=0.1, every node/.style={scale=0.5}]
	\drawHexagon
	\draw (P1) -- (P3);
	\draw (P1) -- (P4);
	\draw (P4) -- (P6);
\end{scope}
\end{tikzpicture}
  \caption{The cluster polytope of $A_3\simeq\Gr(2,6)$, in which each cluster is represented by a triangulation of the hexagon.}
  \label{fig:a3-poly}
\end{figure}

Similarly, the cluster polytope of $\Gr(2,6) \simeq A_3$ is given in figure~\ref{fig:a3-poly}. It has 14 vertices, corresponding to the 14 clusters of $\Gr(2,6)$, each with valency three. These vertices assemble into three square faces and six pentagonal faces---corresponding exactly to  the three $A_1 \times A_1$ subalgebras and the six $A_2$ subalgebras of $A_3$. This makes it easy to read off the subalgebra structure of $A_3$ directly. 

\begin{figure}[t]  \centering
  \includegraphics[scale=0.25]{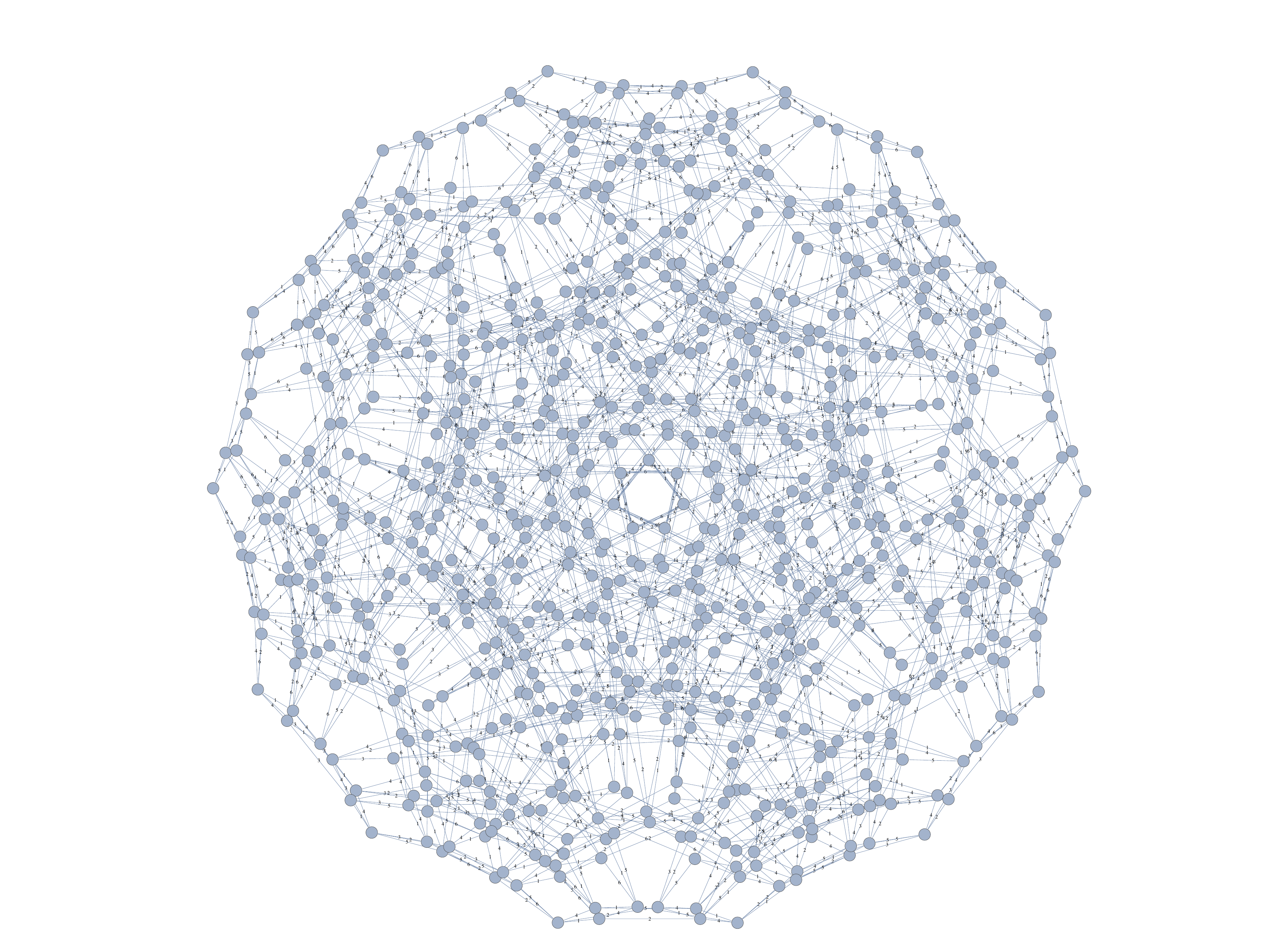}
  \caption{The cluster polytope of $\Gr(4,7) \simeq E_6$.}
  \label{fig:e6-poly}
\end{figure}

In $A_3$, it turns out that each of the faces corresponds to a distinct subalgebra. Cluster polytopes of larger cluster algebras become quite complicated, and it is often the case that distinct faces (or higher-dimensional polytopes) correspond to identical subalgebras. As an example, the cluster polytope of $\Gr(4,7)$, which will be the focus of much of the rest of this paper, is shown in figure~\ref{fig:e6-poly}. It has 833 vertices, each of valence 6, and 1071 pentagons corresponding to $A_2$ subalgebras; however, only 504 of these $A_2$ subalgebras are distinct.

While cluster polytopes give us a nice pictorial way to think about the relations between different clusters in a cluster algebra, their interiors can also be identified with the `positive region' where all (cyclically ordered) \acoords\ (and consequently \xcoords) are positive~\cite{ArkaniHamed:2012nw,Drummond:2018dfd}. Thus, in the case of Grassmannian cluster algebras, the positive Grassmannian $\Gr_+(k,n)$ (restricted to real values) consists of all points in the interior of the cluster polytope. Each cluster, moreover, can be thought of as providing a coordinate system in which the full positive region is spanned by allowing each \xcoord\ to range from $0$ to $\infty$. For a more in-depth discussion of this way of thinking about cluster polytopes, see~\cite{Drummond:2018dfd}.

\subsection{Grassmannian Cluster Algebras and Planar ${\cal N} = 4$ sYM Theory}

So far we have leaned heavily on the correspondence between the triangulations of an $n$-gon and the cluster algebra for $\Gr(2,n)$. Based on the examples of $\Gr(2,5)$ and $\Gr(2,6)$, it is not hard to write down a generic seed cluster for $\Gr(2,n)$ corresponding to the triangulation consisting of all chords $\ket{13},\ldots,\ket{1n{-}1}$:
\begin{equation}\label{eq:g2n-seed}
\begin{gathered}
\begin{tikzpicture}
	\coordinate (P1) at (90:1);
	\coordinate (P2) at (18:1);
	\coordinate (P3) at (320:1.2);
	\coordinate (P4) at (220:1.2);
	\coordinate (P5) at (162:1);
	\draw (0,1.2) node {1};
	\draw (1,.3) node[anchor=west] {2};
	\draw (.9,-.9) node[anchor=west] {3};
	\draw (-.9,-.9) node[anchor=east] {$n-1$};
	\draw (-1,.3) node[anchor=east] {$n$};
	\draw (P1) -- (P2) -- (P3);
	\draw[dotted] (P3) to[out=240,in=300] (P4);
	\draw (P4) -- (P5) -- (P1);
	\draw[red] (P1) -- (P3);
	\draw[red] (P1) -- (P4);
	\draw[red] (P1) -- (290:1.2);
	\draw[red] (P1) -- (250:1.2);
	\draw[dotted] (.2,0) to[out=240,in=300] (-.2,0);
	\draw (2.2,0) node {\scalebox{1.4}{$\Leftrightarrow$}};
\end{tikzpicture}  
\begin{xy} 0;<1pt,0pt>:<0pt,1pt>::
	(25,40) *+{\langle 13\rangle} ="0",
	(75,40) *+{\color{white} \langle 14\rangle} ="1",
	(125,40) *+{\langle 1\ n-1 \rangle} ="2",
	(0,70) *+{\framebox[5ex]{$\langle 12\rangle$}} ="3",
	(25,0) *+{\framebox[5ex]{$\langle 23\rangle$}} ="4",
	(75,0) *+{\color{white} \framebox[5ex]{$\langle 34\rangle$}} ="5",
	(125,0) *+{\framebox[14ex]{$\langle n-2 \ n-1\rangle$}} ="6",
	(190,0) *+{\framebox[10ex]{$\langle n-1 \ n\rangle$}} ="7",
	(190,40) *+{\framebox[5ex]{$\langle 1n \rangle$}} ="8",
	(75,40) *+{\dots} ="01",
	(75,0) *+{\dots} ="05",
	"0", {\ar"1"},
	"1", {\ar"2"},
	"3", {\ar"0"},
	"0", {\ar"4"},
	"5", {\ar"0"},
	"1", {\ar"5"},
	"6", {\ar"1"},
	"2", {\ar"6"},
	"7", {\ar"2"},
	"2", {\ar"8"},
\end{xy}
\end{gathered}.
\end{equation}
Here one sees that $\Gr(2,n) \simeq A_{n-3}$.

For $\Gr(k>2,n)$, there is no longer a simple connection with triangulations or Dynkin diagrams. However, there exists a generalization of eq.~(\ref{eq:g2n-seed}) valid for all $\Gr(k,n)$~\cite{1088.22009}:
\begin{equation}\label{eq:gkn-seed}
\begin{gathered}
\begin{xy} 0;<-.5pt,0pt>:<0pt,-.5pt>::
	(-100,0) *+{\framebox[10ex]{$\ket{1,\ldots,k}$}} ="-1",
	(0,0) *+{f_{1l}} ="0",
	(75,0) *+{\color{white} f_{00}} ="1",
	(150,0) *+{f_{13}} ="2",
	(225,0) *+{f_{12}} ="3",
	(300,0) *+{\framebox[5ex]{$f_{11}$}} ="4",
	(0,75) *+{f_{2l}} ="5",
	(75,75) *+{\color{white} f_{00}} ="6",
	(150,75) *+{f_{23}} ="7",
	(225,75) *+{f_{22}} ="8",
	(300,75) *+{\framebox[5ex]{$f_{21}$}} ="9",
	(0,150) *+{\color{white} f_{00}} ="10",
	(75,150) *+{\color{white} f_{00}} ="11",
	(150,150) *+{\color{white} f_{00}} ="12",
	(225,150) *+{\color{white} f_{00}} ="13",
	(300,150) *+{\color{white} f_{00}} ="14",
	(0,225) *+{\framebox[5ex]{$f_{kl}$}} ="15",
	(75,225) *+{\color{white} f_{00}} ="16",
	(150,225) *+{\framebox[5ex]{$f_{k3}$}} ="17",
	(225,225) *+{\framebox[5ex]{$f_{k2}$}} ="18",
	(300,225) *+{\framebox[5ex]{$f_{k1}$}} ="19",
	(75,0) *+{\cdots} ="01",
	(75,75) *+{\cdots} ="06",
	(0,144) *+{\vdots} ="010",
	(75,144) *+{\ddots} ="011",
	(150,144) *+{\vdots} ="012",
	(225,144) *+{\vdots} ="013",
	(300,144) *+{\vdots} ="014",
	(75,225) *+{\cdots} ="016",
	"-1", {\ar"0"},
	"0", {\ar"1"},
	"0", {\ar"5"},
	"6", {\ar"0"},
	"1", {\ar"2"},
	"2", {\ar"3"},
	"2", {\ar"7"},
	"8", {\ar"2"},
	"3", {\ar"4"},
	"3", {\ar"8"},
	"9", {\ar"3"},
	"5", {\ar"6"},
	"6", {\ar"7"},
	"7", {\ar"8"},
	"8", {\ar"9"},
	"13", {\ar"7"},
	"7", {\ar"12"},
	"8", {\ar"13"},
	"5", {\ar"10"},
	"14", {\ar"8"},
	"10", {\ar"15"},
	"12", {\ar"17"},
	"19", {\ar"13"},
	"18", {\ar"12"},
	"13", {\ar"18"},
	"17", {\ar"11"},
	"16", {\ar"10"},
	"13", {\ar"14"},
	"12", {\ar"13"},
	"11", {\ar"12"},
	"10", {\ar"11"},
	"7", {\ar"1"},
	"11", {\ar"5"},
	"12", {\ar"6"},
	"1", {\ar"6"},
	"6", {\ar"11"},
	"11", {\ar"16"},
\end{xy}
\end{gathered} ,
\end{equation}
where $l=n-k$ and 
\begin{equation}
  f_{i j} =
  \begin{cases}
    \langle i+1, \dotsc, k, k+j, \dotsc, i+j+k-1\rangle, \qquad &i \leq l-j+1,\\
    \langle 1, \dotsc, i+j-l-1, i+1, \dotsc, k, k+j, \dotsc, n\rangle, \qquad &i >l-j+1.
  \end{cases}
\end{equation}
(Note that evaluating the above expression for $k=2$ will not directly give~\eqref{eq:g2n-seed}; the two are equivalent after cyclically rotating the indices in~\eqref{eq:g2n-seed} and flipping the direction of the arrows.) The cluster algebra on $\Gr(k,n)$ is therefore of rank $(n-k-1)(k-1)$, i.e.~the number of mutable nodes in~\eqref{eq:gkn-seed}. 

The cluster algebra on $\Gr(4,n)$ naturally appears in planar $\mathcal{N}=4$ sYM theory, where it parametrizes the space of $n$-particle kinematics. To make this connection, one first decomposes the external momenta $p_i$ (which are endowed with a natural ordering in the planar limit) into a set of spinors $\lambda_i$, $\tilde \lambda_i$ or into a set of dual coordinates $x_i$ as
\begin{equation}
p^\mu_i \sigma_\mu^{\alpha \dot \alpha} = \lambda_i^\alpha \tilde \lambda_i^{\dot \alpha} = x_i^{\alpha \dot \alpha} - x_{i+1}^{\alpha \dot \alpha} \, , \label{eq:dual_coordiantes}
\end{equation}
where $x_{n+1}^{\alpha \dot \alpha} \equiv x_1^{\alpha \dot \alpha}$ (for more background on these spinors and dual coordinates, see for example~\cite{Dixon:2013uaa,Elvang:2013cua}). The dual coordinates describe the cusps of a light-like Wilson loop dual to the amplitude; as a result, the original amplitude respects an additional superconformal symmetry that is associated with these dual coordinates (up to an anomaly associated with the cusps of the Wilson loop, which is accounted for by the BDS ansatz)~\cite{Bern:2005iz,Drummond:2007au,Bern:2008ap,Drummond:2008aq,Drummond:2006rz,Bern:2006ew,Bern:2007ct,Alday:2007hr,Drummond:2008vq}. In terms of the quantities in~\eqref{eq:dual_coordiantes}, we can define momentum twistors 
\begin{equation}
Z^R_i = (\lambda_i^\alpha, x_i^{\beta \dot \alpha} \lambda_{i \beta}) \, ,
\end{equation}
where $R = (\alpha, \dot \alpha)$ is an $SU(2,2)$ index. Momentum twistors are invariant under the little group, which acts as an overall rescaling $Z_i^R \rightarrow  t_i Z_i^R$, and as such represent points in $\mathbb{CP}^3$. 

If we assemble these momentum twistors into a $4 \times n$ matrix in which the $i^\text{th}$ column corresponds to the four $SU(2,2)$ components of $Z_i^R$, invariance under the dual conformal group becomes invariance under $\text{SL}(4)$. The overall rescaling symmetry of one of the momentum twistors can be combined with this $\text{SL}(4)$ invariance to identify this matrix as a point in the (not necessarily positive) Grassmannian $\Gr(4,n)$, modulo the rescaling invariance of the remaining $n-1$ columns. Thus, the kinematic data of an $n$-point scattering process is encoded in a momentum twistor matrix
\begin{equation}
Z \in \Gr(4,n)/\text{GL}(1)^{n-1}. \label{eq:gr4n_momentum_twistor}
\end{equation}
For more details regarding this correspondence, see~\cite{ArkaniHamed:2012nw,Golden:2013xva}. 

To relate the dual-conformal invariants encoded in $Z_n$ to more familiar kinematic quantities, we can translate the (cyclically ordered) Mandelstam invariants into squared differences of dual coordinates,
\begin{equation} 
s_{i,\dots,j-1} \equiv (p_i + \dots p_{j-1})^2= \text{det}(x_i^{\alpha \dot \alpha} - x_j^{\alpha \dot \alpha}) \equiv x_{ij}^2. \label{eq:mandelstam_dual_coord}
\end{equation}
Dual conformal invariants can be constructed out of these objects by putting together combinations that are invariant under the dual conformal inversion generator, which acts on these coordinates as
\begin{equation}
I( x_i^{\alpha \dot \alpha}) = \frac{x_i^{\alpha \dot \alpha}}{x_i^2}, \quad  I(x_{ij}^2) = \frac{x_{ij}^2}{x_i^2 x_j^2}.
\end{equation}
Thus, (regulated) amplitudes in this theory depend only on ratios of squared differences in which the same dual indices appear in both the numerator and denominator. The quantities $x_{ij}^2$ can be translated into momentum twistors using the relation
\begin{equation}
x_{ij}^2 = \frac{\text{det}(Z_{i-1} Z_i Z_{j-1} Z_j)}{(\epsilon_{\alpha \beta} \lambda^\alpha_{i-1} \lambda^\beta_i) (\epsilon_{\gamma \delta} \lambda^\gamma_{j-1} \lambda^\delta_j)},
\end{equation}
where $\epsilon_{\alpha \beta}$ is the Levi-Civita tensor. In dual-conformally invariant quantities, the spinor products $\epsilon_{\alpha \beta} \lambda^\alpha_{i-1} \lambda^\beta_i$ all cancel, leaving only determinants of four-tuples of momentum twistors. These are just minors of the momentum twistor matrix~\eqref{eq:gr4n_momentum_twistor}, which we recognize as the cluster \acoords  
\begin{equation}
\ket{i j k l} = \text{det}(Z_i Z_j Z_k Z_l). \label{eq:def_four_bracket}
\end{equation}
Note that the two-particle Mandelstams $s_{i,i+1}$ correspond to the frozen nodes of~\eqref{eq:gkn-seed}, while higher-particle Mandelstams and more general (polynomials of) Pl\"ucker coordinates can appear as mutable nodes.

By construction, the \xcoords\ on $\Gr(4,n)$ derived from the seed~\eqref{eq:gkn-seed} respect dual-conformal invariance. Both mutation rules~\eqref{eq:a-coord-mutation} and~\eqref{eq:x-coord-mutation} preserve this property (and commute with the translation~\eqref{eq:x_from_a_coordinates}), ensuring that all \xcoords\ are dual conformal invariants. Such invariants cannot be formed in four- or five-particle kinematics, due to an insufficient number of non-lightlike separated points (since we are in massless kinematics, $x_{ii+1}^2 = 0$ for all $i$). This fact shows up in the seed~\eqref{eq:gkn-seed} as $\Gr(4,n<6)$ having no mutable nodes (and therefore no \xcoords). For $n>5$, there are $3(n-5)$ mutable nodes in $\Gr(4,n)$, matching the number of algebraically independent dual conformal invariants that can be formed out of $n$ massless particles. 

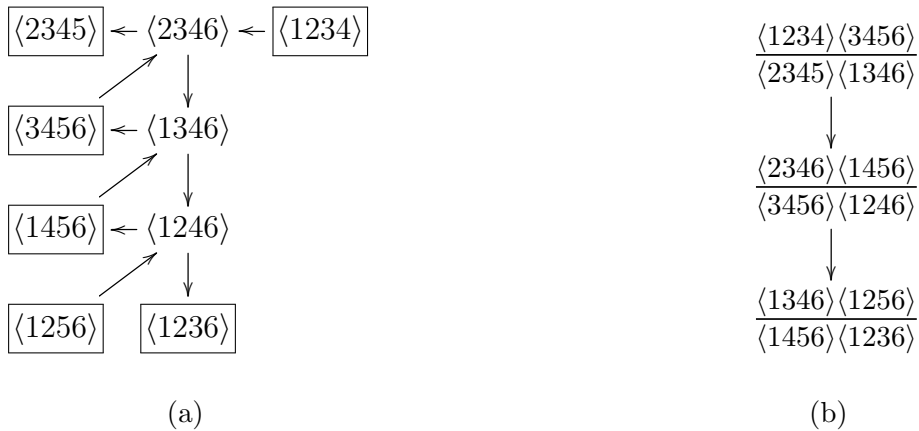
\begin{figure}[t]
\centering
\begin{subfigure}[b]{0.45\textwidth}
\begin{equation*}
\begin{gathered}
\begin{xy} 0;<-.5pt,0pt>:<0pt,-.5pt>::
         (-100,0) *+{\framebox[7ex]{$\ket{1234}$}} ="0",
	(0,0) *+{\ket{2346}} ="1",
	(100,0) *+{\framebox[7ex]{$\ket{2345}$}} ="2",
	(0,75) *+{\ket{1346}} ="3",
	(100,75) *+{\framebox[7ex]{$\ket{3456}$}} ="4",
	(0,150) *+{\ket{1246}} ="5",
	(100,150) *+{\framebox[7ex]{$\ket{1456}$}} ="6",
	(0,225) *+{\framebox[7ex]{$\ket{1236}$}} ="7",
	(100,225) *+{\framebox[7ex]{$\ket{1256}$}} ="8",
	"0", {\ar"1"},
	"1", {\ar"2"},
	"3", {\ar"4"},
	"5", {\ar"6"},
	"1", {\ar"3"},
	"3", {\ar"5"},
	"5", {\ar"7"},
	"4", {\ar"1"},
	"6", {\ar"3"},
	"8", {\ar"5"},
\end{xy}
\end{gathered} 
\end{equation*}
\caption{} \label{fig:g46-a-seed}
\end{subfigure}
\hspace*{\fill} 
\begin{subfigure}[b]{0.45\textwidth}
\begin{equation*}
\begin{gathered}
\begin{xy} 0;<-.5pt,0pt>:<0pt,-.5pt>::
	(0,0) *+\txt{\fontsize{16pt}{16pt} $\frac{\ket{1234}\ket{3456}}{\ket{2345}\ket{1346}}$} ="1",
	(0,100) *+\txt{\fontsize{16pt}{16pt} $\frac{\ket{2346}\ket{1456}}{\ket{3456}\ket{1246}}$} ="3",
	(0,200) *+\txt{\fontsize{16pt}{16pt} $\frac{\ket{1346}\ket{1256}}{\ket{1456}\ket{1236}}$} ="5",
	"1", {\ar"3"},
	"3", {\ar"5"},
\end{xy}
\end{gathered} 
\end{equation*}
\caption{} \label{fig:g46-x-seed}
\end{subfigure}
\caption{The \acoord\ seed quiver (a) and \xcoord\ seed quiver (b) for $\Gr(4,6)$.} 
\label{fig:g46-seed}
\end{figure}

The \acoord\ and \xcoord\ seed clusters of the first nontrivial example, $\Gr(4,6)$, are shown in figure~\ref{fig:g46-seed}. As discussed above, the three \xcoords\ in this cluster furnish us with a chart that covers the space of (dual-conformally invariant) six-particle kinematics. Moreover, we can generate new charts by mutation---every \xcoord\ cluster of $\Gr(4,n)$ provides a valid chart for $n$-particle kinematics. As explored in great depth in~\cite{ArkaniHamed:2012nw}, these charts are especially well suited to describing the boundaries of the positive Grassmannian $\Gr_+(4,n)$, where the integrands of $n$-particle amplitudes can develop physical singularities. In particular, every such boundary occurs at the vanishing locus of an \acoord\ of $\Gr(4,n)$, which implies it also occurs at the vanishing locus of some set of \xcoords.  

This fact is especially propitious for loop-level amplitudes (and integrals) that only have branch points on the boundaries of the positive Grassmannian. In such cases, the symbol alphabet encoding the polylogarithmic part of these amplitudes is naturally given in terms of cluster coordinates. We defer discussion of the coaction and symbol alphabets to section~\ref{sec:coproduct}, but here note that it is multiplicative independence, rather than algebraic independence, that is relevant in the context of symbol alphabets. Thus, while it is not possible to realize all boundaries of the positive Grassmannian as the vanishing loci of either type of cluster coordinate in a single chart~\cite{ArkaniHamed:2012nw}, all boundaries are exposed as the vanishing of some symbol letter if cluster \acoords\ or \xcoords\ on $\Gr(4,n)$ are adopted as a symbol alphabet.

While amplitudes in planar ${\cal N}=4$ are not generically expected to have this property (and indeed, certain Feynman integrals have been computed that do not~\cite{Bourjaily:2018aeq,Henn:2018cdp}), an infinite class of amplitudes do---namely, all two-loop MHV amplitudes~\cite{CaronHuot:2011ky}, and all six- and seven-particle amplitudes computed to date~\cite{CaronHuot:2011kk,Dixon:2014iba,Drummond:2014ffa,Dixon:2015iva,Caron-Huot:2016owq,Dixon:2016nkn,Dixon:2016apl,Drummond:2018dfd,Caron-Huot:2019vjl,Caron-Huot:2019bsq}. The significance of this property is illustrated by the two-loop, six-particle remainder function, which encodes the MHV amplitude. Namely, this function can be put in the form
\begin{equation} \label{eq:R26cobracket}
	R^{(2)}_6 = -\sum_{\text{cyclic}} \left[ \text{Li}_4\left(-\frac{\langle 1234 \rangle \langle 3456 \rangle}{\langle 2345 \rangle \langle 1346 \rangle}\right) - \frac{1}{4} \text{Li}_4 \left(-\frac{\langle 1234 \rangle \langle 1456 \rangle}{\langle 1246 \rangle \langle 1345 \rangle}\right) \right] + \dots,
\end{equation}
where the cyclic sum is over all rotations of the four-bracket indices $i \rightarrow i+j$ for $0\leq j <6$, and the dots indicate this equality only holds up to products of lower-weight polylogarithms. This projection is well-defined and will be introduced, along with the $n$-particle remainder function, in section~\ref{sec:cluster_polylog_MHV_review}. Here we just emphasize the simplicity of this expression, which takes the form of classical polylogarithms with negative \xcoord\ arguments. (In particular, the argument of the first polylogarithm is the top node in figure~\ref{fig:g46-x-seed}, while the argument of the second polylogarithm appears in the cluster generated by mutating on that node.) Moreover, the part of the expression we have dropped in~\eqref{eq:R26cobracket} can also be expressed entirely in terms of products of classical polylogarithms with negative \xcoord\ arguments~\cite{Golden:2014xqf}. 

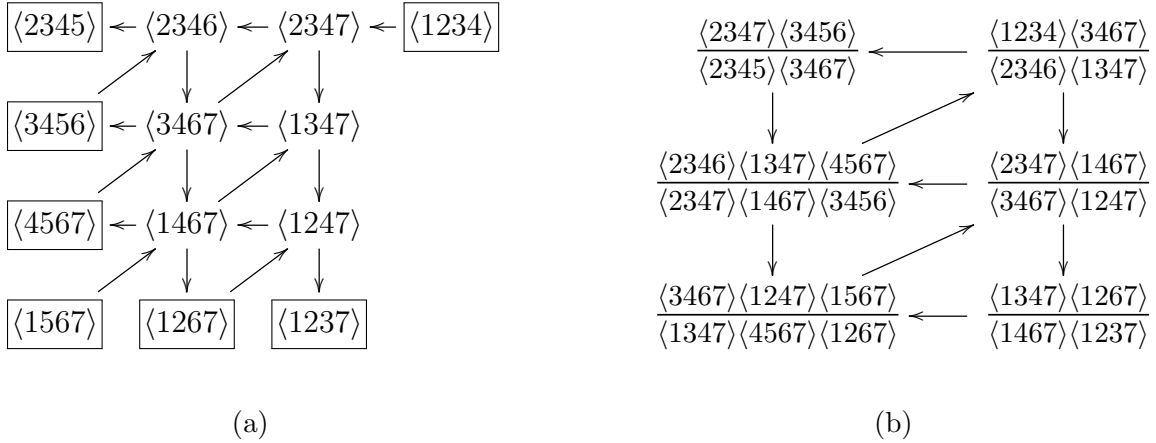
\begin{figure}
\centering
\begin{subfigure}[b]{0.45\textwidth}
\begin{equation*}
\begin{gathered}
\begin{xy} 0;<-.5pt,0pt>:<0pt,-.5pt>::
         (-100,0) *+{\framebox[7ex]{$\ket{1234}$}} ="0",
	(0,0) *+{\ket{2347}} ="1",
	(100,0) *+{\ket{2346}} ="2",
	(200,0) *+{\framebox[7ex]{$\ket{2345}$}} ="3",
	(0,75) *+{\ket{1347}} ="4",
	(100,75) *+{\ket{3467}} ="5",
	(200,75) *+{\framebox[7ex]{$\ket{3456}$}} ="6",
	(0,150) *+{\ket{1247}} ="7",
	(100,150) *+{\ket{1467}} ="8",
	(200,150) *+{\framebox[7ex]{$\ket{4567}$}} ="9",
	(0,225) *+{\framebox[7ex]{$\ket{1237}$}} ="10",
	(100,225) *+{\framebox[7ex]{$\ket{1267}$}} ="11",
	(200,225) *+{\framebox[7ex]{$\ket{1567}$}} ="12",
	"0", {\ar"1"},
	"1", {\ar"2"},
	"2", {\ar"3"},
	"4", {\ar"5"},
	"5", {\ar"6"},
	"7", {\ar"8"},
	"8", {\ar"9"},
	"1", {\ar"4"},
	"2", {\ar"5"},
	"4", {\ar"7"},
	"5", {\ar"8"},
	"7", {\ar"10"},
	"8", {\ar"11"},
	"5", {\ar"1"},
	"6", {\ar"2"},
	"8", {\ar"4"},
	"9", {\ar"5"},
	"11", {\ar"7"},
	"12", {\ar"8"},
\end{xy}
\end{gathered} 
\end{equation*} 
\caption{} \label{fig:g47-a-seed}
\end{subfigure}
\hspace*{\fill} 
\begin{subfigure}[b]{0.45\textwidth}
\begin{equation*}
\begin{gathered}
\begin{xy} 0;<-.5pt,0pt>:<0pt,-.5pt>::
	(0,0) *+\txt{\fontsize{16pt}{16pt} $\frac{\ket{1234}\ket{3467}}{\ket{2346}\ket{1347}}$} ="1",
	(220,0) *+\txt{\fontsize{16pt}{16pt} $\frac{\ket{2347}\ket{3456}}{\ket{2345}\ket{3467}}$} ="2",
	(0,100) *+\txt{\fontsize{16pt}{16pt} $\frac{\ket{2347}\ket{1467}}{\ket{3467}\ket{1247}}$} ="4",
	(220,100) *+\txt{\fontsize{16pt}{16pt} $\frac{\ket{2346}\ket{1347}\ket{4567}}{\ket{2347}\ket{1467}\ket{3456}}$} ="5",
	(0,200) *+\txt{\fontsize{16pt}{16pt} $\frac{\ket{1347}\ket{1267}}{\ket{1467}\ket{1237}}$} ="7",
	(220,200) *+\txt{\fontsize{16pt}{16pt} $\frac{\ket{3467}\ket{1247}\ket{1567}}{\ket{1347}\ket{4567}\ket{1267}}$} ="8",
	"1", {\ar"2"},
	"4", {\ar"5"},
	"7", {\ar"8"},
	"1", {\ar"4"},
	"2", {\ar"5"},
	"4", {\ar"7"},
	"5", {\ar"8"},
	"5", {\ar"1"},
	"8", {\ar"4"},
\end{xy}
\end{gathered} 
\end{equation*} 
\caption{} \label{fig:g47-x-seed}
\end{subfigure}
\caption{The \acoord\ seed quiver (a) and \xcoord\ seed quiver (b) for $\Gr(4,7)$.} 
\label{fig:g47-seed}
\end{figure}

This surprising property---of being expressible as polylogarithms with negative \xcoord\ arguments---is enjoyed by the two-loop MHV amplitude at all $n$. However, for $n>6$ these amplitudes have a nonclassical component, so generalized polylogarithms with negative \xcoord\ arguments also appear~\cite{Golden:2013xva}. Although this component represents the mathematically most complicated part of the remainder function, it was shown in~\cite{Golden:2014xqa} that it is decomposable into building blocks related to the $A_2$ and $A_3$ subalgebras of $\Gr(4,n)$. This allows the all-$n$ symbol computed in~\cite{CaronHuot:2011ky} to be systematically upgraded to a function, as was done for seven particles in~\cite{Golden:2014xqf}. In the later sections of this paper, we demonstrate the existence of further subalgebra structure in the nonclassical part of the seven-particle MHV amplitude (leaving higher-point kinematics to future work~\cite{cluster_subalgebras_ii}). The \acoord\ and \xcoord\ seeds for the cluster algebra relevant to seven-particle scattering, $\Gr(4,7)$, are presented in figure~\ref{fig:g47-seed}.

Before turning to the remaining aspects of cluster algebras that we wish to develop, we note that plabic graphs---which are dual to the clusters we've been describing---encode a great deal more about planar ${\cal N}=4$ than we have had reason to touch on. We refer interested readers to the exposition of this rich structure given in~\cite{ArkaniHamed:2012nw}.

\subsection{Finite Cluster Algebras}\label{sec:finite-algebras}

The procedure of writing down an oriented quiver, dressing it with coordinates, and iteratively mutating on all non-frozen nodes using either the $\a$-coordinate or $\x$-coordinate mutation rule will always produce a cluster algebra. However, generic quivers give rise to exceedingly complicated cluster algebras---in fact, for a wide class of seeds, mutation will generate an infinite numbers of clusters. For the remainder of this paper we will mostly restrict our attention to finite cluster algebras, leaving the consideration of infinite algebras to future work. 

Fortunately, all finite cluster algebras were classified in~\cite{1054.17024}. In particular, it was shown that a cluster algebra is of finite type if and only if the mutable part of at least one of its clusters takes the form of an oriented, simply-laced Dynkin diagram: $A_n$, $D_n$, or $E_{n\le8}$.  As we will primarily be interested in subalgebras of the cluster algebra on $\Gr(4,7)$, we here focus on the cases where $n < 6$ (and on the case of $E_6 \simeq \Gr(4,7)$ itself). 

As mentioned above, cluster algebras of type $A_n$ can be generated by the seed
\begin{equation}\label{def:An}
  x_1\to x_2\to \ldots \to x_n \ ,
\end{equation}
which corresponds to the cluster algebra on $\Gr(2,n{+}3)$. Each of the clusters in these algebras can be though as triangulating an $(n+3)$-gon, where the \acoords\ correspond to chords and the \xcoords\ to quadrilateral faces. This makes the counting easy: the number of clusters for $A_n$ is given by the Catalan number $C(n+1)$, the number of distinct (mutable) \acoords\ is $\binom{n+3}{2}-n$, and the number of distinct \xcoords\ is $2\binom{n+3}{4}$. Any smaller polygon embedded into the $(n+3)$-gon gives rise to a subalgebra; for example, there are $56=\binom{8}{5}$ pentagonal embeddings in an octagon, so there are 56 $A_2$ subalgebras in $A_5$. 

Of particular interest is the cluster algebra generated by $A_3 \simeq \Gr(4,6)$, which describes six-particle scattering. By comparison with figure~\ref{fig:g46-x-seed}, we see that the \xcoords\ in the quiver~\eqref{def:An} act as coordinates on the space of momentum twistors, where they correspond to the functions
\begin{equation}\label{eq:a3-seed-def}
x_1 = \frac{\ket{1234}\ket{3456}}{\ket{2345}\ket{1346}}, \quad x_2 = \frac{\ket{2346}\ket{1456}}{\ket{3456}\ket{1246}}, \quad x_3 = \frac{\ket{1346}\ket{1256}}{\ket{1456}\ket{1236}}.
\end{equation}
More generally, any \xcoord\ in $\Gr(4,6)$ can be expressed in terms of the variables $x_1$, $x_2$, and $x_3$ by evaluating its four-brackets on the momentum twistor matrix
\begin{equation} \label{eq:a3_momentum_twistor_matrix}
Z_{A_3} = 
\begingroup
\setlength\arraycolsep{4pt}
\begin{pmatrix} 
1 & 0 & 0 & 0 & -1 & -1 \\
1 & 1 & 0 & 0 & x_1 & 0 \\
0 & 1 & 1 & 0 & -x_1 x_2 & 0 \\
0 & 0 & 1 & 1 & x_1 x_2 x_3 & 0
\end{pmatrix}
\endgroup .
\end{equation}
The chief advantage of working directly in terms of cluster \xcoords\ such as $x_1$, $x_2$, and $x_3$ is that they trivialize all Pl\"ucker relations. Furthermore, cluster coordinates rationalize many of the square roots that appear when amplitudes and integrals are expressed in terms of dual-conformally-invariant cross ratios~\cite{Bourjaily:2018aeq}. For instance, in this chart the dual conformal cross ratios commonly used to express the six-particle amplitude evaluate to
\begin{align} \label{eq:uvw_in_x}
u &= \frac{\ket{6123}\ket{3456}}{\ket{6134}\ket{2356}} = \frac{1}{1+x_2+x_2 x_3}, \\ 
v &= \frac{\ket{1234} \ket{4561}}{\ket{1245}\ket{3461}} =\frac{x_1 x_2}{1+x_1+x_1 x_2}, \\ 
w &= \frac{\ket{2345} \ket{5612}}{\ket{2356}\ket{4512}} =\frac{x_2 x_3}{(1+x_1+x_1 x_2)(1+x_2+x_2 x_3)},
\end{align}
which rationalizes the well-known square root that appears in these cross ratios 
\begin{equation}
\sqrt{(1 - u - v - w)^2 - 4 u v w} = \frac{x_2 \left(1-x_1 x_3\right)}{\left(1+x_1+x_1
   x_2\right) \left(1+x_2+x_2 x_3\right)} .
\end{equation}
Note that the cluster coordinate expressions~\eqref{eq:uvw_in_x} are rotated compared to those given elsewhere in the literature (for example,~\cite{Golden:2013xva,Parker:2015cia}) even though both arise from an \xcoord\ seed of the form~\eqref{def:An}; this reflects a differing convention for the seed of $\Gr(k,n)$.

The first nondegenerate Dynkin diagram of type $D_n$ is $D_4$, corresponding to the seed quiver
\begin{equation} \label{eq:D4_quiver}
    \begin{gathered}
    \begin{xy} 0;<1pt,0pt>:<0pt,-1pt>::
      (0,20) *+{x_1} ="1",
      (30,20) *+{x_2} ="2",
      (60,0) *+{x_3} ="3",
      (60,40) *+{x_4} ="4",
      "1", {\ar"2"},
      "2", {\ar"3"},
      "2", {\ar"4"},
    \end{xy}
    \end{gathered} .
\end{equation}
(Note here that the variables $x_1$, $x_2$, and $x_3$ are not the same as those defined by \eqref{eq:a3-seed-def}; we apologize for using the variables $x_i$ ubiquitously, but hope their meaning is always clear from context.) This seed turns out to generate the same cluster algebra as $\Gr(3,6)$; in particular, starting from the seed in~\eqref{eq:gkn-seed} and mutating on the nodes initially labeled by $f_{13}$ and then $f_{23}$, one arrives at the \xcoord\ quiver~\eqref{eq:D4_quiver}, where
\begin{equation}\label{eq:d4-seed-def}
x_1 = \frac{\ket{123}\ket{345}}{\ket{234}\ket{135}}, \quad x_2 = \frac{\ket{156}\ket{235}}{\ket{125}\ket{356}}, \quad x_3 = \frac{\ket{135}\ket{456}}{\ket{156}\ket{345}}, \quad x_4 = \frac{\ket{126}\ket{135}}{\ket{123}\ket{156}}.
\end{equation}
The corresponding momentum twistor matrix is given by
\begin{equation}
Z_{D_4} = 
\begingroup
\setlength\arraycolsep{4pt}
\begin{pmatrix} 
 1 & 0 & 0 & x_2 & x_2 & 1+x_2+x_2 x_4 \\
 0 & 1 & 0 & -(1+x_1) & -1 & - (1+x_4) \\
 0 & 0 & 1 & 1 + x_1 + x_1 x_2 + x_1 x_2 x_3 & 1 & x_4
 \end{pmatrix}
\endgroup .
\end{equation}
Conversely, the cluster algebra generated by $D_5$,
\begin{equation}\label{def:D5}
    \begin{gathered}
    \begin{xy} 0;<1pt,0pt>:<0pt,-1pt>::
      (0,20) *+{x_1} ="1",
      (30,20) *+{x_2} ="2",
      (60,20) *+{x_3} ="3",
      (90,0) *+{x_4} ="4",
      (90,40) *+{x_5} ="5",
      "1", {\ar"2"},
      "2", {\ar"3"},
      "3", {\ar"4"},
      "3", {\ar"5"},
    \end{xy}
    \end{gathered} ,
\end{equation}
is not equivalent to the cluster algebra on (the top cell of) any Grassmannian. However, it appears as a subalgebra of any $\Gr(k,n)$ with rank greater than five. The $D_4$ cluster algebra consists of 50 clusters, 16 \acoords, and 104 \xcoords, while $D_5$ has 182 clusters, 25 \acoords, and 260 \xcoords. More generally, there are $\frac{n}{3}(n-1)(n^2+4n-6)$ \xcoords\ in cluster algebras of type $D_n$~\cite{2018arXiv180302492S}. 

Finally, the cluster algebra $E_6$ is generated by the quiver
\begin{equation}\label{def:E6}
    \begin{gathered}
    \begin{xy} 0;<1pt,0pt>:<0pt,-1pt>::
      (0,0) *+{x_1} ="1",
      (30,0) *+{x_2} ="2",
      (60,0) *+{x_3} ="3",
      (60,-25) *+{x_4} ="4",
      (90,0) *+{x_5} ="5",
      (120,0) *+{x_6} ="6",
      "1", {\ar"2"},
      "2", {\ar"3"},
      "3", {\ar"4"},
      "5", {\ar"3"},
      "6", {\ar"5"},
    \end{xy}
    \end{gathered} .
\end{equation}
This cluster algebra is equivalent to $\Gr(4,7)$, as can be seen by mutating the seed~\eqref{eq:gkn-seed} on nodes $f_{12}$, $f_{13}$, $f_{23}$, $f_{12}$, $f_{22}$, and then $f_{32}$. By comparison with~\eqref{def:E6}, we then have 
\begin{align} \label{eq:E6_x_def}
x_1 &= \frac{\ket{1234}\ket{1267}}{\ket{1237} \ket{1246}}, \hspace{13.03ex} \quad \quad 
x_2 = - \frac{\ket{1247} \ket{3456}}{\ket{4(12)(35)(67)}}, \quad \nonumber \\
x_3 &= \frac{\ket{1246} \ket{5(12)(34)(67)}}{\ket{1245} \ket{1267} \ket{3456}}, \hspace{4.12ex} \quad \quad 
x_4 = -\frac{\ket{4(12)(35)(67)}}{\ket{1234} \ket{4567}}, \quad \\
x_5 &= -\frac{\ket{1267} \ket{1345}\ket{4567}}{\ket{1567}\ket{4(12)(35)(67)}}, \quad \quad \quad 
x_6 = \frac{\ket{1567}\ket{2345}}{\ket{5(12)(34)(67)}}, \quad \nonumber
\end{align}
where we have made use of the notation
\begin{equation} \label{eq:twistor_intersection}
\ket{a(bc)(de)(fg)} \equiv \ket{abde}\ket{acfg}-\ket{abfg}\ket{acde}.
\end{equation}
Any \xcoord\ on $\Gr(4,7)$ can be expressed in terms of these \xcoords\ using the momentum twistor matrix 
\begin{equation}
Z_{E_6} = 
\begingroup
\setlength\arraycolsep{3pt}
\begin{pmatrix} 
 x_3 x_5 & x_3 x_5 x_6 & x_1 & 0 & 1 & 0 & -1 \\
 1+x_5 & 1 + x_6 + x_5 x_6 & -x_1 & 0 & -1+x_4 & x_4 & x_2 x_4 \\
 0 & 0 & x_1 & 0 & 1 & 1 & x_2 \\
 0 & 0 & -1 & 1 & 1+x_2+x_2 x_3 & 1 & 0 \\
\end{pmatrix} .
\endgroup
\end{equation}
The cluster polytope of $\Gr(4,7) \simeq E_6$ was displayed in figure~\ref{fig:e6-poly}; it contains 833 clusters, 42 \acoords, and 770 \xcoords. The subalgebras of $E_6$, as well as those of its subalgebras, are tabulated in appendix~\ref{appendix:subalgebras}.

\subsection{Cluster Automorphisms}\label{sec:automorphisms}

Cluster algebras come equipped with an automorphism group that maps the set of cluster coordinates (but not necessarily the set of clusters) back to itself. We introduce here only what we need to elucidate the automorphisms of the cluster algebras introduced in the last section, and refer the interested reader to~\cite{Chang:2015} for a more thorough mathematical introduction. Note that we describe automorphisms in terms of \xcoords, whereas~\cite{Chang:2015} works in the \acoord\ language. 

The simplest example of a cluster automorphism is what we call a direct automorphism. Let $\a$ be a cluster algebra equipped with a mutation rule $\mu(x_i, \bf{X})$ that mutates the cluster $\bf{X}$ on node $x_i$. Then, we can define:
\begin{quote}
{\bf Direct Automorphism}: The map $f: \a \to \a$ is a direct automorphism of $\a$ if 
\vspace{-.2cm}
 \begin{itemize}
 \item[(i)] for every cluster $\mathbf{X}$ of $\a$, $f(\mathbf{X})$ is also a cluster of $\a$, 
 \item[(ii)] $f$ respects mutations, i.e. $f(\mu(x_i,\mathbf{X})) = \mu(f(x_i),f(\mathbf{X}))$.
 \end{itemize}
\end{quote}
An example of a direct automorphism on $A_2$ is given by
\begin{equation}
  \sigma_{A_2}:\quad \mathcal{X}_i \to \mathcal{X}_{i+1},
\end{equation}
where we are using the coordinates introduced in ~\eqref{def:a2-xcoords}. This automorphism cycles the five clusters $1/\x_i\to \x_{i+1}$ amongst themselves. The action of this automorphism can also be recast as
\begin{equation} \label{eq:a2_sigma_x}
  \sigma_{A_2}:\quad x_1\to \frac{1}{x_2},~~ x_2\to x_1(1+x_2),
\end{equation}
using the \xcoords\ $x_1$ and $x_2$ that appear in~\eqref{def:An}. This is of course equivalent to the cyclic symmetry of the pentagon. 

Cluster algebras are also endowed with what we call indirect automorphisms, which respect mutations but do not map the set of clusters back to itself. Instead, indirect automorphisms map the clusters in $\a$ to clusters in $\a'$, where $\a'$ is constructed from $\a$ by multiplicatively inverting all cluster \xcoords\ and reversing the direction of all quiver arrows. Then we have:
\begin{quote}
{\bf Indirect Automorphism}: The map $f: \a \to \a'$ is an indirect automorphism if 
\vspace{-.2cm}
\begin{itemize}
  \item[(i)] for every cluster $\mathbf{X}$ of $\a$, $f(\mathbf{X})$ is a cluster of $\a'$ 
  \item[(ii)] $f$ respects mutations, i.e. $f(\mu(x_i,\mathbf{X})) = \mu(f(x_i),f(\mathbf{X}))$.
\end{itemize}
\end{quote}
$A_2$ is also equipped with an indirect automorphism generated by
\begin{equation} \label{eq:a2_tau_X}
  \tau_{A_2}:\quad \mathcal{X}_i \to \mathcal{X}_{6-i},
\end{equation}
where indices are understood to be mod 5. This can be recast in term of $x_1$ and $x_2$ as
\begin{equation} \label{eq:a2_tau_x}
  \tau_{A_2}:\quad  x_1 \to \frac{x_1 x_2}{1 + x_1}, ~~x_2 \to \frac{1 + x_1 + x_1 x_2}{x_2}.
\end{equation}
To see that this is an indirect automorphism, consider $\tau_{A_2}(1/\x_1 \to \x_2) = 1/\x_5 \to \x_4$. Inverting the cluster coordinates on the right hand side and reversing the arrow, we get back to $\x_5 \leftarrow 1/\x_4$, which was one of the original clusters of $A_2$. The operation generated by $\tau_{A_2}$ can be interpreted as the dihedral flip of the pentagon. 

It is useful to think of indirect automorphisms as generating a ``mirror'' or ``flipped'' version of the original cluster algebra, where the total collection of $\x$-coordinates is the same, but their Poisson bracket has flipped sign. The existence of this flip then can be seen as resulting from the arbitrary choice of overall sign for the exchange matrix $b_{ij}$; picking the other sign would have generated the same cluster-algebraic structure, but with different labels for the nodes. Indirect automorphisms capture the superficiality of this notation change.

The automorphisms $\sigma_{A_2}$ and $\tau_{A_2}$ generate the complete automorphism group for $A_2$, namely the dihedral group ${\mathfrak D}_5$ (we denote the dihedral group in this font throughout so as not to confuse with the Dynkin diagram $D_n$). More generally, cluster algebras of type $A_n$ have as their automorphism group the dihedral group ${\mathfrak D}_{n+3}$, which is generated by a cyclic generator
\begin{equation} \label{eq:def_An_automorphic_cycle}
  \sigma_{A_n}:\quad x_{k<n} \to \frac{x_{k+1}(1+x_{1,\ldots,k-1})}{1+x_{1,\ldots,k+1}},~~x_n\to\frac{1+x_{1,\ldots,n-1}}{\prod_{i=1}^n x_i} ,
\end{equation}
of length $n+3$ and a flip generator 
\begin{equation} \label{eq:def_An_automorphic_flip}
  \tau_{A_n}: \quad x_1 \to \frac{1}{x_n},~~x_2 \to \frac{1}{x_{n-1}},~\ldots~, \quad x_n\to\frac{1}{x_1}
\end{equation}
of length $2$. (Note that these definitions don't exactly match~\eqref{eq:a2_sigma_x} or~\eqref{eq:a2_tau_x} when $n$ is 2, but produce a pair of equally valid generators.) In $\sigma_{A_n}$ we have introduced the notation
\begin{equation} \label{eq:compound_x_def}
	x_{i_1,\ldots, i_k} \equiv \sum_{a=1}^k \prod_{b=1}^a x_{i_b} = x_{i_1}+x_{i_1}x_{i_2} + \ldots + x_{i_1}\cdots x_{i_k},
\end{equation}
which we will also use below. The operator $\sigma_{A_n}$ generates a direct automorphism while $\tau_{A_n}$ generates an indirect automorphism.

The cluster algebra $D_4$ has automorphism group ${\mathfrak D}_4\times S_3$. Both ${\mathfrak D}_4$ and $S_3$ come with a cyclic (direct automorphism) generator,
\vspace{.1cm}
\begin{align}
  \sigma^{({\mathfrak D}_4)}_{D_4}:\quad
    &x_1\to\frac{x_2}{1+x_{1,2}}, \quad
    x_2\to\frac{\left(1+x_1\right)x_1 x_2 x_3 x_4}{\left(1+x_{1,2,3}\right) \left(1+x_{1,2,4}\right)},  \nonumber \\
    &x_3\to\frac{1+x_{1,2}}{x_1 x_2 x_3},
    x_4\to\frac{1+x_{1,2}}{x_1 x_2 x_4}, \\[2ex]
  \sigma^{(S_3)}_{D_4}:\quad& 
    x_1\to \frac{1}{x_3}, \quad
    x_2\to \frac{x_1 x_2 \left(1+x_3\right)}{1+x_1}, \quad
    x_3\to x_4, \quad
    x_4\to \frac{1}{x_1} , \nonumber
\end{align}
where $\sigma^{({\mathfrak D}_4)}_{D_4}$ has length four, and $\sigma^{(S_3)}_{D_4}$ has length three. Then there are two flip generators
\vspace{.1cm}
\begin{equation}
\begin{split}
  \tau^{({\mathfrak D}_4)}_{D_4}:\quad& 
    x_2\to \frac{1+x_1}{x_1 x_2 \left(1+x_3\right) \left(1+x_4\right)}, \\[2ex]
  \tau^{(S_3)}_{D_4}:\quad& 
    x_3\to x_4,~~
    x_4\to x_3 , \\[1ex]
\end{split}  
\end{equation}
where $\tau^{({\mathfrak D}_4)}_{D_4}$ generates an indirect automorphism, and $\tau^{(S_3)}_{D_4}$ generates a direct automorphism.

The cluster algebras on $D_{n>4}$ with defining quiver
\begin{equation}
    \begin{gathered}
    \begin{xy} 0;<1pt,0pt>:<0pt,-1pt>::
      (0,20) *+{x_1} ="1",
      (30,20) *+{x_2} ="2",
      (60,20) *+{\ldots} ="3",
      (90,20) *+{x_{n-2}} ="4",
      (120,0) *+{x_{n-1}} ="5",
      (120,40) *+{x_{n}} ="6",
      "1", {\ar"2"},
      "2", {\ar"3"},
      "3", {\ar"4"},
      "4", {\ar"5"},
      "4", {\ar"6"},
    \end{xy}
    \end{gathered},
\end{equation}
have the automorphism group ${\mathfrak D}_n \times \mathbb{Z}_2$, with generators $\sigma_{D_n}$ (length $n$, direct), $\tau_{D_n}$ (length 2, indirect), and $\mathbb{Z}_{2,D_n}$ (length 2, direct). In the case of $D_5$, these generators can be chosen to be
\vspace{.1cm}
\begin{align}
  \sigma_{D_5}:\quad 
    &x_1\to \frac{x_2}{1+x_{1,2}},~~
    x_2\to \frac{(1+x_1) x_3}{1+x_{1,2,3}},~~
    x_3\to \frac{x_1 x_2 x_3 x_4 x_5 (1+x_{1,2})}{(1+x_{1,2,3,4}) (1+x_{1,2,3,5})}, \nonumber \\
    &x_4\to \frac{1+x_{1,2,3}}{x_1 x_2 x_3 x_4},~~
    x_5\to \frac{1+x_{1,2,3}}{x_1 x_2 x_3 x_5}, \nonumber \\[2ex]
  \tau_{D_5}:\quad 
    &x_2\to \frac{1+x_1}{x_1 x_2 (1+x_3 x_5+x_{3,4,5})},~~
    x_3\to \frac{x_3 x_4 x_5}{(1+x_{3,4}) (1+x_{3,5})},\\
    &x_4\to \frac{1+x_3 x_5+x_{3,4,5}}{x_4},~~
    x_5\to \frac{1+x_3 x_5+x_{3,4,5}}{x_5}, \nonumber \\[2ex]
    \mathbb{Z}_{2,D_n}:\quad &x_4 \to x_5,~~ x_5 \to x_4. \nonumber
\end{align}
More generally, for $D_n$ cluster algebras, the action of $\mathbb{Z}_2$ is always realized by the exchange $x_{n-1} \leftrightarrow x_n$. 

Finally, the automorphism group of $E_6 \simeq \Gr(4,7)$ is the dihedral group ${\mathfrak D}_{14}$. This group has generators $\sigma_{E_6}$ (length 7, direct), $\tau_{E_6}$ (length 2, indirect), and $\mathbb{Z}_{2,E_6}$ (length 2, direct). In the coordinates of the quiver~\eqref{def:E6}, these can be chosen to be 
\begin{align}
  \sigma_{E_6}:\quad 
    &x_1\to \frac{1}{x_6 (1+x_{5,3,4})},~~
    x_2\to \frac{1+x_{6,5,3,4}}{x_5 (1+x_{3,4})},~~
    x_3\to \frac{(1+x_{2,3,4}) (1+x_{5,3,4})}{x_3 (1+x_4)}, \nonumber \\
    &x_4\to \frac{1+x_{3,4}}{x_4},~~
    x_5\to \frac{1+x_{1,2,3,4}}{x_2 (1+x_{3,4})},~~
    x_6\to \frac{1}{x_1 (1+x_{2,3,4})},\nonumber \\[2ex]
  \tau_{E_6}:\quad
    &x_1\to \frac{x_5}{1+x_{6,5}},~~
    x_2\to (1+x_5) x_6,~~
    x_3\to \frac{(1+x_{1,2}) (1+x_{6,5})}{x_1 x_2 x_3 x_5 x_6 (1+x_4)},  \\
    &x_5\to x_1 (1+x_2),~~
    x_6\to \frac{x_2}{1+x_{1,2}}, \nonumber \\[2ex]
  \mathbb{Z}_{2,E_6}:\quad 
    &x_1\to x_6, \quad x_2\to x_5, \quad x_5\to x_2, \quad x_6\to x_1. \nonumber
\end{align}
In the language of $\Gr(4,7)$, these generators correspond to cycling momentum twistor indices $Z_i \to Z_{i+1}$, flipping momentum twistor indices $Z_i \to Z_{8-i}$, and parity conjugation. 

The polylogarithmic part of the $n$-particle MHV amplitude is invariant under parity transformations, as well as the dihedral group that represents Bose symmetry. These symmetries directly translate to automorphisms of the cluster algebra on $\Gr(4,n)$. However, it turns out the nonclassical part of these amplitudes can also be decomposed into building blocks that respect the automorphism group of certain subalgebras of $\Gr(4,n)$. Making this statement precise will be the focus of much of the remainder of this paper. 

\section{Cluster Polylogarithms and MHV Amplitudes} \label{sec:cluster_polylog_MHV_review}

The BDS ansatz captures the infrared structure of planar ${\cal N} = 4$ sYM to all orders in the coupling~\cite{Bern:2005iz}. In four- and five-particle kinematics it also furnishes the complete finite part of the amplitude, while for six or more particles it must be corrected by a finite dual-conformally invariant function~\cite{Drummond:2007au,Bern:2008ap,Drummond:2008aq}. In the case of the MHV amplitude, this correction is often computed in the form of the $n$-particle remainder function $R_n$, defined by
\begin{equation} \label{eq:remainder_function}
\mathcal{A}_n^{\rm MHV}\ =\ \mathcal{A}_n^{\rm BDS}  \times \exp(R_n)\,,
\end{equation}
where $\mathcal{A}_n^{\rm BDS}$ is the BDS ansatz for $n$ particles~\cite{Bern:2005iz}. Like the amplitude, the remainder function can be expanded in the coupling
\begin{equation}
R_n = g^4 R_n^{(2)} + g^6 R_n^{(3)} + g^8 R_n^{(4)} + \dots ,
\end{equation}
where $g^2 = \frac{g_{\text{YM}}^2 N_c}{16 \pi^2}$. In this expansion we have used the fact that $R_n^{(1)} = 0$, since the BDS ansatz encodes the complete one loop MHV amplitude at all $n$. 

The remaining $L$-loop contributions to the remainder function are expected to be expressible in terms of generalized polylogarithms~\cite{Chen,FBThesis,Gonch} of uniform transcendental weight $2L$. This space is spanned by (products of) the functions
\begin{equation} \label{eq:multiple_polylogs}
G(a_1, \dots, a_k; z) \equiv \int_0^z \frac{dt}{t-a_1} G(a_2,\dots,a_k; t), \hspace{0.8cm} G(\underbrace{0,\dots,0}_k;z) \equiv \frac{\log^k z}{k!}\, ,
\end{equation}
where $G(;z) \equiv 1$, and the transcendental weight of each function corresponds to its number of indices $k$. In particular, the remainder function is expected to be a pure function of this type, meaning that its kinematic dependence appears in the indices and arguments $a_i$ and $z$, but not in the rational prefactors multiplying these functions. This is known to be true at two loops, due to an impressive all-$n$ computation that leveraged the superconformal symmetry of this theory~\cite{CaronHuot:2011ky}, as well as through six loops in six-particle kinematics~\cite{Dixon:2013eka,Dixon:2014voa,Caron-Huot:2016owq,Caron-Huot:2019vjl} and through four loops in seven-particle kinematics~\cite{Drummond:2014ffa,Dixon:2016nkn}. 

In addition to being generalized polylogarithms, loop-level contributions to the remainder function exhibit a great deal of cluster-algebraic structure. In particular, they are members of the space of `cluster polylogarithms' studied in~\cite{Golden:2014xqa}, indicating that their symbol is naturally expressible in terms of cluster \acoords, while their Lie cobracket is naturally expressible in terms of cluster \xcoords\ (in a way that will be made precise below). Their cobracket, moreover, has been shown to be decomposable into simple functions associated with their $A_2$ and $A_3$ subalgebras~\cite{Golden:2014xqa}. As will be shown in the next section, these functions (the `$A_2$ function' and the `$A_3$ function') are invariant under the automorphism group of the algebras on which they are defined, up to an overall sign. This encodes the fact that these functions are well-defined under coordinate relabelings (or, are well-defined functions of oriented graphs). We correspondingly propose that the space of cluster polylogarithms be refined to include only functions that respect the automorphism group of the cluster algebra on which they are defined. In addition to this subalgebra structure, the symbols of these amplitudes have been found to satisfy a `cluster adjacency' principle~\cite{Drummond:2017ssj}, and their cobracket takes a similarly restricted form~\cite{Golden:2014xqf}. The rest of this section is devoted to making these properties precise, for which purpose we first describe the motivic structure of polylogarithms. 

\subsection{The Symbol and Cobracket} \label{sec:coproduct}

The space of generalized polylogarithms defined by~\eqref{eq:multiple_polylogs} is colossally overcomplete. This is because $a_i$ and $z$ are allowed to be arbitrarily complicated algebraic functions, and because these polylogarithms satisfy a shuffle and stuffle algebra. The shuffle algebra represents the fact that unordered integrations can be triangulated into a sum over iterated integrals~\cite{Duhr:2011zq,Duhr:2014woa}. In general, this means that when two polylogarithms share an argument $z$, their product can be re-expressed as the sum of functions 
\begin{equation} \label{eq:shuffle_relation}
G(a_1,\dots,a_{k_1};z)\ G(a_{{k_1}+1},\dots,a_{k_1+k_2};z)  = \sum_{\sigma \in \Sigma(k_1,k_2)} G(a_{\sigma(1)},\dots,a_{\sigma(k_1+k_2)};z),
\end{equation}
where $\Sigma(k_1,k_2)$ denotes the set of all shuffles between the sets of integers $\{1,\dots,k_1\}$ and $\{k_1+1,\dots,k_1+k_2\}$ (that is, all ways of interleaving these two sets such that the ordering of the elements within each of the original sets is maintained). The stuffle algebra naturally arises when generalized polylogarithms are re-expressed as infinite sums, 
\begin{align} \label{eq:Li_notation}
\text{Li}_{n_1,\dots,n_d}(z_1,\dots, z_d) &\equiv \sum_{0 < m_1 < \dots < m_d} \frac{z_1^{m_1} \cdots z_d^{m_d}}{m_1^{n_1} \cdots m_d^{n_d}} \\
&= (-1)^d G(\underbrace{0,\dots,0}_{n_d-1},\frac{1}{z_d},\dots,\underbrace{0,\dots,0}_{n_1-1},\frac{1}{z_1 \cdots z_d}; 1) \nonumber
\end{align}
where $d$ is called the depth of the polylogarithm. Stuffle identities represent the freedom to split up unordered summation indices (arising from products of polylogarithms) into nested sums where these indices are ordered, as in~\eqref{eq:Li_notation}.

This overcompleteness gives rise to a rich space of identities. This can already be seen at the level of classical polylogarithms, which correspond to the instances of~\eqref{eq:Li_notation} with depth one,
\begin{equation}
\text{Li}_k(z) = - G(\underbrace{0,\dots,0}_{k-1},1;z) .
\end{equation}
For instance, classical polylogarithms satisfy Abel's identity, which itself can be expressed in the language of the $A_2$ cluster algebra. Using the definition of the \xcoords\ $\x_i$ given in eq.~(\ref{eq:exchange-relation}), we have
\begin{equation}\label{eq:abels-identity}
	\sum_{i=1}^5 \Li_2(-\x_i)+\log\x_i\log\x_{i+1} = -\frac{\pi^2}{2}.
\end{equation}
Additional cluster-algebraic identities, including an identity involving $\Li_3$ evaluated on the cluster $\x$-coordinates of $D_4$, are discussed in \cite{Golden:2013xva,GanglPolylogIdentities}.

Fortunately, all identities between polylogarithms are trivialized (up to algebraic identities between symbol letters) by the symbol map, for generic indices $a_i$ and argument $z$~\cite{2011arXiv1101.4497D,2011arXiv1102.1312B,2015arXiv151206409B}. This map can be defined in terms of derivatives; for instance, taking the total derivative of~\eqref{eq:multiple_polylogs}, we have
\begin{equation} \label{eq:symbol_def}
d G(a_1, \dots, a_k; z) = \sum_{i=1}^{k} G(a_1,\dots,\hat{a}_i,\dots,a_k; z)\ d\log \left( \frac{a_{k-i+1} - a_{k-i}}{a_{k-i+1} - a_{k-i+2}} \right),
\end{equation} 
where $a_0 \equiv z$ and $a_{k+1} \equiv 0$, and the notation $\hat{a}_i$ indicates this index should be omitted~\cite{GoncharovMixedTate}. The symbol map is then defined recursively by~\cite{Goncharov:2010jf}
\begin{equation} \label{eq:symbol_def}
\mathcal{S}\big(G(a_1, \dots, a_k; z)\big) \equiv \sum_{i=1}^{k} \mathcal{S}\big(G(a_1,\dots,\hat{a}_i,\dots,a_k; z) \big) \otimes \left( \frac{a_{k-i+1} - a_{k-i}}{a_{k-i+1} - a_{k-i+2}} \right).
\end{equation} 
The entries of the resulting $k$-fold tensor product are referred to as symbol letters. These symbol letters inherit the distributive properties of (arguments of) logarithms, and can therefore be expanded into a multiplicatively independent basis of symbol letters (the `symbol alphabet'). Complicated polylogarithmic identities are thereby reduced to identities between logarithms, at the cost of losing information about the boundary of integration in~\eqref{eq:multiple_polylogs}.
 
The symbol also captures the analytic structure of polylogarithms, insofar as it encodes their (iterated) discontinuity structure. Namely, for generic indices $a_i$ and argument $z$, these functions have nonzero monodromy only where the letters in the first entry of their symbol vanish or become infinite. These monodromies can themselves have branch cuts that the original function did not have, when new symbol letters appear in the second entry of the symbol (and similarly for iterated monodromies, when new symbol letters appear at higher weights). For special values of $a_i$ and $z$, some of this information is lost due to the fact that higher-weight transcendental constants (such as Riemann $\zeta$ values) are in the kernel of the symbol~\cite{Gonch}. However, this information is retained by the coaction~\cite{Brown:2011ik,Duhr:2012fh}, and can be recovered by specifying an integration constant at each weight. We defer further consideration of these transcendental constants to a follow-up paper~\cite{cluster_subalgebras_ii}.

In addition to the symbol, polylogarithms come equipped with a Lie cobracket structure~\cite{Golden:2013xva}. The cobracket $\delta$ can be calculated using the symbol projection operator 
\begin{equation}
\rho(s_1 \otimes \cdots \otimes s_k ) = \frac{k-1}{k} \Big(\rho(s_1 \otimes \cdots \otimes s_{k-1}) \otimes s_k - \rho(s_2 \otimes \cdots \otimes s_{k}) \otimes s_1 \Big),
\end{equation}
where $\rho(s_1) \equiv s_1$. This projects onto the component of a symbol that cannot be written as a product of lower-weight polylogarithms (for instance, via the shuffle relations~\eqref{eq:shuffle_relation}). The action of $\rho$ can be lifted to a projection on functions, up to terms proportional to transcendental constants; since we will not be concerned with these terms in what follows, we will abuse notation by applying $\rho$ to functions directly. The cobracket $\delta$ of a weight $k$ polylogarithm $f$ can then be calculated as
\begin{align} \label{eq:cobracket_def}
\delta(f) &\equiv \sum_{i=1}^{k-1} (\rho_i \wedge \rho_{k-i})\rho(f) .
\end{align}
This notation indicates that the projection operator $\rho$ is first applied to $f$, after which each term in the resulting sum is partitioned into a wedge product of weight $i$ and weight $k-i$ functions (either by splitting up the symbol into its first $i$ and last $k-i$ entries, or by taking the `$i,k-i$ component' of the coproduct); the projection operator is then applied to the entries in this wedge product separately. In general, the wedge product in~\eqref{eq:cobracket_def} involves spaces of different weight. Without loss of generality, we can put the cobracket of any function into a form where the first factor of the wedge product has weight equal to or higher than that of the second factor (exchanging the order of factors when needed, at the cost of a minus sign). We denote by $\delta_{i,j}(f)$ the component of the cobracket of $f$ that involves a wedge product of weight $i$ and $j$ functions, in that order---but we emphasize that this includes contributions from all terms in~\eqref{eq:cobracket_def} that involve these weights in either order.

In the context of two-loop amplitudes, the salient property of the cobracket is that it isolates the component of weight four polylogarithms that cannot be written in terms of classical polylogarithms. Under the action of $\rho$, classical polylogarithms are mapped to elements of the Bloch group $\text{B}_k$~\cite{Bloch:2000, Suslin:1990}, namely the algebra of polylogarithms modulo identities between classical polylogarithms. Following~\cite{Golden:2013xva}, we denote these elements by
\begin{align}
 \{ z \}_k  &\equiv \rho(-\text{Li}_k(-z)) \in \text{B}_k, \quad k>1, \\
 \{ z \}_1  &\equiv \rho(\log(z)) \hspace{.675cm} \in \text{B}_1.
\end{align}
For instance, Abel's identity, eq.~(\ref{eq:abels-identity}), can be expressed (and easily checked) when written in terms of Bloch group elements:
\begin{equation}
	\sum_{i=1}^5 \{\x_i\}_2 = 0.
\end{equation}
In this language, the action of the cobracket on classical polylogarithms is given by
\begin{align}
\delta \big( \text{Li}_k(-z) \big) &= - \{ z \}_{k-1} \wedge \{ z \}_1 , \quad k>2 , \\
\delta \big( \text{Li}_2(-z) \big) &= - \{ 1 + z \}_{1} \wedge \{ z \}_1.
\end{align}
Note that, for the first time at weight four, there exists a component of the cobracket that is not mapped to by classical polylogarithms---namely, $\delta_{2,2}(\text{Li}_4(-z))=0$. This is not true of of weight four polylogarithms in general, and in particular $\delta_{2,2} \big(R_n^{(2)} \big)$ is nonzero for $n\ge7$. However, it has been shown that any weight four function that \emph{is} annihilated by $\delta_{2,2}$ can be written in terms of classical polylogarithms (with potentially complicated arguments)~\cite{G91a,2008arXiv0809.3984D,GanglPolylogIdentities,2018arXiv180107816G,2018arXiv180308585G}. The converse is worth stating as well: any function with non-zero $\delta_{2,2}$ must involve a nonclassical polylogarithm of weight 4. 

Lastly, it is often quite useful to employ the fact that the trivial cohomology of $\delta$ gives us an integrability condition,
\begin{equation}\label{eq:def-cobracket-integrability}
	\delta^2\big(f\big) = 0,
\end{equation}
for any polylogarithm $f$. This condition implies an intricate relationship between the arguments of the Bloch group elements appearing in $\delta(f)$. For example, at weight 4 the relation~\eqref{eq:def-cobracket-integrability} translates to
\begin{equation}
	\delta\big(\delta_{2,2}\big(f\big)\big) + \delta\big(\delta_{3,1}\big(f\big)\big) = 0,
\end{equation}
where
\begin{align}
\delta\big( \{x\}_3 \wedge \{y\}_1 \big)&= \{ x\}_2 \wedge \{x \}_1 \wedge \{y\}_1 \\
\delta\big( \{x\}_2 \wedge \{y\}_2 \big) &= \{ y\}_2 \wedge \{1+x\}_1 \wedge  \{x\}_1 \\
&\hspace{1cm} - \{x\}_2 \wedge  \{1+y\}_1 \wedge  \{y\}_1. \nonumber 
\end{align}
So, while the $\delta_{2,2}$ component alone captures the nonclassical contribution to a function, this nonclassical contribution is linked to the classical contribution through its relation with the $\delta_{3,1}$ component. Cobracket integrability is discussed in much greater detail in~\cite{Golden:2014xqa}.

The symbol and cobracket naturally stratify the study of two-loop amplitudes and integrals that can be expressed as polylogarithms. The operator $\delta_{2,2}$ isolates the nonclassical component of these functions, while the symbol captures their analytic structure up to terms proportional to transcendental constants. In the case of MHV amplitudes in planar ${\cal N} = 4$ sYM theory, both objects turn out to distill intriguing cluster-algebraic structure that would otherwise be hard to see at the level of full functions. It is to this structure that we now turn.

\subsection{Cluster-Algebraic Structure at Two Loops}\label{sec:cluster-algebra-R2n}

As outlined in the introduction, the two-loop MHV amplitudes of this theory exhibit different forms of cluster-algebraic structure at the level of their cobracket, their symbol, and as full functions~\cite{Golden:2013xva,Golden:2014xqa,Golden:2014xqf,Golden:2014pua}. The first facet of this structure concerns the building blocks that appear at each level, which are found to lie within restricted classes:
\begin{itemize}
\item[$\bullet$] The cobracket of $R^{(2)}_n$ can be written in terms of elements of the Bloch group taking the form $\{x_i \}_k$, where $x_i$ is an \xcoord\ on $\Gr(4,n)$
\item[$\bullet$] The symbol of $R^{(2)}_n$ can be expressed in terms of symbol letters drawn from the set of \acoords\ on $\Gr(4,n)$
\item[$\bullet$] The function $R^{(2)}_n$ can expressed entirely in terms of (products of) polylogarithms taking the form $\text{Li}_{n_1,\dots,n_d}(-x_i,\dots,-x_j)$, where each $x_p$ is again an \xcoord\ on $\Gr(4,n)$
\end{itemize}
The physical meaning of the symbol alphabet restriction is clear, if unilluminating---the kinematic configurations in which these amplitudes are singular (due to internal propagators going on-shell in the Feynman diagram expansion) coincide with the vanishing loci of certain cluster \acoords. Even so, it is not clear how (or if) this restriction follows from physical principles. In this respect, the positive Grassmannian formulation of this theory is suggestive, insofar as the integrands of these amplitudes are seen to develop physical singularities where certain \acoords\ vanish or become infinite~\cite{ArkaniHamed:2012nw}. However, this does not preclude the emergence of new singularities during integration that are not rationally expressible in terms of \acoords, as has already been explicitly observed in Feynman integrals contributing to non-MHV amplitudes in this theory~\cite{Bourjaily:2018aeq,Henn:2018cdp,Prlina:2017azl}. The physical meaning of the restricted functional form and cobracket structure exhibited by these amplitudes remains even more obscure. 

In the polylogarithmic amplitudes where cluster coordinates do prove sufficient as a symbol alphabet (in particular, in all two-loop MHV amplitudes), cluster algebras also seem to play a role in how these building blocks are assembled. In particular, it was recently observed that---when these amplitudes are normalized appropriately---the only cluster \acoords\ that appear in adjacent entries of their symbol are those that appear together in at least one cluster of $\Gr(4,n)$~\cite{Drummond:2017ssj}. This `cluster adjacency' principle is not enjoyed by the remainder function~\eqref{eq:remainder_function}, but by BDS-like normalized amplitudes ${\cal E}_n$~\cite{Alday:2009dv,Yang:2010as,Caron-Huot:2016owq,Dixon:2016nkn}. These are defined by
\begin{equation} \label{eq:bds_like_normalized_amplitude}
\mathcal{A}_n^{\rm MHV}\ =\ \mathcal{A}_n^{\rm BDS-like}  \times  {\cal E}_n,
\end{equation}
where $\mathcal{A}_n^{\rm BDS-like}$ is related to $\mathcal{A}_n^{\rm BDS}$ by the cusp anomalous dimension~\cite{Beisert:2006ez} 
\begin{equation}
\Gamma_{\text{cusp}} = 4 g^2- 8 \zeta_2 g^4 + \mathcal{O}(g^6)
\end{equation}
and a simple weight two polylogarithm $Y_n$ via
\begin{equation} \label{eq:bds_like_ansatz}
\mathcal{A}_n^{\rm BDS-like} = \mathcal{A}_n^{\rm BDS} \times \exp\left( \frac{\Gamma_{\text{cusp}}}{4} Y_n \right).
\end{equation}
The function $Y_n$ corresponds to the part of the one-loop MHV amplitude that depends on three- and higher-particle Mandelstam invariants, where these invariants have been assembled into dual-conformally-invariant cross ratios (with the help of two-particle invariants). The BDS-like ansatz that remains only depends on two-particle invariants, yet accounts for the full infrared structure of these amplitudes. 

The motivation for switching to the BDS-like normalization is precisely this restricted kinematic dependence. Since the BDS-like ansatz depends only on two-particle invariants, the functions ${\cal E}_n$ directly inherit the Steinmann relations between three- and higher-particle invariants that are obeyed by the full amplitude~\cite{Steinmann,Steinmann2,Cahill:1973qp,Caron-Huot:2016owq,Dixon:2016nkn}. These relations tell us that
\begin{equation} \label{eq:steinmann}
\begin{rcases}
\text{Disc}_{s_{j,\dots,j+p+q}}\left[\text{Disc}_{s_{i,\dots,i+p}} \big({\cal E}_n \big) \right] &\!\!= \ 0, \hspace{.3cm} \\
\text{Disc}_{s_{i,\dots,i+p}}\left[\text{Disc}_{s_{j,\dots,j+p+q}} \big({\cal E}_n \big) \right] &\!\!= \ 0, 
\end{rcases} \quad 
\begin{gathered} 0 < j\!-\!i \leq p \text{\ \ or\ \ } q < i\!-\!j  \leq p\!+\!q, \end{gathered}
\end{equation}
for positive $p$ and nonnegative $q$. Formulated in terms of symbol entries, this implies that the cluster \acoords\ $\ket{j-1,j,j+p+q-1,j+p+q}$ and $\ket{i-1,i,i+p-1,i+p}$ never appear next to each other in the first two entries of the symbol (when the conditions in~\eqref{eq:steinmann} are met). In fact, it is believed that these constraints can be applied at all depths in the symbol, as these letters are never seen to appear next to each other~\cite{Dixon:2016nkn,Caron-Huot:2018dsv,Caron-Huot:2019vjl,Caron-Huot:2019bsq}. These generalized constraints have been termed the extended Steinmann relations, as they amount to applying the relations~\eqref{eq:steinmann} to all discontinuities of the amplitude in addition to the amplitude itself. 

The fact that the functions ${\cal E}_n$ also obey the cluster adjacency principle in all known cases is far more surprising. The constraints that follow from this principle take a form similar to the extended Steinmann relations (insofar as they restrict which symbol letters can appear in adjacent entries), and in fact turn out to be equivalent in six-particle kinematics when applied to functions with physical branch cuts. It is therefore tempting to believe cluster adjacency follows from some set of physical principles that includes the extended Steinmann relations. However, it is not yet known whether these two conditions are equivalent at all $n$, even in polylogarithmic cases where algebraic roots are absent. (Once functions more complicated than polylogarithms (such as elliptic polylogarithms~\cite{Broedel:2017kkb,Broedel:2018qkq}) start to appear, it is not even known whether anything resembling cluster adjacency can be formulated.)

Seeming to complicate this question of equivalence is the fact that the BDS-like ansatz \eqref{eq:bds_like_ansatz} only exists when $n$ is not a multiple of four. This is because no function satisfying the above description of $Y_n$ exists for these particle multiplicities~\cite{Yang:2010az,Dixon:2016nkn}. (Such a function not only exists for all other $n$, but is uniquely picked out by this description.) Stated another way, when $n$ is a multiple of four, any normalization that absorbs the infrared-divergent part of the amplitude either depends on some set of three- or higher-particle Mandelstam invariants, or spoils the dual conformal invariance of the resulting normalized amplitude. However, this shows this complication is superficial---for the purpose of understanding the relationship between the extended Steinmann relations and the cluster adjacency principle, we need merely choose the latter horn of this dilemma, and give up dual conformal invariance. 

This issue first arises in eight-particle kinematics. There it can be seen---by direct computation---that both the extended Steinmann and cluster adjacency conditions are obeyed when the amplitude is normalized by a `generalized BDS-like ansatz' whose kinematic dependence is restricted to two-particle Mandelstam invariants~\cite{cluster_subalgebras_ii}. (It is always possible to formulate such a normalization, as only two-particle Mandelstams appear in the infrared-divergent terms of the one-loop amplitude~\cite{Bern:1994zx}.) This provides further evidence that the cluster adjacency principle and the extended Steinmann relations (when combined with physical branch cuts) are equivalent, in the cases where cluster adjacency can be unambiguously applied.

While cluster adjacency was first observed in \acoords, it can also be formulated in terms of \xcoords. Unlike \acoords, which are multiplicatively independent, \xcoords\ satisfy numerous multiplicative identities. Thus, there will exist many representations of the same symbol in terms of \xcoords, whereas its representation in terms of \acoords\ is unique. In general, only a subset of these \xcoord\ representations will satisfy cluster adjacency (even if the \acoord\ representation does). Moreover, while \xcoord\ adjacency trivially implies \acoord\ adjacency by the relation~\eqref{eq:x_from_a_coordinates}, the converse cannot be true in general. This is because \acoords\ can express a larger class of functions than \xcoords, insofar as the latter necessarily respect dual conformal symmetry while the former do not. However, it is easy to check in simple symbol alphabets, for example $\Gr(4,6)$, that dual conformal invariance and \acoord\ adjacency together imply the existence of at least one \xcoord\ adjacent representation. We conjecture that this remains true for all dual-conformally-invariant symbols constructed on $\Gr(k,n)$, for general $k$ and $n$. 

The cobracket $\delta\big(R^{(2)}_n\big)$ also satisfies its own form of cluster adjacency~\cite{Golden:2014pua}. Namely, $\delta\big(R^{(2)}_n\big)$ can be expressed as a linear combination of terms $\{x_i\}_2 \wedge \{x_j\}_2$ and $\{x_k\}_3 \wedge \{x_l\}_1$ where $x_i$ and $x_j$ appear together in a cluster of $\Gr(4,n)$, and similarly for $x_k$ and $x_l$. For example, $\delta_{2,2}\big(R^{(2)}_n\big)$ can be expressed as a sum of terms of the form $\{ v_{ijk} \}_2 \wedge \{ z^{\pm}_{pqr} \}_2$, where (using the notation defined in~\eqref{eq:twistor_intersection})
\begin{equation}
v_{ijk} = - \frac{\ket{i\!+\!1 (i,i\!+\!2) (j,j\!+\!1)(k,k\!+\!1)}}{\ket{i,i\!+\!1,k,k\!+\!1}\ket{i\!+\!1,i\!+\!2,j,j\!+\!1}},
\end{equation}
and
\begin{align}
z_{ijk}^+ &= \frac{\ket{i,j\!-\!1,j,j\!+\!1}\ket{i\!+\!1,k\!-\!1,k,k\!+\!1}  -  \ket{i\!+\!1,j\!-\!1,j,j\!+\!1}\ket{i,k\!-\!1,k,k\!+\!1}}{\ket{i,k\!-\!1,k,k\!+\!1}\ket{i\!+\!1,j\!-\!1,j,j\!+\!1}}, \nonumber \\
z_{ijk}^- &= \frac{\ket{i,i\!+\!1,j,k}\ket{i\!-\!1,i,i\!+\!1,i\!+\!2}}{\ket{i\!-\!1,i,i\!+\!1,k}\ket{i,i\!+\!1,i\!+\!2,j}}.
\end{align}
As long as $i<j<k$ (considered mod $n$), the quantities $v_{ijk}$, $z_{ijk}^+$, and $z_{ijk}^-$ each constitute \xcoords, where $z_{ijk}^+$ and $z_{ijk}^-$ are also parity conjugate to each other. The coordinates $v_{ijk}$ were originally motivated by consideration of the location of physical branch cuts, while the $z_{ijk}^\pm$ were motivated by consideration of the final symbol entry of the remainder function, which is constrained to take the form $\ket{i-1,i,i+1,j}$ by dual superconformal symmetry~\cite{CaronHuot:2011kk}. However, for the present discussion, the pertinent point is that $\delta_{2,2}\big(R^{(2)}_n\big)$ can be expressed as a sum of terms $\{ v_{ijk} \}_2 \wedge \{ z^{\pm}_{pqr} \}_2$ in which $v_{ijk}$ and $z^{\pm}_{pqr}$ also occur together in a cluster. Importantly, this `cobracket-level cluster adjacency' isn't implied by the cluster \acoord\ adjacency observed in~\cite{Drummond:2017ssj}, as can be seen in the $A_2$ function we will define in the next subsection---the symbol of this function satisfies cluster \acoord\ adjacency, but the \xcoords\ appearing in its $\delta_{2,2}$ cobracket component cannot be chosen to satisfy cobracket-level cluster adjacency.

While this cobracket-level structure was originally observed in the remainder function, the same characterization carries over to the BDS-like normalized amplitude. This can be seen by combining equations~\eqref{eq:remainder_function},~\eqref{eq:bds_like_normalized_amplitude}, and~\eqref{eq:bds_like_ansatz} to express ${\cal E}_n$ in terms of $R_n$:
\begin{align}
{\cal E}_n &= \exp \left(R_n- \frac{\Gamma_{\text{cusp}}}{4} Y_n \right) \\
&= 1 - Y_n g^2 + \left(R_n^{(2)} + 2 \zeta_2 Y_n + \frac{1}{2} Y_n^2 \right) g^4 + \mathcal{O}(g^6). \nonumber
\end{align} 
Thus, $\rho \big({\cal E}_n^{(2)}\big) = \rho \big(R_n^{(2)} \big)$, since these two functions differ only by products and terms involving transcendental constants, and the remainder function and the BDS-like normalized amplitude have identical cobrackets. At both the level of its cobracket and symbol, ${\cal E}^{(2)}_n$ can therefore be expressed in terms of the same restricted building blocks as $R^{(2)}_n$. But in the case of ${\cal E}^{(2)}_n$ we can now add:
\begin{itemize}
\item[$\bullet$] The cobracket $\delta\big({\cal E}^{(2)}_n\big)$ can be expressed as a linear combination of terms $\{x_i\}_2 \wedge \{x_j\}_2$ and $\{x_k\}_3 \wedge \{x_l \}_1$ where $x_i$ and $x_j$ appear together in a cluster of $\Gr(4,n)$, and similarly for $x_k$ and $x_l$. 
\item[$\bullet$] Pairs of \acoords\ only appear in adjacent entries of the symbol ${\cal S} \big({\cal E}^{(2)}_n\big)$ when they also appear together in at least one cluster of $\Gr(4,n)$.
\end{itemize}
As discussed above, the last statement can also be applied at $n$ that are multiples of four by going to a generalized BDS-like normalization in which only two-particle invariants appear in the normalizing function.

As it turns out, there exists yet more structure in $\delta_{2,2}\big(R^{(2)}_n\big) = \delta_{2,2}\big({\cal E}^{(2)}_n\big)$. In particular, it was shown in~\cite{Golden:2014xqa} that this cobracket component can be decomposed into a sum over various $A_2$ subalgebras of $\Gr(4,n)$, by defining an $A_2$ function that can be evaluated on each of these subalgebras. Moreover, this $A_2$ function can be assembled into an $A_3$ function, in terms of which this cobracket component can similarly be decomposed. As this subalgebra decomposability will play a central role in what follows, we devote the next subsection to its description.

\subsection{Subalgebra Structure and Cluster Polylogarithms}

As we saw in section~\ref{sec:subalgebras_cluster_polytopes}, cluster algebras are endowed with subalgebras that can be generated by mutating on restricted sets of nodes. This motivates looking for physically relevant cluster polylogarithms on algebras other than $\Gr(4,n)$, when these algebras appear as subalgebras of the latter. Before we do so, let us return to the definition of these objects, which we are now in a position to make precise. Following~\cite{Golden:2014xqa}, we define cluster polylogarithms (at least through weight four) to have the following properties:
\begin{quote}
{\bf Cluster Polylogarithm}: A generalized polylogarithm $f$ is a cluster polylogarithm on a cluster algebra $\a$ if
\vspace{-.2cm}
 \begin{itemize}
 \item[(i)] the symbol alphabet of $f$ is composed of only \acoords\ on $\a$, 
 \item[(ii)] the cobracket of $f$ can be expressed in terms of Bloch group elements $\{x_i \}_k$, where $x_i$ is an \xcoord\ of $\a$,
 \item[(iii)] the function $f$ is invariant under the automorphisms of $\a$, up to a sign.
 \end{itemize}
\end{quote}
Property (iii) can be thought of as the requirement that cluster polylogarithms be well-defined functions on the cluster algebra, or more specifically of the oriented graph representing that cluster algebra. For instance, if we wish to define a function on the $A_2$ cluster algebra, as we shall do shortly, it should satisfy the property that $f_{A_2}(1/\x_1\to\x_2)=\pm f_{A_2}(1/\x_2\to\x_3)$, and similarly when the function is evaluated on the other automorphic images of this cluster. The ambiguity in the overall sign, which will be discussed below, reflects the fact that some automorphisms flip the orientation of cluster algebras while others do not.

The first nontrivial cluster polylogarithm is found on the rank-two cluster algebra associated with $A_2$. Given that $A_2$ subalgebras are generated by any pair of connected nodes, they appear ubiquitously in $\Gr(4,n)$. A central fact about the `$A_2$ function' is that it is uniquely determined by the cluster polylogarithm conditions, up to products of classical polylogarithms with weight $\le3$ (see~\cite{Golden:2014xqa} for an in-depth discussion of the $A_2$ function). However, its analytic properties can be tuned by adding and subtracting these products of lower-weight polylogarithms (which should only be done in a way that respects the automorphisms of $A_2$). We choose to define the $A_2$ function (which can be thought of as a `function on an oriented graph') as
\begin{align}\label{def:a2-function}
        f_{A_2}(x_1 \to x_2)  &= \sum_{\text{skew-dihedral}} \bigg[ \Li_{2,2}\left(-\frac{1}{\x_{i-1}},-\frac{1}{\x_{i+1}}\right) - \Li_{1,3}\left(-\frac{1}{\x_{i-1}},-\frac{1}{\x_{i+1}}\right)  \\
        &\hspace{1.1cm} -6 \Li_3\left(-\x_{i-1}\right) \log \left(\x_{i+1}\right) -\frac{1}{2} \log \left(\x_{i-2}\right) \log^2\! \left(\x_i\right) \log \left(\x_{i+1}\right)  {\color{white} \bigg|} \nonumber \\
        &\hspace{1.1cm} +\Li_2(-\x_{i-1}) \Big(3 \log (\x_{i-1})\log (\x_{i+1})+\log \big(\x_{i-2}/\x_{i+2}\big)\log (\x_{i+2})\Big) \bigg], \nonumber
\end{align}
where the $\x_i$ are defined in terms of $x_1$ and $x_2$ as in eq.~(\ref{def:a2-xcoords}), and the skew-dihedral sum indicates subtracting the dihedral flip $\x_i \to \x_{6-i}$ and summing over $i$ from 1 to 5. 

The function defined in~\eqref{def:a2-function} necessarily has the same cobracket as the $A_2$ function considered in~\cite{Golden:2014xqa},
\begin{align} \label{eq:A2_cobracket}
	\delta\big(f_{A_2}\big) &= -\! \! \! \sum_{\text{skew-dihedral}} \Big[ \{\x_{i-1}\}_2 \wedge \{\x_{i+1}\}_2 + 3\{\x_{i}\}_2 \wedge \{\x_{i+1}\}_2 \\
	&\hspace{7.4cm} + 10\{\x_{i}\}_3 \wedge \{\x_{i+1} \}_1 \Big], \nonumber
\end{align}
but differs from the function considered there in some salient respects. Recalling that the clusters of $A_2$ all take the form $1/\x_i\to \x_{i+1}$, it is easy to see that the symbol
\begin{align}
   {\cal S}(f_{A_2}) = \!\! \sum_{\text{skew-dihedral}} &\Big[ 2 \ \x_{i}\otimes\x_{i+1}\otimes\x_{i}\otimes\x_{i+1} + 2 \ \x_{i}\otimes\x_{i+1}\otimes\x_{i+2}\otimes\x_{i+1} \nonumber \\[-.3cm]
&\ + \x_{i+1}\otimes\x_{i}\otimes\x_{i+1}\otimes\x_{i+2} - \x_{i}\otimes\x_{i}\otimes\x_{i+1}\otimes\x_{i+1}  \\
&\ - 2 \ \x_{i}\otimes\x_{i+1}\otimes\x_{i+1}\otimes\x_{i+2}  - \x_{i}\otimes\x_{i+1}\otimes\x_{i+1}\otimes\x_{i} \Big], \nonumber
\end{align}
satisfies cluster \xcoord\ adjacency, and therefore also \acoord\ adjacency (recall that $\x_i$ and $1/\x_i$ can always be exchanged in the symbol at the cost of a minus sign). As a function, $f_{A_2}$ is also smooth and real-valued in the positive domain $x_1, x_2>0$. The $A_2$ cluster algebra plays a crucial role in endowing $f_{A_2}$ with this analytic behavior, as $\Li_{2,2}(x,y)$ and $\Li_{1,3}(x,y)$ have branch cuts in three locations, namely $x=1$, $y=1$, and $xy=1$. The first two branch cuts are trivially avoided as $-1/\x_i<0$ for $x_1,x_2>0$. However, the last branch cut is also avoided because of the $A_2$ exchange relation: 
\begin{equation}
	0<\left(-\frac{1}{\x_{i-1}}\right)\left(-\frac{1}{\x_{i+1}}\right) = \frac{1}{1+\x_i}<1.
\end{equation}
Note that any sum of $A_2$ functions evaluated on the subalgebras of some larger cluster algebra will also inherit these cluster adjacency and smoothness properties. 

Remarkably, it was shown in~\cite{Golden:2014xqa} that all of the information contained in $\delta_{2,2}\big(R^{(2)}_n\big)$ is encoded in the $A_2$ function, when this function is evaluated on some collection of the $A_2$ subalgebras of $\Gr(4,n)$. That is,
\begin{equation}
\delta_{2,2}\big( R_n^{(2)} \big) = \! \sum_{(x_i\to x_j) \subset \Gr(4,n)} \! c_{ij}\ \delta_{2,2} \big(f_{A_2}(x_i \to x_j)\big) 
\end{equation}
for some rational coefficients $c_{ij}$. Moreover, the terms in this sum can themselves be arranged into $A_3$ subalgebras, giving rise to a natural $A_3$ function of the form
\begin{equation} \label{eq:a3_func_tilde}
 f_{A_3}(x_1 \to x_2 \to x_3) \ \sim \sum_{(x_i\to x_j) \subset A_3} d_{ij}\ f_{A_2}(x_i \to x_j), 
\end{equation}
where the $d_{ij}$ are some rational coefficients that we treat in full detail in the next section. For now, we merely highlight the fact that these coefficients can be chosen in such a way that $\delta_{2,2}\big(f_{A_3} \big)$ contains only terms $\{x_i\}_2 \otimes \{x_j \}_2$ in which $x_i$ and $x_j$ appear together in at least one cluster of $A_3$. That is, there exists a decomposition
\begin{equation} \label{eq:remainder_A3_deomposition}
 \delta_{2,2}\big( R_n^{(2)} \big) = \! \sum_{(x_i\to x_j \to x_k) \subset \Gr(4,n)}  \! c_{ijk}\ \delta_{2,2} \big(f_{A_3}(x_i \to x_j \to x_k)\big) 
\end{equation}
that makes the cobracket-level cluster adjacency enjoyed by these amplitudes manifest term-by-term.

In the following sections we analyze these types decompositions systematically, and find they can be extended to much larger subalgebras. As in the $A_3$ decomposition, the $A_2$ function will continue to play a privileged role.

\section{Nonclassical Cluster Polylogarithms}\label{sec:sub-constructibility} 

We now turn to a more systematic exploration of the space of weight-four nonclassical cluster polylogarithms. We restrict our attention to functions that can be defined on $\Gr(4,7) \simeq E_6$ and its subalgebras, as this space is complex enough to give rise to an interesting collection of functions, yet can be explored exhaustively (this space was also explored in~\cite{Harrington:2015bdt}, using a slightly different approach). The techniques we utilize can also be applied to infinite cluster algebras (see for instance~\cite{Golden:2014xqa}), but we leave further exploration of this kind to future work~\cite{cluster_subalgebras_ii}. Before exploring the properties of this space of functions on $\Gr(4,7)$, we describe an efficient method for its generation using the $A_2$ function~\eqref{def:a2-function}. We then describe how this method can be extended to more restricted spaces of functions---those that are constructible out of cluster polylogarithms associated with the higher-rank subalgebras of a given cluster algebra. It is hoped that these subalgebra-constructible functions will prove to be of interest to physicists and mathematicians beyond the role they play in MHV amplitudes in planar ${\cal N}=4$ sYM theory. 

\subsection{\pdfeq{A_2} Functions as a Basis}

A remarkable (conjectured) property of $f_{A_2}$ is that it forms a complete basis for the $\delta_{2,2}$ component of any weight-four cluster polylogarithm \cite{Golden:2014xqa}. (This property can be constructively realized for all cluster polylogarithms defined on $E_6$ or its subalgebras, but has not been proven beyond these cases.\footnote{This was first checked in $E_6$ by Daniel Parker and
Adam Scherlis.}) That is, given a nonclassical weight-four cluster polylogarithm $F$ defined on a cluster algebra $\a$, there always exists some decomposition
\begin{equation}\label{eq:a2-decomp}
	\delta_{2,2} \big(F\big) = \sum_{(x_i\to x_j) \subset \a} c_{ij} ~\delta_{2,2}\big(f_{A_2}(x_i \to x_j) \big) 
\end{equation}
involving rational coefficients $c_{ij}$, where the sum ranges over all the $A_2$ subalgebras of $\a$. Given the symbol (or cobracket) of $F$, this decomposition is simple to compute (but may not be unique). This decomposition can also be phrased as 
\begin{equation}\label{eq:a2-decomp-full}
	F = \sum_{(x_i\to x_j) \subset \a} c_{ij} f_{A_2}(x_i \to x_j) + \dots, \vspace{-.1cm}
\end{equation}
where the dots indicate a residual, purely classical contribution to this function (also expressible in terms of cluster coordinates on $\a$). Since the $f_{A_2}$ function is smooth, real-valued, and satisfies cluster adjacency, this classical contribution will inherit these properties whenever they are also respected by $F$. In other words, this $f_{A_2}$ decomposition cleanly separates the nonclassical from the classical components without disrupting any of the other salient properties of $F$. 

We can glean some intuition for the meaning of this decomposition from the cobracket integrability condition~\eqref{eq:def-cobracket-integrability}. This condition implies specific algebraic relationships between the arguments of the $\delta_{2,2}$ and $\delta_{3,1}$ cobracket components. In particular, the $A_2$ exchange relation
\begin{equation}
	1+\x_i = \x_{i-1}\x_{i+1}
\end{equation}
is sufficient to generate one solution to integrability, namely $f_{A_2}$. The decomposition~\eqref{eq:a2-decomp} thus tells us that \emph{all} solutions to the cobracket integrability relation can be interpreted as linear combinations of exchange relations on $A_2$ subalgebras. This hardly comes as a surprise, given that the only algebraic relations between cluster coordinates are generated by mutation (assuming that the seed cluster is composed of algebraically independent coordinates), and that an $A_2$ subalgebra can be generated on any pair of connected nodes.

In practice, then, the existence (and uniqueness) of $f_{A_2}$ solves the problem of writing down weight-four cluster polylogarithms. To illustrate this, let us consider the space of nonclassical $A_3$ cluster polylogarithms. This space could be computed by forming an ansatz out of all possible cobracket components constructed from cluster \xcoords, and then imposing the integrability condition~\eqref{eq:def-cobracket-integrability}. However, the decomposability conjecture~\eqref{eq:a2-decomp} reduces this computation to the simpler question: how many cluster polylogarithms can be constructed out of the $A_2$ subalgebras of $A_3$? 

Recall from section~\ref{sec:subalgebras_cluster_polytopes} that there are six $A_2$ subalgebras in $A_3$. To label these subalgebras, we choose a representative cluster from each one; in terms of $A_3$ coordinates (defined by the initial seed $x_1 \to x_2 \to x_3$), we take the set
\begin{gather}
	x_1 \to x_2, \hspace{2cm}
	x_2 \to x_3, \hspace{1cm}
	\frac{x_2}{1+x_{1,2}} \to \frac{\left(1+x_1\right) x_3}{1+x_{1,2,3}},  \\
	\frac{x_1 x_2}{1+x_1} \to x_3, \hspace{.5cm}
	x_1 \left(1+x_2\right) \to \frac{x_2 x_3}{1+x_2}, \hspace{1.3cm}
	x_1 \to x_2 \left(1+x_3\right). \hspace{.8cm}  \label{eq:a2-in-a3}
\end{gather}
We then construct an ansatz for the space of nonclassical $A_3$ functions by considering the span of $f_{A_2}$ evaluated on each of these subalgebras:
\begin{equation}\label{eq:fa3-ansatz}
	f_{A_3}\big(x_1\to x_2\to x_3 \big) = c_1 f_{A_2}\big(x_1 \to x_2\big) \ +\ \ldots \ +\ c_6 f_{A_2}\big(x_1\to x_2 \left(1+x_3\right)\big).
\end{equation}	
All that remains is to find values for the coefficients $c_i$ such that $f_{A_3}$ is invariant (up to an overall sign) under the automorphisms of $A_3$. In this case, there are two automorphism group generators, $\sigma_{A_3}$ and $\tau_{A_3}$, which were defined in~\eqref{eq:def_An_automorphic_cycle} and~\eqref{eq:def_An_automorphic_flip}.  

We first consider the case where $f_{A_3}$ is invariant under both $A_3$ automorphism generators, namely
\begin{equation}\label{eq:fA3++}
\begin{gathered}
	\sigma_{A_3}\big(f_{A_3} (x_1 \to x_2\to x_3) \big) = f_{A_3}(x_1 \to x_2\to x_3),\\ \\[-1.2em]
	\tau_{A_3}\big(f_{A_3} (x_1 \to x_2\to x_3) \big) = f_{A_3}(x_1 \to x_2\to x_3). \\ \\[-1.4em]
\end{gathered}
\end{equation}
In practice, this means we take the ansatz for $f_{A_3}$ in eq.~(\ref{eq:fa3-ansatz}) and solve the constraints
\begin{equation}
\begin{gathered}
	f_{A_3}\left(\frac{x_2}{1+ x_{1,2}}\to \frac{x_3(1+x_1)}{1+x_{1,2,3}}\to \frac{1+x_{1,2}}{x_1x_2x_3}\right) = f_{A_3}(x_1\to x_2\to x_3),\\ \\[-1.4em]
	f_{A_3}\left(\frac{1}{x_3} \to \frac{1}{x_2} \to \frac{1}{x_1}\right) = f_{A_3}(x_1\to x_2\to x_3).  
\end{gathered}
\end{equation}
It turns out there is no nontrivial solution to this set of constraints. Conversely, allowing $f_{A_3}$ to pick up an overall minus sign when acted upon by $\tau_{A_3}$, we solve the constraints
\begin{equation}
\begin{gathered}
	\sigma_{A_3}\big(f_{A_3}(x_1\to x_2\to x_3)\big) = f_{A_3}\big(x_1\to x_2\to x_3 \big),\\ \\[-1.2em]
	 \tau_{A_3}\big(f_{A_3}(x_1\to x_2\to x_3)\big) = -f_{A_3}\big(x_1\to x_2\to x_3 \big),  \\ \\[-1.4em]
\end{gathered}
\end{equation}
and find the solution 
\begin{equation}
	c_i = 1
\end{equation}
(which can always be rescaled by an overall constant). We label this particular solution
\begin{align}
	f_{A_3}^{+-}(x_1\to x_2\to x_3) &= f_{A_2}(x_1 \to x_2) + \ldots + f_{A_2}(x_1\to x_2 \left(1+x_3\right)) \nonumber \\
	&= \sum_{i=1}^6 \sigma_{A_3}^i\big(f_{A_2}(x_1\to x_2)\big),
\end{align}
where the superscripts on $f_{A_3}^{+-}$ label its behavior under $\sigma$ and $\tau$, respectively. (The superscript on $\sigma_{A_3}^i$ indicates how many times the operator $\sigma_{A_3}$ should be applied.) There are two remaining sign choices to check: $\sigma^-_{A_3}\tau^+_{A_3}$ and $\sigma^-_{A_3}\tau^-_{A_3}$. We find only the trivial solution in the first case, while		
\begin{equation} \label{eq:fA3mm}
	f_{A_3}^{--}(x_1\to x_2\to x_3) =\sum_{i=1}^6(-1)^i\sigma_{A_3}^i \big(f_{A_2}(x_1\to x_2)\big)
\end{equation}
turns out to be the unique $A_3$ function that picks up a minus sign under the action of both $A_3$ automorphism generators.

Therefore, unlike the case of $A_2$, there are two functions one can associate with the $A_3$ cluster algebra: $f_{A_3}^{+-}$ and $f_{A_3}^{--}$. These functions arise purely from the interplay between the overall symmetries of the $A_3$ cluster algebra and the structure of the $A_2$ subalgebras in $A_3$, i.e. there has been no physics input so far. However, motivated by the properties of the two-loop MHV amplitudes, we can check one further aspect of these functions---whether or not they respect cobracket-level cluster adjacency. Recall that the function $f_{A_2}$ does not satisfy cobracket-level cluster adjacency, while it was stated around equation~\eqref{eq:a3_func_tilde} that there exists an $f_{A_3}$ that does have this property---in fact, it was pointed out in~\cite{Golden:2014xqa} that there exists only one such function. It is easy to check that only $f_{A_3}^{--}$ can be expressed in terms of cluster-adjacent cobrackets, and that it matches the function reported in~\cite{Golden:2014xqa}. 

\subsection{Constructing all Cluster Polylogarithms in $\Gr(4,7)$} \label{sec:all_nonclassical_polylogs_gr47}

\begin{table}
\begin{center}
\vspace{.2cm}
Nonclassical $A_n$ Polylogarithms
\vspace{.2cm}

\begin{tabular}{ l | c | c | c | c }			
\tikzmark{anTopLeft0}  & $\ \sigma^+\tau^+ \ $ & $ \ \sigma^+\tau^- \ $ & $ \ \sigma^-\tau^+ \ $ & $\ \sigma^-\tau^- \ $ \\
  \hline
  $A_2$ & 0 & 1 \ (0) & 0 & 0 \\  
  \hline
  $A_3$ & 0 & 1 \ (0) & 0 & 1\ (1) \\  
  \hline
  $A_4$ & 0 & 3 \ (0) & 0 & 0 \\
  \hline
  $A_5$ & 2 \ (1) & 5 \ (1) & 2 \ (0) & \ 5 \ (3) \tikzmark{anBottomRight0}
\end{tabular} 
\end{center}
\begin{tikzpicture}[overlay, remember picture,decoration={markings,mark=at position .99 with {\arrow[scale=1.4,>=stealth]{>}}}]
    \draw [dashed,rounded corners=1cm] ([yshift=1.2cm,xshift=-.8cm]{pic cs:anTopLeft0}) rectangle ([yshift=-.4cm,xshift=.8cm]{pic cs:anBottomRight0}) ;
\end{tikzpicture}
\caption{The number of nonclassical weight-four cluster polylogarithms on the $A_{n\le5}$ cluster algebra, for each of the four possible automorphism signatures. The number of functions that additionally respect cobracket-level cluster adjacency is given in parentheses.} 
\label{table:nonclassical_An_counts}
\end{table}

It is a straightforward exercise to extend the construction outlined in the last subsection to any finite algebra (as well as to infinite algebras by specifying a set of $A_2$ subalgebras). In the case of $A_n$, there are just the two automorphism generators $\sigma_n$ and $\tau_n$, which were given in~\eqref{eq:def_An_automorphic_cycle} and~\eqref{eq:def_An_automorphic_flip}. The results are summarized in table~\ref{table:nonclassical_An_counts}, where each entry denotes the number of nonclassical weight-four cluster polylogarithms with a given automorphism signature, and the numbers in parentheses indicate how many of these functions additionally satisfy cobracket-level cluster adjacency. It is interesting to note that the automorphism signature $\sigma^+\tau^- $ admits at least one solution for each of these $A_n$, and in particular that it gives rise to the only nontrivial solutions in both $A_2$ and $A_4$. 

This procedure gets only slightly more complicated in the $D_n$ algebras, due to their larger automorphism groups. As discussed in section~\ref{sec:automorphisms}, the automorphism group of $D_4$ is a product of dihedral and symmetric groups, $\smash{{\mathfrak D}_4\times S_3}$. The dihedral group $\smash{{\mathfrak D}_4}$ is generated by a pair of operators $\smash{\sigma_{D_4}^{({\mathfrak D}_4)}}$ and $\tau^{({\mathfrak D}_4)}_{D_4}$\!, while the symmetric group is generated by a pair of operators $\smash{\sigma_{D_4}^{(S_3)}}$ and $\smash{\tau^{(S_3)}_{D_4}}$\!. This gives rise to 16 possible automorphism signatures, which we impose on a general ansatz of $A_2$ functions; the results are presented in table~\ref{table:nonclassical_Dn_counts}. In the same table we give the results for $D_5$, which has only the three automorphism generators $\sigma_{D_5}$, $\tau_{D_5}$, and $\mathbb{Z}_{2,D_5}$, corresponding to the automorphism group ${\mathfrak D}_5 \times \mathbb{Z}_2$. In both $D_4$ and $D_5$, we again see that the space of functions respecting automorphisms is remarkably constrained. There are no functions with odd signature under $\sigma^{({\mathfrak D}_4)}_{D_4}$\!, and only a single $D_4$ automorphism signature gives rise to more than one solution. In $D_5$, there are no functions that have opposite signature in $\sigma_{D_5}$ and $\mathbb{Z}_{2,D_5}$.

\begin{table}
\begin{center}
\vspace{.2cm}
Nonclassical $D_4$ Polylogarithms
\vspace{.2cm}

\begin{tabular}{ c | c | c |}
\multicolumn{1}{c}{\tikzmark{d4TopLeft0}} &\multicolumn{2}{c}{$\underline{\ \sigma_{\mathfrak{D}_4}^+ \tau_{\mathfrak{D}_4}^+ \ }$} \\[-1em]
\multicolumn{1}{c}{} & \multicolumn{1}{c}{} & \multicolumn{1}{c}{} \\
& $\sigma_{S_3}^+$ & \multicolumn{1}{c}{$\sigma_{S_3}^-$} \\[-1em]
&  & \multicolumn{1}{c}{} \\
\cline{1-3} $\tau_{S_3}^+$ & 0 & \multicolumn{1}{c}{0} \\[.05cm]
\cline{1-3} $\tau_{S_3}^-$ & 1 (0) & \multicolumn{1}{c}{0} 
\end{tabular} 
\hspace{.6cm}
\begin{tabular}{ c | c | c |}
\multicolumn{1}{c}{} &\multicolumn{2}{c}{$\underline{\ \sigma_{\mathfrak{D}_4}^+ \tau_{\mathfrak{D}_4}^- \ }$} \\[-1em]
\multicolumn{1}{c}{} & \multicolumn{1}{c}{} & \multicolumn{1}{c}{}\\
 & $\sigma_{S_3}^+$ & \multicolumn{1}{c}{$\sigma_{S_3}^-$} \\[-1em]
 & & \multicolumn{1}{c}{} \\
\cline{1-3} $\tau_{S_3}^+$ & 2 (0) & \multicolumn{1}{c}{0} \\[.05cm]
\cline{1-3} $\tau_{S_3}^-$ & 0 & \multicolumn{1}{c}{0}
\end{tabular}
\hspace{.6cm}
\begin{tabular}{ c | c | c |}
\multicolumn{1}{c}{} &\multicolumn{2}{c}{$\underline{\ \sigma_{\mathfrak{D}_4}^- \tau_{\mathfrak{D}_4}^+ \ }$} \\[-1em]
\multicolumn{1}{c}{} & \multicolumn{1}{c}{} & \multicolumn{1}{c}{}\\
 & $\sigma_{S_3}^+$ & \multicolumn{1}{c}{$\sigma_{S_3}^-$} \\[-1em]
 & & \multicolumn{1}{c}{} \\
\cline{1-3} $\tau_{S_3}^+$ & 1 (0) & \multicolumn{1}{c}{0} \\[.05cm]
\cline{1-3} $\tau_{S_3}^-$ & 1 (1) & \multicolumn{1}{c}{0} 
\end{tabular}
\hspace{.6cm}
\begin{tabular}{ c | c | c |}
\multicolumn{1}{c}{} &\multicolumn{2}{c}{$\underline{\ \sigma_{\mathfrak{D}_4}^- \tau_{\mathfrak{D}_4}^- \ }$} \\[-1em]
\multicolumn{1}{c}{} & \multicolumn{1}{c}{} & \multicolumn{1}{c}{}\\
 & $\sigma_{S_3}^+$ & \multicolumn{1}{c}{$\sigma_{S_3}^-$} \\[-1em]
 & & \multicolumn{1}{c}{} \\
\cline{1-3} $\tau_{S_3}^+$ & 1 (0) & \multicolumn{1}{c}{0} \\[.05cm]
\cline{1-3} $\tau_{S_3}^-$ & 0 & \multicolumn{1}{c}{\ 0 \tikzmark{d4BottomRight0}} 
\end{tabular}

\vspace{.6cm}
Nonclassical $D_5$ Polylogarithms
\vspace{.2cm}

\begin{tabular}{ c | c |}
\multicolumn{2}{c}{\tikzmark{d5TopLeft0} $\underline{\ \sigma_{\mathfrak{D_5}}^+ \tau_{\mathfrak{D_5}}^+\ }$} \\[-1em]
\multicolumn{1}{c}{} & \multicolumn{1}{c}{} \\
 $\mathbb{Z}_2^+$ & \multicolumn{1}{c}{$\mathbb{Z}_2^-$} \\[-1em]
 & \multicolumn{1}{c}{} \\
\hline
5 (2) & \multicolumn{1}{c}{0} 
\end{tabular} 
\hspace{1.2cm}
\begin{tabular}{ c | c |}
\multicolumn{2}{c}{$\underline{\ \sigma_{\mathfrak{D_5}}^+ \tau_{\mathfrak{D_5}}^-\ }$} \\[-1em]
\multicolumn{1}{c}{} & \multicolumn{1}{c}{} \\
 $\mathbb{Z}_2^+$ & \multicolumn{1}{c}{$\mathbb{Z}_2^-$} \\[-1em]
 & \multicolumn{1}{c}{} \\
\hline
9 (2) & \multicolumn{1}{c}{0} 
\end{tabular} 
\hspace{1.2cm}
\begin{tabular}{ c | c |}
\multicolumn{2}{c}{$\underline{\ \sigma_{\mathfrak{D_5}}^- \tau_{\mathfrak{D_5}}^+ \ }$} \\[-1em]
\multicolumn{1}{c}{} & \multicolumn{1}{c}{} \\
 $\mathbb{Z}_2^+$ & \multicolumn{1}{c}{$\mathbb{Z}_2^-$} \\[-1em]
 & \multicolumn{1}{c}{} \\
\hline
0 & \multicolumn{1}{c}{3 (1)}
\end{tabular} 
\hspace{1.2cm}
\begin{tabular}{ c | c |}
\multicolumn{2}{c}{$\underline{\ \sigma_{\mathfrak{D_5}}^- \tau_{\mathfrak{D_5}}^-\ }$} \\[-1em]
\multicolumn{1}{c}{} & \multicolumn{1}{c}{} \\
 $\mathbb{Z}_2^+$ & \multicolumn{1}{c}{$\mathbb{Z}_2^-$} \\[-1em]
 & \multicolumn{1}{c}{} \\
\hline
0 & \multicolumn{1}{c}{\ 7 (5) \tikzmark{d5BottomRight0}} \\
\end{tabular} 
\end{center}
\begin{tikzpicture}[overlay, remember picture,decoration={markings,mark=at position .99 with {\arrow[scale=1.4,>=stealth]{>}}}]
    \draw [dashed,rounded corners=1cm] ([yshift=1.2cm,xshift=-1cm]{pic cs:d4TopLeft0}) rectangle ([yshift=-.4cm,xshift=.7cm]{pic cs:d4BottomRight0}) ;
    \draw [dashed,rounded corners=1cm] ([yshift=1.2cm,xshift=-.8cm]{pic cs:d5TopLeft0}) rectangle ([yshift=-.4cm,xshift=.7cm]{pic cs:d5BottomRight0}) ;
\end{tikzpicture}
\caption{The number of nonclassical weight-four cluster polylogarithms on the $D_4$ and $D_5$ cluster algebras, with each possible automorphism signature. The number of functions that additionally respect cobracket-level cluster adjacency is given in parentheses.}
\label{table:nonclassical_Dn_counts}
\end{table}

Finally, we turn to $E_6$, which has the automorphism group ${\mathfrak D}_{14}$. This group has three generators---$\sigma_{E_6}$, $\tau_{E_6}$, and $\mathbb{Z}_{2,E_6}$. $E_6$ is much larger than any of the cluster algebras considered above, with 504 distinct $A_2$ subalgebras. Even so, the spaces of automorphic functions on it remains surprisingly small, as shown in table~\ref{table:nonclassical_E6_counts}. It is especially surprising here that there are no odd solutions in $\sigma_{E_6}$. Of primary interest is the space with automorphism signature $\sigma_{E_6}^+ \tau_{E_6}^+ \mathbb{Z}_{2,E_6}^+$, as this space contains $R^{(2)}_7$, which will be our primary object of interest in section~\ref{sec:r27-sub-constructibility}. In particular, $R^{(2)}_7$ must be a linear combination of the 6 cluster polylogarithms in this space that respect cobracket-level cluster adjacency.

\begin{table}
\begin{center}
\vspace{.2cm}
Nonclassical $E_6$ Polylogarithms
\vspace{.2cm}

\begin{tabular}{ c | c |}
\multicolumn{2}{c}{\tikzmark{e6TopLeft0} $\underline{\ \sigma_{\mathfrak{D}_{14}}^+ \tau_{\mathfrak{D}_{14}}^+\ }$} \\[-1em]
\multicolumn{1}{c}{} & \multicolumn{1}{c}{} \\
 $\mathbb{Z}_2^+$ & \multicolumn{1}{c}{$\mathbb{Z}_2^-$} \\[-1em]
 & \multicolumn{1}{c}{} \\
\hline
12 (6) & \multicolumn{1}{c}{14 (6) }
\end{tabular} 
\hspace{1.2cm}
\begin{tabular}{ c | c |}
\multicolumn{2}{c}{$\underline{\ \sigma_{\mathfrak{D}_{14}}^+ \tau_{\mathfrak{D}_{14}}^- \ }$} \\[-1em]
\multicolumn{1}{c}{} & \multicolumn{1}{c}{} \\
 $\mathbb{Z}_2^+$ & \multicolumn{1}{c}{$\mathbb{Z}_2^-$} \\[-1em]
 & \multicolumn{1}{c}{} \\
\hline
21 (6) & \multicolumn{1}{c}{17 (9)}
\end{tabular} 
\hspace{1.2cm}
\begin{tabular}{ c | c |}
\multicolumn{2}{c}{$\underline{\ \sigma_{\mathfrak{D}_{14}}^- \tau_{\mathfrak{D}_{14}}^+ \ }$} \\[-1em]
\multicolumn{1}{c}{} & \multicolumn{1}{c}{} \\
 $\mathbb{Z}_2^+$ & \multicolumn{1}{c}{$\mathbb{Z}_2^-$} \\[-1em]
 & \multicolumn{1}{c}{} \\
\hline
0 & \multicolumn{1}{c}{0} 
\end{tabular} 
\hspace{1.2cm}
\begin{tabular}{ c | c |}
\multicolumn{2}{c}{$\underline{\ \sigma_{\mathfrak{D}_{14}}^- \tau_{\mathfrak{D}_{14}}^- \ }$} \\[-1em]
\multicolumn{1}{c}{} & \multicolumn{1}{c}{} \\
 $\mathbb{Z}_2^+$ & \multicolumn{1}{c}{$\mathbb{Z}_2^-$} \\[-1em]
 & \multicolumn{1}{c}{} \\
\hline
0 & \multicolumn{1}{c}{\ 0 \tikzmark{e6BottomRight0}}
\end{tabular} 
\end{center}
\begin{tikzpicture}[overlay, remember picture,decoration={markings,mark=at position .99 with {\arrow[scale=1.4,>=stealth]{>}}}]
    \draw [dashed,rounded corners=1cm] ([yshift=1.2cm,xshift=-1.2cm]{pic cs:e6TopLeft0}) rectangle ([yshift=-.4cm,xshift=.7cm]{pic cs:e6BottomRight0}) ;
=\end{tikzpicture}
\caption{The number of nonclassical weight-four cluster polylogarithms on the $E_6$ cluster algebra, with each of eight possible automorphism signatures. The number of functions that additionally respect cobracket-level cluster adjacency is given in parentheses.} \label{table:nonclassical_E6_counts}
\end{table}

We pause at this point to emphasize that the $A_2$ constructibility of all nonclassical cluster polylogarithms has been verified on each of the finite cluster algebras considered above, and its completeness only becomes conjectural on infinite cluster algebras. While it is clearly not possible to form an ansatz out of all $A_2$ subalgebras in an infinite cluster algebra such as $\Gr(4,8)$, this method can still be used to generate all nonclassical polylogarithms that are constructible out of (the automorphic completion of) any finite set of $A_2$ subalgebras. This proves sufficient in the case of the two-loop remainder function to all $n$~\cite{Golden:2014xqa}, and may prove sufficient in other cases of physical interest. 

\subsection{Nested Cluster Polylogarithms} \label{sec:nested_cluster_polylogs}

Given that the space of nonclassical cluster polylogarithms coincides with the space of $A_2$-constructible functions, it is natural to ask whether other (even more special) spaces of functions are constructible out of other cluster polylogarithms. In fact, this has already been shown to be the case, since the nonclassical part of all two-loop MHV amplitudes can be constructed out of $f_{A_3}^{--}$ as defined in~\eqref{eq:fA3mm}~\cite{Golden:2014xqa}. This gives rise to an interesting nested structure, since $f_{A_3}^{--}$ is itself constructible out of $f_{A_2}$. More generally, in the functions constructed in section~\ref{sec:all_nonclassical_polylogs_gr47} there will be many instances of $A_2$ subalgebras assembling into larger subalgebras, and it remains an open question how intrinsically interesting such nested constructibility might be. 

The procedure for constructing such spaces clearly proceeds just as in the case of $A_2$ constructibility. There are many spaces one can consider constructing on each cluster algebra, corresponding to the inclusion of different sets of functions defined on (possibly) different subalgebras. We leave the exploration of compositely-constructible spaces to future work. For now, we just consider the example of functions constructible out of $f_{A_3}^{--}$, since all $R_n^{(2)}$ are believed to be in this class. We tabulate the space of $f_{A_3}^{--}$-constructible functions in table~\ref{table:nonclassical_A3_constructible_dimensions}.

\begin{table}
\begin{center}
\vspace{.2cm}
$A_3^{--}$-constructible $A_n$ Polylogarithms
\vspace{.2cm}

\begin{tabular}{ c |  c | c | c | c }      
\tikzmark{anTopLeft1}  & $\sigma^+\tau^+$ & $\sigma^+\tau^-$ & $\sigma^-\tau^+$ & $\sigma^-\tau^-$ \\
\hline
$f_{A_4} \in \text{span}\big(f_{A_3}^{--}\big)$ & 0 & 0 & 0 & 0 \\
\hline
$f_{A_5}  \in \text{span}\big(f_{A_3}^{--}\big)$ & 1 & 1 & 0 & \ 3 \tikzmark{anBottomRight1}
\end{tabular} 

\vspace{.6cm}
$A_3^{--}$-constructible $D_4$ Polylogarithms
\vspace{.2cm}

\begin{tabular}{ c | c | c |}
\multicolumn{1}{c}{\tikzmark{d4TopLeft1}} &\multicolumn{2}{c}{$\underline{\ \sigma_{\mathfrak{D}_4}^+ \tau_{\mathfrak{D}_4}^+ \ }$} \\[-1em]
\multicolumn{1}{c}{} & \multicolumn{1}{c}{} & \multicolumn{1}{c}{} \\
 & $\sigma_{S_3}^+$ & \multicolumn{1}{c}{$\sigma_{S_3}^-$} \\[-1em]
 &  & \multicolumn{1}{c}{} \\
\cline{1-3} $\tau_{S_3}^+$ & 0 & \multicolumn{1}{c}{0} \\[.05cm]
\cline{1-3} $\tau_{S_3}^-$ & 0 & \multicolumn{1}{c}{0} 
\end{tabular} 
\hspace{.6cm}
\begin{tabular}{ c | c | c |}
\multicolumn{1}{c}{} &\multicolumn{2}{c}{$\underline{\ \sigma_{\mathfrak{D}_4}^+ \tau_{\mathfrak{D}_4}^- \ }$} \\[-1em]
\multicolumn{1}{c}{} & \multicolumn{1}{c}{} & \multicolumn{1}{c}{}\\
 & $\sigma_{S_3}^+$ & \multicolumn{1}{c}{$\sigma_{S_3}^-$} \\[-1em]
 & & \multicolumn{1}{c}{} \\
\cline{1-3} $\tau_{S_3}^+$ & 0 & \multicolumn{1}{c}{0} \\[.05cm]
\cline{1-3} $\tau_{S_3}^-$ & 0 & \multicolumn{1}{c}{0}
\end{tabular}
\hspace{.6cm}
\begin{tabular}{ c | c | c |}
\multicolumn{1}{c}{} &\multicolumn{2}{c}{$\underline{\ \sigma_{\mathfrak{D}_4}^- \tau_{\mathfrak{D}_4}^+ \ }$} \\[-1em]
\multicolumn{1}{c}{} & \multicolumn{1}{c}{} & \multicolumn{1}{c}{}\\
 & $\sigma_{S_3}^+$ & \multicolumn{1}{c}{$\sigma_{S_3}^-$} \\[-1em]
 & & \multicolumn{1}{c}{} \\
\cline{1-3} $\tau_{S_3}^+$ & 0 & \multicolumn{1}{c}{0} \\[.05cm]
\cline{1-3} $\tau_{S_3}^-$ & 1 & \multicolumn{1}{c}{0} 
\end{tabular}
\hspace{.6cm}
\begin{tabular}{ c | c | c |}
\multicolumn{1}{c}{} &\multicolumn{2}{c}{$\underline{\ \sigma_{\mathfrak{D}_4}^- \tau_{\mathfrak{D}_4}^- \ }$} \\[-1em]
\multicolumn{1}{c}{} & \multicolumn{1}{c}{} & \multicolumn{1}{c}{}\\
 & $\sigma_{S_3}^+$ & \multicolumn{1}{c}{$\sigma_{S_3}^-$} \\[-1em]
 & & \multicolumn{1}{c}{} \\
\cline{1-3} $\tau_{S_3}^+$ & 0 & \multicolumn{1}{c}{0} \\[.05cm]
\cline{1-3} $\tau_{S_3}^-$ & 0 & \multicolumn{1}{c}{\ 0 \tikzmark{d4BottomRight1}} 
\end{tabular}

\vspace{.6cm}
$A_3^{--}$-constructible $D_5$ Polylogarithms
\vspace{.2cm}

\begin{tabular}{c | c |}
\multicolumn{2}{c}{\tikzmark{d5TopLeft1} $\underline{\ \sigma_{\mathfrak{D_5}}^+ \tau_{\mathfrak{D_5}}^+ \ }$} \\[-1em]
\multicolumn{1}{c}{} & \multicolumn{1}{c}{} \\
 $\mathbb{Z}_2^+$ & \multicolumn{1}{c}{$\mathbb{Z}_2^-$} \\[-1em]
 & \multicolumn{1}{c}{} \\
\hline
2 & \multicolumn{1}{c}{0} 
\end{tabular} 
\hspace{1.2cm}
\begin{tabular}{ c | c |}
\multicolumn{2}{c}{$\underline{\ \sigma_{\mathfrak{D_5}}^+ \tau_{\mathfrak{D_5}}^-\ }$} \\[-1em]
\multicolumn{1}{c}{} & \multicolumn{1}{c}{} \\
 $\mathbb{Z}_2^+$ & \multicolumn{1}{c}{$\mathbb{Z}_2^-$} \\[-1em]
 & \multicolumn{1}{c}{} \\
\hline
2 & \multicolumn{1}{c}{0} 
\end{tabular} 
\hspace{1.2cm}
\begin{tabular}{ c | c |}
\multicolumn{2}{c}{$\underline{\ \sigma_{\mathfrak{D_5}}^- \tau_{\mathfrak{D_5}}^+\ }$} \\[-1em]
\multicolumn{1}{c}{} & \multicolumn{1}{c}{} \\
 $\mathbb{Z}_2^+$ & \multicolumn{1}{c}{$\mathbb{Z}_2^-$} \\[-1em]
 & \multicolumn{1}{c}{} \\
\hline
0 & \multicolumn{1}{c}{1} 
\end{tabular} 
\hspace{1.2cm}
\begin{tabular}{ c | c |}
\multicolumn{2}{c}{$\underline{\ \sigma_{\mathfrak{D_5}}^- \tau_{\mathfrak{D_5}}^-\ }$} \\[-1em]
\multicolumn{1}{c}{} & \multicolumn{1}{c}{} \\
 $\mathbb{Z}_2^+$ & \multicolumn{1}{c}{$\mathbb{Z}_2^-$} \\[-1em]
 & \multicolumn{1}{c}{} \\
\hline
0 & \multicolumn{1}{c}{\ 5 \tikzmark{d5BottomRight1}} 
\end{tabular} 

\vspace{.6cm}
$A_3^{--}$-constructible $E_6$ Polylogarithms
\vspace{.2cm}

\begin{tabular}{ c | c |}
\multicolumn{2}{c}{\tikzmark{e6TopLeft1} $\underline{\ \sigma_{\mathfrak{D}_{14}}^+ \tau_{\mathfrak{D}_{14}}^+\ }$} \\[-1em]
\multicolumn{1}{c}{} & \multicolumn{1}{c}{} \\
 $\mathbb{Z}_2^+$ & \multicolumn{1}{c}{$\mathbb{Z}_2^-$} \\[-1em]
& \multicolumn{1}{c}{} \\
\hline
6 & \multicolumn{1}{c}{6} 
\end{tabular} 
\hspace{1.2cm}
\begin{tabular}{ c | c |}
\multicolumn{2}{c}{$\underline{\ \sigma_{\mathfrak{D}_{14}}^+ \tau_{\mathfrak{D}_{14}}^-\ }$} \\[-1em]
\multicolumn{1}{c}{} & \multicolumn{1}{c}{} \\
 $\mathbb{Z}_2^+$ & \multicolumn{1}{c}{$\mathbb{Z}_2^-$} \\[-1em]
 & \multicolumn{1}{c}{} \\
\hline
6 & \multicolumn{1}{c}{9}
\end{tabular} 
\hspace{1.2cm}
\begin{tabular}{ c | c |}
\multicolumn{2}{c}{$\underline{\ \sigma_{\mathfrak{D}_{14}}^- \tau_{\mathfrak{D}_{14}}^+\ }$} \\[-1em]
\multicolumn{1}{c}{} & \multicolumn{1}{c}{} \\
 $\mathbb{Z}_2^+$ & \multicolumn{1}{c}{$\mathbb{Z}_2^-$} \\[-1em]
& \multicolumn{1}{c}{} \\
\hline
0 & \multicolumn{1}{c}{0} 
\end{tabular} 
\hspace{1.2cm}
\begin{tabular}{ c | c |}
\multicolumn{2}{c}{$\underline{\ \sigma_{\mathfrak{D}_{14}}^- \tau_{\mathfrak{D}_{14}}^-\ }$} \\[-1em]
\multicolumn{1}{c}{} & \multicolumn{1}{c}{} \\
 $\mathbb{Z}_2^+$ & \multicolumn{1}{c}{$\mathbb{Z}_2^-$} \\[-1em]
 & \multicolumn{1}{c}{} \\
\hline 
0 & \multicolumn{1}{c}{\ 0 \tikzmark{e6BottomRight1}} 
\end{tabular} 
\end{center}
\begin{tikzpicture}[overlay, remember picture,decoration={markings,mark=at position .99 with {\arrow[scale=1.4,>=stealth]{>}}}]
    \draw [dashed,rounded corners=1cm] ([yshift=1.2cm,xshift=-2.2cm]{pic cs:anTopLeft1}) rectangle ([yshift=-.4cm,xshift=.9cm]{pic cs:anBottomRight1}) ;
    \draw [dashed,rounded corners=1cm] ([yshift=1.2cm,xshift=-1cm]{pic cs:d4TopLeft1}) rectangle ([yshift=-.4cm,xshift=.7cm]{pic cs:d4BottomRight1}) ;
    \draw [dashed,rounded corners=1cm] ([yshift=1.2cm,xshift=-.8cm]{pic cs:d5TopLeft1}) rectangle ([yshift=-.4cm,xshift=.7cm]{pic cs:d5BottomRight1}) ;
    \draw [dashed,rounded corners=1cm] ([yshift=1.2cm,xshift=-.8cm]{pic cs:e6TopLeft1}) rectangle ([yshift=-.4cm,xshift=.7cm]{pic cs:e6BottomRight1}) ;
\end{tikzpicture}

\caption{The number of $f_{A_3}^{--}$-constructible cluster polylogarithms on the $E_6$ cluster algebra and its subalgebras, with each possible automorphism signature.}
\label{table:nonclassical_A3_constructible_dimensions}
\end{table}

Since $f_{A_3}^{--}$ satisfies cobracket-level cluster adjacency, all functions constructed out of it also have this property. As can be seen by comparing table~\ref{table:nonclassical_A3_constructible_dimensions} to tables~\ref{table:nonclassical_An_counts}-\ref{table:nonclassical_E6_counts}, the converse is also true---\emph{all} cluster polylogarithms that satisfy cobracket-level cluster adjacency are $f_{A_3}^{--}$-constructible. While we have checked this explicitly on $E_6$ and all its subalgebras, it has been conjectured to hold more generally~\cite{Golden:2014xqa}. 

It is worth highlighting that the dimensions we have tabulated in this (and the previous) section have taken into account polylogarithmic identities that reduce the difference between two nonclassical functions to something purely classical. For instance, in $E_6$ there are identities relating different instances of $f_{A_2}$ and $f_{A_3}^{--}$ such as
\begin{align}
\sum_{E_6^{+++}} \Bigg[ \delta_{2,2}\Bigg(&2 f_{A_3}^{--}\left(x_1 \to x_2 (1 + x_3)\to \frac{x_3 x_4 x_5}{1 + x_{5,3}} \right)  \\
& \ - f_{A_3}^{--}\left(x_1\to x_2 (1 + x_3)\to \frac{x_3 x_4 x_5 x_6}{1 + x_{6,5,3}} \right) \nonumber  \\ 
&\ - f_{A_3}^{--}\left(x_1\to \frac{x_2 (1 + x_{6,5,3,4})}{1 + x_{6,5}}\to \frac{x_3 x_5}{(1 + x_5) (1 + x_{6,5,3})}\right)  \Bigg) \Bigg] = 0, \nonumber
\end{align}
where we have introduced the notation
\begin{equation}\label{eq:e6-symmetric-sum}
\sum_{E_6^{+++}} f = \sum_{i=0}^6\sum_{j=0}^1\sum_{k=0}^1 \ \tau_{E_6}^k \circ \mathbb{Z}_{2,E_6}^j \circ \sigma_{E_6}^i \big( f \big).
\end{equation}
Although we won't discuss these spaces of identities further, it would be interesting to study them (and their geometry on cluster polytopes) systematically.

\section{Subalgebra Constructibility and \pdfeq{R^{(2)}_7}} \label{sec:r27-sub-constructibility}

Having seen the extent to which the $E_6$ cluster algebra admits nontrivial functional embeddings, we turn to the study of $R^{(2)}_{7}$\!. Specifically, we explore the ways in which the nonclassical part of this function is subalgebra-constructible, thus probing the extent to which $\mathcal{N}=4$ sYM theory ``knows'' about the rich subalgebra structure of $E_6$. It is already known that $R^{(2)}_7$ is $A_2$- and $A_3$-constructible in terms of the functions defined in~\eqref{def:a2-function} and~\eqref{eq:fA3mm}~\cite{Golden:2014xqa}. Using the techniques of section~\ref{sec:sub-constructibility}, we can ask this question systematically for any set of cluster polylogarithms defined on a subalgebra of $E_6$.

We will in particular be interested in the corank-one subalgebras of $E_6$---namely, $D_5$ and $A_5$---as $D_5$- and $A_5$-constructible functions are not subject to the same amount of representational ambiguity as functions that are constructible out of smaller subalgebras. For instance, while there are 1071 subpolytopes of the $E_6$ cluster polytope that correspond to $A_2$ subalgebras, only 504 of these give rise to distinct cluster algebras. Moreover, there exist 56 identities between the instances of $f_{A_2}$ evaluated on these 504 subalgebras~\cite{Golden:2014xqa}. Similarly, out of 476 distinct $A_3$ subpolytopes, only 364 give rise to distinct cluster algebras, and there are $169$ identities between the instances of $f_{A_3}^{--}$ evaluated on these subalgebras. These redundancies make the representation of $\smash{\delta_{2,2} \big(R^{(2)}_7\big)}$ in terms of either $f_{A_2}$ or $f_{A_3}^{--}$ far from unique. Correspondingly, it is hard to assign a clear geometric interpretation to these decompositions on the $E_6$ cluster polytope, since the $A_2$ function associated with any specific subpolytope can be traded for $A_2$ functions associated with other subpolytopes (and similarly for $f_{A_3}^{--}$). 

On the other hand, there are only 14 $D_5$ subpolytopes and 7 $A_5$ subpolytopes of the $E_6$ cluster polytope, each of which gives rise to a distinct subalgebra. These collections of subalgebras respectively form complete orbits under the automorphism group of $E_6$, implying that all $D_5$- and $A_5$-constructible functions in $E_6$ take the form 
\begin{align}
\sum_{D_5\subset E_6} f_{D_5}(x_i \to \ldots) &= \sum_{i=0}^6\sum_{j=0}^1 \ (\pm1)^i (\pm1)^j \ \mathbb{Z}_{2,E_6}^j \circ \sigma_{E_6}^i \Big(f_{D_5}(x_i \to \ldots) \Big) \label{eq:D5_in_E6_sum} 
\end{align}
and
\begin{align}
\sum_{A_5\subset E_6} f_{A_5}(x_i \to \ldots) &= \sum_{i=0}^6 \ (\pm1)^i \ \sigma_{E_6}^i \Big(f_{A_5}(x_i \to \ldots) \Big), \label{eq:A5_in_E6_sum} \end{align}
where each sign choice determines the signature of the resulting $E_6$ polylogarithm under the associated automorphism generator. In the case of $R_7^{(2)}$\!, the positive sign should be chosen in each case. Thus, $D_5$ and $A_5$ functions can only be embedded in $E_6$ with the right automorphism signature a single way---making $D_5$ and $A_5$ decompositions of $R_7^{(2)}$ (if they exist) canonical in a way that decompositions into smaller subalgebras are not. 

We pause to note that the sum notation used in~\eqref{eq:D5_in_E6_sum} and~\eqref{eq:A5_in_E6_sum} differs from the notation used in~\eqref{eq:e6-symmetric-sum}. The two notations are in practice equivalent, since---as we require cluster polylogarithms to respect automorphisms (up to a sign)---summing over indirect automorphisms (with the appropriate sign) only rescales these functions by a multiplicative factor. We will adopt notation such as in~\eqref{eq:D5_in_E6_sum} and~\eqref{eq:A5_in_E6_sum} when it is useful to think of summing over all subalgebras corresponding to subpolytopes of a given type in the cluster polytope. The downside of this representation is that it obfuscates the behavior of the resulting object under the indirect automorphisms of the parent algebra. We will thus make use of notation similar to~\eqref{eq:e6-symmetric-sum} when we deem it more important to make the behavior under all automorphisms explicit. Adding such a sum over indirect automorphisms amounts to summing over subalgebras that appear in the ``flipped'' cluster polytope as well as the original.

\subsection{The \pdfeq{D_5} Constructibility of \pdfeq{R^{(2)}_7}}\label{sec:d5-func}

We begin with $D_5$, the largest subalgebra of $E_6$ in terms of its number of clusters. For each of the automorphism signatures in $D_5$ that admits a nontrivial space of (nonclassical) cluster polylogarithms, we insert a general ansatz of such functions into the sum~\eqref{eq:D5_in_E6_sum} with positive signs chosen for both direct automorphism generators. Referring back to table~\ref{table:nonclassical_Dn_counts}, we see there are four nontrivial cases to consider. Note that we do not restrict our ans\"atze to $D_5$ functions that respect cobracket-level cluster adjacency, because it is possible for functions that don't have this property to assemble into linear combinations that do when evaluated on the subalgebras of $E_6$ (as happens when $f_{A_3}^{--}$ is constructed out of $A_2$ functions). We then check to see if the nonclassical part of $\smash{R_7^{(2)}}$ is in the span of any of these ansatz sums.

Amazingly, there exists precisely one $D_5$ automorphism signature in terms of which $R_7^{(2)}$ can be decomposed---the totally odd signature $\sigma_{D_5}^- \tau_{D_5}^- \mathbb{Z}_{2,D_5}^-$. Two free parameters remain in this decomposition, corresponding to two degrees of freedom that cancel in the sum~\eqref{eq:D5_in_E6_sum}. More concretely, we can decompose 
\begin{align}
\delta_{2,2}\big(R^{(2)}_7 \big) = \frac{1}{20} \sum_{D_5\subset \Gr(4,7)} \delta_{2,2} \Big( f_{D_5}^{---}\left(
\begin{gathered}
    \begin{xy} 0;<1pt,0pt>:<0pt,-1pt>::
      (0,20) *+{x_1} ="1",
      (30,20) *+{x_2} ="2",
      (80,20) *+{\color{white} xxxxxxx} ="3",
      (114,0) *+{\color{white} x} ="4",
      (114,40) *+{\color{white} x} ="5",
      (72,20) *+{\begin{aligned} \frac{x_3 x_5}{1+x_5} \end{aligned}} ="6",
      (118,0) *+{x_4} ="7",
      (118,34) *+{\begin{aligned} \frac{1}{x_5}\end{aligned}} ="8",
      "1", {\ar"2"},
      "2", {\ar"3"},
      "3", {\ar"4"},
      "3", {\ar"5"},
    \end{xy}
\end{gathered}
\right) \Big), \label{eq:R27_D5_decomposition}
\end{align}
where the sum over all $D_5$ subalgebras is taken according to~\eqref{eq:D5_in_E6_sum}, and we define
\begin{align}\label{eq:fd5-init}
f_{D_5}^{---}\left(
\begin{gathered}
    \begin{xy} 0;<1pt,0pt>:<0pt,-1pt>::
      (0,15) *+{x_1} ="1",
      (30,15) *+{x_2} ="2",
      (60,15) *+{x_3} ="3",
      (90,-2) *+{\color{white} x_i} ="4",
      (90,32) *+{\color{white} x_i} ="5",
      (92,0) *+{x_4} ="6",
      (92,30) *+{x_5} ="7",
      "1", {\ar"2"},
      "2", {\ar"3"},
      "3", {\ar"4"},
      "3", {\ar"5"},
    \end{xy}
    \end{gathered} 
\right) &\equiv \sum_{D_5^{---}} \bigg[
	\frac{1}{2} c_1 f_{A_2}\big(x_1\to x_2 \left(1+x_{3,4}\right)\big)\nl
	&\hspace{-4cm} - \left(\frac{1}{2}-\frac{c_1}{2}\right) f_{A_2}\left(\frac{x_1 x_2}{1+x_1}\to x_3 \left(1+x_4\right)\right) + 
	\frac{1}{4} c_2 f_{A_2}\left(\frac{x_1 x_2 x_3}{1+x_{1,2}}\to x_4\right)\nl
	&\hspace{-4cm}+ \left(c_1-c_2\right) f_{A_2}\left(\frac{x_2 x_3}{1+x_2}\to x_4\right)+
	\left(1-c_1 \right) f_{A_2}\big(x_2\to x_3 \left(1+x_4\right)\big) \\
	&\hspace{-4cm}+ \left(\frac{1}{2}-c_1+\frac{3 c_2}{4}\right) f_{A_2}\left(x_2 \left(1+x_3\right)\to \frac{x_3 x_4}{1+x_3}\right) +
	\frac{1}{2} c_2 f_{A_2}\big(x_3\to x_4\big) \bigg] \nonumber
\end{align}
using the notation
\begin{equation}
	\sum_{D_5^{---}} f = \sum_{i=0}^4\sum_{j=0}^1\sum_{k=0}^1 (-1)^{i+j+k}\  \sigma_{D_5}^i \circ \tau_{D_5}^j \circ \mathbb{Z}_{2,D_5}^k \big(f\big).
\end{equation}
Unlike in the sums~\eqref{eq:D5_in_E6_sum} and~\eqref{eq:A5_in_E6_sum}---where the sum was over all subalgebras of a given type in $E_6$ (but not its flipped twin)---we now include a sum over the indirect automorphism $\tau_{D_5}$ so as to make explicit the totally odd automorphism signature. The \xcoords\ in~\eqref{eq:R27_D5_decomposition} should be understood to be the $E_6$ coordinates defined in~\eqref{eq:E6_x_def}, while the \xcoords\ in~\eqref{eq:fd5-init} should be thought of as $D_5$ coordinates that take different values when evaluated on the $D_5$ subalgebras of $E_6$. (It is easy to see that the $E_6$ seed~\eqref{def:E6} doesn't contain any $D_5$ sub-quivers of the form~\eqref{def:D5}, but that mutating on the $x_5$ node will give rise to a quiver containing the $D_5$ quiver seen in~\eqref{eq:R27_D5_decomposition}.) The coefficients $c_1$ and $c_2$ represent the two remaining degrees of freedom in our ansatz, which drop out of the sum~\eqref{eq:R27_D5_decomposition}.

It is natural to ask whether the parameters $c_1$ and $c_2$ can be chosen in such a way that $f_{D_5}^{---}$ is endowed with additional nice properties. For instance, $f_{D_5}^{---}$ does not satisfy cobracket-level cluster adjacency for generic values of $c_1$ and $c_2$, but it can be given this property by choosing $c_2 = -\frac{6}{5} + \frac{8}{5} c_1$. As discussed in section~\ref{sec:nested_cluster_polylogs}, this necessarily makes $f_{D_5}^{---}$ itself an $A_3$-constructible function:
\begin{align}
	f_{A_3\subset D_5}^{---} = - \frac{1}{10}\sum_{D_5^{---}}\bigg[ 
	&\left(6-3 c_1\right) f_{A_3}^{--}\left(x_1\to x_2 \left(1+x_3\right)\to \frac{x_3 x_4}{1+x_3}\right)  \nonumber \\[-.34cm]
	&+ \left(-3+4 c_1\right) f_{A_3}^{--}\big(x_1\to x_2\to x_3 \left(1+x_4\right)\big)  \nl
	&+ \left(2-c_1\right) f_{A_3}^{--}\big(x_2\to x_3 \left(1+x_4\right)\to x_5\big) \\
	&+ \left(\frac{1}{2}-\frac{3 c_1}{2}\right) f_{A_3}^{--}\left(\frac{1}{x_4}\to x_3 \left(1+x_4\right)\to x_5\right)  \nl
	&+ \left(-1+\frac{c_1}{2}\right) f_{A_3}^{--}\left(\frac{1}{x_4}\to \frac{x_2 x_3 \left(1+x_4\right)}{1+x_2}\to x_5\right) \bigg]. \nonumber
\end{align}
As in the (more general) $f_{A_2}$ representation~\eqref{eq:fd5-init}, we have made the full automorphism signature of the $D_5$ function manifest at the level of the sum. 

It turns out there is one additional way $f_{D_5}^{---}$ can be decomposed. An $A_4$ decomposition is obtainable if we choose $c_1 = \frac{3}{5}$ and $c_2 = 0$. Moreover, although there exists no canonical decomposition such as~\eqref{eq:D5_in_E6_sum} or~\eqref{eq:A5_in_E6_sum} for $A_4$-constructible $E_6$ polylogarithms, there does exist such a decomposition for $A_4$-constructible $D_5$ polylogarithms. That is, all 10 $A_4$ subpolytopes of the $D_5$ cluster polytope are in the same orbit of the automorphism group of $D_5$. Thus, such a decomposition must take the form
\begin{align}
\sum_{A_4\subset D_5} f_{A_4}(x_i \to \ldots) &= \sum_{i=0}^4\sum_{j=0}^1 \ (\pm1)^i (\pm1)^j \ \mathbb{Z}_{2,D_5}^j \circ \sigma_{D_5}^i \Big(f_{A_4}(x_i \to \ldots) \Big). \label{eq:A4_in_D5_sum} 
\end{align}
Choosing both signs to be negative to match the automorphism signature of $f_{D_5}^{---}$, we find the decomposition
\begin{align}
	f_{A_4\subset D_5}^{---} &= \sum_{A_4 \subset D_5 } f_{A_4}^{+-}(x_1\to x_2\to x_3 \to x_4)
 \end{align}
in terms of a new $A_4$ function
\begin{equation}\label{eq:fa4-def}
	f_{A_4}^{+-} = \frac{1}{10}\sum_{A_4^{+-}} \Big( 3 f_{A_2} \big(x_1 \to x_2(1+x_3) \big) - f_{A_2} \big(x_2\to x_3(1+x_4) \big)  \Big),
\end{equation}
where we have made use of the notation
\begin{equation}
 	\sum_{A_4^{+-}}f = \sum_{i=0}^6\sum_{j=0}^1(-1)^j \ \tau_{A_4}^j \circ \sigma_{A_4}^i \big(f \big).
\end{equation} 
The function $\smash{f_{A_4}^{+-}}$ itself is not $A_3$-constructible, as it does not satisfy cobracket-level adjacency. However, the $\smash{f_{A_4\subset D_5}^{---}}$ representation of $\smash{R^{(2)}_7}$,
\begin{equation}\label{eq:r27-d5-a4-decomp1}
	\delta_{2,2} \big(R^{(2)}_7\big) = \frac{1}{20} \sum_{D_5\subset \Gr(4,7)} \sum_{A_4\subset D_5} \delta_{2,2} \big(f_{A_4}^{+-}(x_1\to x_2 \to x_3 \to x_4)\big) ,
\end{equation}
is still quite remarkable---by evaluating the simple function~\eqref{eq:fa4-def} on all $A_4$ subalgebras of $\Gr(4,7)$ and arranging these functions into $D_5$ polylogarithms (as uniquely dictated by the choice of a totally odd $D_5$ automorphism signature), $\smash{\delta_{2,2}\big(R^{(2)}_7\big)}$ is recovered by simply summing over these $D_5$ polylogarithms. Note that every $A_4$ subalgebra in $\Gr(4,7)$ appears at least once in this sum, however some instances of $f_{A_4}^{+-}$ appear multiple times and with different sign, leaving only 56 $A_4$ subalgebras which contribute.

Having considered some cluster-algebraic motivations for choosing certain values of $c_1$ and $c_2$, we now investigate whether collinear limits provide us with a natural choice for these parameters. In particular, we know that the nonclassical portion of $R^{(2)}_7$ must vanish under the $7\to6$ collinear limit, as $R^{(2)}_6$ is purely classical. We can therefore ask whether the 14 instances of $f_{D_5}^{---}$ in $E_6$ can be made to separately vanish in the collinear limit. It turns out this is not possible---at best, by choosing $c_1 = 0$ and $c_2 = -\frac{6}{5}$, one can get 10 of the 14 $f_{D_5}^{---}$ functions to separately vanish (at the nonclassical level). The nonclassical part of the remaining 4 $f_{D_5}^{---}$ functions then cancel off pairwise. 

\subsection{The \pdfeq{A_5} Constructibility of \pdfeq{R^{(2)}_7}}\label{sec:a5-func}

We now turn our attention to $A_5$ decompositions of $R^{(2)}_7$, using the ansatz sum~\eqref{eq:A5_in_E6_sum}. Intriguingly, there is (again) precisely one automorphism signature in terms of which $\delta_{2,2}\big(R^{(2)}_7\big)$ can be decomposed---and it is (again) the totally antisymmetric signature $\sigma_{A_5}^-\tau_{A_5}^-$. This time there is a single internal degree of freedom that cancels in the sum~\eqref{eq:A5_in_E6_sum}. This decomposition can be written explicitly as
\begin{align}
\delta_{2,2}\big(R^{(2)}_7 \big) &= \label{eq:R27_A5_decomposition}\\ 
&\hspace{-1cm}\frac{1}{20} \sum_{A_5\subset E_6} \delta_{2,2} \Big(f_{A_5}^{--} \left(x_1 \to x_2 \to \frac{x_3 x_5 x_6}{1+x_{6,5}} \to \frac{1}{x_6 (1+x_5)}  \to \frac{1+x_{6,5}}{x_5} \right) \Big), \nonumber 
\end{align}
where the $A_5$ sum~\eqref{eq:A5_in_E6_sum} is taken over the function
\begin{align}
	f_{A_5}^{--}\big( x_1\to x_2\to x_3 \to x_4  \to x_5 \big) \equiv \sum_{A_5^{--}} 
	& \bigg[ \frac{1}{2} c_1 f_{A_2}\big(x_2\to x_3 \left(1+x_4\right)\big)  \label{eq:fa5-init}  \\[-.4cm]
	&\ - \left(1+ c_1\right) f_{A_2}\big(x_2\to x_3 \left(1+x_{4,5}\right)\big)  \nonumber \\ 
	&\ - \left(\frac{1}{2} + c_1\right) f_{A_2}\left(\frac{x_1 x_2}{1+x_1}\to x_3 \left(1+x_4\right)\right) \bigg], \nonumber
\end{align}
which makes use of the notation
\begin{equation}
\sum_{A_5^{--}} f = \sum_{i=0}^7 \sum_{j=0}^1 (-1)^{i+j} \ \sigma_{A_5}^i \circ \tau_{A_5}^j \big(f \big).
\end{equation}
As in the last subsection, the \xcoords\ in~\eqref{eq:R27_A5_decomposition} are the $E_6$ coordinates defined in~\eqref{eq:E6_x_def}, while the \xcoords\ in~\eqref{eq:fa5-init} are $A_5$ coordinates that take different values when evaluated on different $A_5$ subalgebras of $E_6$. The initial $A_5$ quiver in~\eqref{eq:R27_A5_decomposition} can be generated by mutating on the nodes associated with $x_5$ and then $x_6$ in~\eqref{def:E6}, and then freezing the node associated with $x_4$. 

Let us first address $f_{A_5}^{--}$'s collinear limits. In this case, the nonclassical part of all 7 instances of $f_{A_5}^{--}$ on $E_6$ separately vanish in each $7\to6$ collinear limit, for all values of $c_1$. This makes the collinear behavior of $\smash{R^{(2)}_7}$ considerably cleaner in the $A_5$ decomposition than in the $D_5$ decomposition. However, it also gives us no hints for how to choose $c_1$.

Luckily, we can find preferred values for $c_1$ by requiring $f_{A_5}^{--}$ to be either $A_3$- or $A_4$-constructible. It becomes $A_3$-constructible if we choose $c_1 = -\frac{1}{2}$, but this time in terms of the function $f_{A_3}^{+-}$. Specifically, we find the decomposition
\begin{equation}
	f_{A_3\subset A_5}^{--} = - \frac{1}{8}\sum_{A_5^{--}} f_{A_3}^{+-}\left(x_2\to x_3(1+x_4)\to \frac{x_4 x_5}{1+x_4}\right).
\end{equation}
If we instead choose $c_1 = -1$, $\smash{f_{A_5}^{--}}$ becomes $A_4$-constructible. Analogous to the $D_5$ case, there is only a single orbit of $A_4$ subalgebras in $A_5$, implying that this decomposition must take the form 
\begin{equation}
\sum_{A_4 \subset A_5} f_{A_4}(x_i \to \ldots) = \sum_{i=0}^7 \ (\pm1)^i  \ \sigma_{A_5}^i \Big(f_{A_4}(x_i \to \ldots) \Big) \label{eq:A4_in_A5_sum}.
\end{equation}
Choosing the minus sign in this sum, we find the unique decomposition
\begin{equation}
	f_{A_4\subset A_5}^{--} = - \sum_{A_4 \subset A_5} f_{A_4}^{+-}(x_1\to x_2 \to x_3 \to x_4),
\end{equation}
where $\smash{f_{A_4}^{+-}}$ is the \emph{same} function that appeared in the $A_4$ decomposition of $\smash{f_{D_5}^{---}}$\!. Therefore we find an $A_4\subset A_5$ representation of $R^{(2)}_7$ that bears a striking resemblance to the $A_4\subset D_5$ representation found in~\eqref{eq:r27-d5-a4-decomp1}:
\begin{equation}\label{eq:r27-a5-a4-decomp1}
	\delta_{2,2} \big(R^{(2)}_7\big) = -\frac{1}{20} \sum_{A_5\subset \Gr(4,7)} \sum_{A_4\subset A_5} \delta_{2,2} \big(f_{A_4}^{+-}(x_1\to x_2 \to x_3 \to x_4)\big).
\end{equation}
In fact, the two sums are identical. Each of the seven $A_5$ subalgebras contains eight $A_4$ subalgebras, resulting in 56 instances of $f_{A_4}^{+-}$. These are the same 56 instances of $f_{A_4}^{+-}$ that contributed in~\eqref{eq:r27-d5-a4-decomp1} (the overall sign difference between these two decompositions results from the orientations of the initial $D_5$ and $A_5$ subalgebras chosen in \eqref{eq:R27_D5_decomposition} and \eqref{eq:R27_A5_decomposition}, respectively). We now turn to a discussion of these overlapping decompositions of $R^{(2)}_7$. 

\subsection{The Many Facets of \pdfeq{R^{(2)}_7}}

\begin{figure}[t] \centering
  \begin{tikzpicture}
  state/.style={circle, draw, minimum size=3cm}
	\node (H1) at (0cm,.2cm) {\color{white} ${}^i_j F^i_j$};
	\node (P1) at (0cm,.2cm) {$R_7^{(2)}\!\!$};
	\node (H2) at (-2cm,-1.5cm) {\color{white} ${}^i_j F^i_j$};
	\node (P2) at (-1.9cm,-1.5cm) {$f_{D_5}^{---}$};
	\node (H3) at (2cm,-1.5cm) {\color{white} ${}^i_j F^i_j$};
	\node (P3) at (2cm,-1.5cm) {$f_{A_5}^{--}$};
	\node (H4) at (0cm,-3cm) {\color{white} ${}^i_j F^i_j$};
	\node (P4) at (0cm,-3cm) {$f_{A_4}^{+-}$};
	\node (H5) at (-3cm,-3.4cm) {\color{white} ${}^i_j F^i_j$};
	\node (P5) at (-2.9cm,-3.4cm) {$f_{A_3}^{--}$};
	\node (H6) at (3cm,-3.4cm) {\color{white} ${}^i_j F^i_j$};
	\node (P6) at (3cm,-3.4cm) {$f_{A_3}^{+-}$};
	\node (H7) at (0cm,-5.4cm) {\color{white} ${}^i_j F^i_j$};
	\node (P7) at (0cm,-5.4cm) {$f_{A_2}^{--}$};
	\draw[->] (H1) -- (H2);
	\draw[->] (H1) -- (H3);
	\draw[->] (H2) -- (H4);
	\draw[->] (H3) -- (H4);
	\draw[->] (H2) -- (H5);
	\draw[->] (H3) -- (H6);
	\draw[->] (H4) -- (H7);
	\draw[->] (H5) -- (H7);
	\draw[->] (H6) -- (H7);
\end{tikzpicture}
  \caption{The various pathways along which $R^{(2)}_7$ can be decomposed.}
\label{fig:R27_decompositions}
\end{figure}
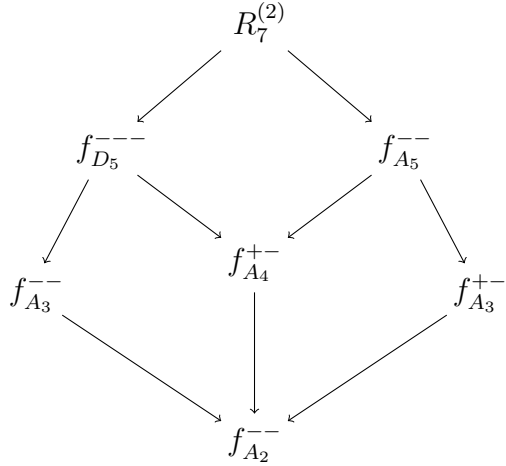

The results of the previous sections are summarized in figure \ref{fig:R27_decompositions}. The fact that the two most canonical decompositions of $R^{(2)}_7$ involve functions that are themselves canonically decomposable in terms of the same $A_4$ function is highly unexpected. In equation form, this interlinked structure can be summarized as
\begin{align}
	R^{(2)}_7 &=  {\color{white} - } \frac{1}{20} \sum_{D_5\subset \Gr(4,7)} \sum_{A_4\subset D_5} f_{A_4}^{+-}(x_1\to x_2 \to x_3 \to x_4) + \dots \label{eq:r27_a4_d5_e6_decomposition} \\
	& = - \frac{1}{20}  \sum_{A_5\subset \Gr(4,7)} \sum_{A_4\subset A_5} f_{A_4}^{+-}(x_1\to x_2 \to x_3 \to x_4) + \dots \label{eq:r27_a4_a5_e6_decomposition}
\end{align}
where the trailing dots indicate that this equality holds up to the contribution of purely classical polylogarithms, and we have left the signs in these sums implicit. As emphasized previously, the required classical contribution is the \emph{same} in both lines, as strict equality holds between the two nested sums over $f_{A_4}^{+-}$ (including the minus sign in the second line, for the orientations chosen above). This equality does not stem from any complicated polylogarithm identities, but is rather a direct consequence of the arrangement of these subalgebras within $\Gr(4,7)$. It is perhaps suggestive to interpret the equivalence of these two decompositions as coming from the fact that they are different subalgebraic ``coverings'' of the same object, namely $\Gr(4,7)$.

In addition to this tightly interlinked pair of decompositions, both $f_{D_5}^{---}$ and $f_{A_5}^{--}$ admit $A_3$ decompositions. As discussed above, each of these decompositions makes some property of the remainder function manifest at the expense of others---for instance, the $f_{A_5}^{--}$ representation makes the vanishing of the nonclassical component of this amplitude in collinear limits manifest term-by-term, whereas the $A_2 \subset A_3 \subset D_5 \subset \Gr(4,7)$ representation manifestly satisfies cobracket-level cluster adjacency when phrased in terms of $f_{A_3}^{--}$ or $f_{A_3\subset D_5}^{---}$. While it might have been hoped that a single decomposition would exhibit all the the nice mathematical features of the remainder function at once, it is perhaps better that $R^{(2)}_7$ ``requires'' nearly the full breadth of the subalgebra structure of $\Gr(4,7)$ in order to express all of its intricate behavior. 

The only subalgebra of $\Gr(4,7)$ we have not had reason to touch on yet is $D_4$. Interestingly, there exists exactly one $D_4$-constructible function in $E_6$ that has signature $\sigma_{E_6}^+ \tau_{E_6}^+ \mathbb{Z}_{2,E_6}^+$ and respects cobracket-level cluster adjacency. However, this function fails to vanish (or be well-defined) in the $7\to6$ collinear limit, and so cannot be directly related to $R_{7}^{(2)}\!$. Still, the uniqueness of this function is itself intriguing and may merit further study.

While the mathematical features of each nonclassical decomposition are easy to check, their physical meaning remains more obscure. It would be interesting to find kinematic limits that cleanly isolate individual instances of the cluster polylogarithms these decompositions make use of, but---as the number of subalgebras in every decomposition is greater than the dimension of seven-particle kinematics---this would require intricate cancellations between the remaining cluster polylogarithms, making such limits hard to engineer. Soft and collinear limits are of particularly little use here, as the nonclassical part of the six-particle amplitude is zero. Moreover, while it is in principle no problem to consider singular limits, only the finite term will survive, as large logs will get projected out by the cobracket. (While we could also study these cluster polylogarithms as full functions, this doesn't seem particularly meaningful since all the expressions presented in this paper only hold up to undetermined classical contributions.)  We correspondingly defer the consideration of further kinematic limits to future work involving the full analytic form of the eight-particle amplitude~\cite{cluster_subalgebras_ii}.

\section{Conclusion}

We have explored the space of nonclassical cluster polylogarithms on $\Gr(4,7)$ and its subalgebras, and have shown that---while this space is relatively small---it admits a natural geometric construction that gives rise to interesting sequences of nested functional embeddings. The seven-particle two-loop MHV amplitude seems almost specially engineered to take advantage of this subalgebra structure, insofar as its nonclassical component can be decomposed into cluster polylogarithms defined on almost every type of subalgebra appearing in $\Gr(4,7)$, excluding only $D_4$. The decompositions associated with the corank-one subalgebras $D_5$ and $A_5$ are especially canonical, as are the further decompositions of the associated $D_5$ and $A_5$ cluster polylogarithms into their $A_4$ subalgebras, as the form of each of these decompositions is uniquely dictated by the automorphism group of the parent algebra. These decompositions identify new cluster polylogarithms that are of special physical interest, supplementing the $A_2$ and $A_3$ cluster polylogarithms previously identified in~\cite{Golden:2014xqa}. It is natural to ask whether higher-point MHV amplitudes can also be decomposed into these new cluster polylogarithms, as is known to be the case for the $A_2$ and $A_3$ functions; we will take one step towards addressing this question in a forthcoming paper~\cite{cluster_subalgebras_ii}.

It would be interesting to try and extend this type of construction to the NMHV sector, where R-invariants also appear in the amplitude. There is a natural way to associate a subalgebra of $\Gr(4,n)$ to each R-invariant~\cite{Drummond:2018dfd}, allowing these subalgebras to (potentially) enter the decomposition of the polylogarithms these R-invariants multiply. More generally, although we have not attempted to explore (or even define) the space of cluster polylogarithms beyond weight four, it is hoped that some form of subalgebra constructibility can be extended to higher weight (and higher loops). 

While we have almost entirely worked at the level of the nonclassical component of the cobracket, it can also be asked whether these amplitudes are decomposable into functions defined on their subalgebras before applying this projection. It is not hard to show that the symbol of $R_6^{(2)}$ (as well as that of ${\cal E}_6^{(2)}$) is not subalgebra-constructible in this way---the six-particle two-loop MHV amplitude is intrinsically an $A_3$ polylogarithm. While it is possible that the $A_2$ and $A_1 \times A_1$ subalgebras of $\Gr(4,6)$ are too simple to permit such a decomposition while larger subalgebras in $\Gr(4,n>6)$ would prove sufficiently complex, it seems more likely that the cobracket projection is essential to distilling the part of the amplitude that can be decomposed in this way. The fact that it is the mathematically most complicated component of these amplitudes that is easiest to compute to all $n$ (as was done in~\cite{Golden:2014pua}) still calls out for some physical explanation.

More speculatively, it can be hoped that the pervasiveness of cluster-algebraic structure exhibited by polylogarithmic amplitudes in planar ${\cal N} = 4$ sYM theory points to the existence of some more general combinatorial structure underlying all amplitudes in this theory, including those that involve algebraic roots and/or functions beyond polylogarithms. Much of the cluster-algebraic structure seen in the polylogarithmic case is exposed by the coaction, which (especially in the guise of the symbol) distills these functions down to information about their integration kernels. Notably then, it is possible to formulate coactions on more general periods that appear in quantum field theory, such as elliptic polylogarithms~\cite{2015arXiv151206410B,Abreu:2017enx,Abreu:2017mtm,Broedel:2017kkb,Caron-Huot:2018dsv,Broedel:2018iwv,Broedel:2018qkq}. While much of the loop integration technology required for dealing with generic Feynman integrals remains to be developed, we can already begin to study the higher-genus and higher-dimensional varieties that appear in the integration contours contributing to planar ${\cal N} = 4$ sYM theory (see for example~\cite{2009arXiv0910.0114B,Bourjaily:2017bsb,Bourjaily:2018ycu,Bourjaily:2018yfy}). Hopefully, understanding these (seemingly always Calabi-Yau) geometries will provide salient hints for how the geometric and algebraic picture provided by cluster algebras can be generalized beyond the case of polylogarithms.

\section*{Acknowledgements}

We thank Jacob Bourjaily, Lance Dixon, James Drummond, Henriette Elvang, Matt von Hippel, Marcus Spradlin, and Matthias Wilhelm for illuminating discussion. We also thank Jacob Bourjaily, James Drummond, Andrea Orta, and Marcus Spradlin for comments on the manuscript. This project has received funding from an ERC Starting Grant \mbox{(No.\ 757978)}, a grant from the Villum Fonden (AJM), and the support of a Van Loo Postdoctoral Fellowship (JG). The authors are grateful to the Kavli Institute for Theoretical Physics (National Science Foundation grant NSF PHY11-25915) for hospitality.

\appendix

\newpage

\section{Counting Subalgebras of Finite Cluster Algebras}\label{appendix:subalgebras}
In this appendix we catalog the subalgebra structure of $\Gr(4,7) \simeq E_6$ and all its finite connected subalgebras, namely $A_2$, $A_3$, $A_4$, $D_4$, $A_5$, and $D_5$. We include both the number of subalgebras and subpolytopes (of the cluster polytope) that occur in each algebra. We only consider subalgebras to be distinct when they differ in at least one of their mutable nodes, but count identical subalgebras as distinct subpolytopes when they are connected to different frozen nodes. Note also that we consider the \xcoords\ $x$ and $1/x$ to be distinct.

\enlargethispage{\baselineskip}
\begin{center}
\vspace{.1cm}
{\qquad \qquad \Large{\(A_2\)} \hfill} 

\qquad Clusters: 5 \quad --- \quad  \acoords: 5 \quad --- \quad \xcoords: 10 \hfill

\vspace{.5cm}

{ \qquad \qquad \Large{\(A_3\)} \hfill} 

\vspace{.2cm}
\qquad Clusters: 14 \quad --- \quad \acoords: 9 \quad --- \quad \xcoords: 30 \hfill \\[1em]

\begin{tabular}{ | l | l | l |}
\multicolumn{1}{c}{Type} &  \multicolumn{1}{c}{Subpolytopes}  &  \multicolumn{1}{c}{Subalgebras} \\
\hline \(A_2\) & 6 & 6 \\ 
\hline \(A_1 \times A_1\) & 3 & 3 \\ 
\hline
\end{tabular} 
\vspace{.5cm}

{ \qquad \qquad \Large{\(A_4\)} \hfill}

\vspace{.2cm}
\qquad Clusters: 42 \quad --- \quad \acoords: 14 \quad --- \quad \xcoords: 70 \hfill \\[1em]

\begin{tabular}{ | l | l | l |}
\multicolumn{1}{c}{Type} &  \multicolumn{1}{c}{Subpolytopes}  &  \multicolumn{1}{c}{Subalgebras} \\
\hline \(A_2\) & 28 & 21 \\ 
\hline \(A_1 \times A_1\) & 28 & 28 \\ \hline 
\hline \(A_3\) & 7 & 7 \\ 
\hline \(A_2 \times A_1\) & 7 & 7 \\ 
\hline \(A_1 \times A_1 \times A_1\) & 0 & 0 \\ 
\hline
\end{tabular}
\vspace{.5cm}

{ \qquad \qquad \Large{\(D_4\)} \hfill} 

\vspace{.2cm}
\qquad Clusters: 50 \quad --- \quad \acoords: 16 \quad --- \quad \xcoords: 104 \hfill \\[1em]

\begin{tabular}{ | l | l | l |}
\multicolumn{1}{c}{Type} &  \multicolumn{1}{c}{Subpolytopes}  &  \multicolumn{1}{c}{Subalgebras} \\
\hline \(A_2\) & 36 & 36 \\ 
\hline \(A_1 \times A_1\) & 30 & 18 \\ \hline 
\hline \(A_3\) & 12 & 12 \\ 
\hline \(A_2 \times A_1\) & 0 & 0 \\ 
\hline \(A_1 \times A_1 \times A_1\) & 4 & 4 \\ 
\hline
\end{tabular}

\newpage 

{ \qquad \qquad \Large{\(A_5\)} \hfill}

\vspace{.2cm}
\qquad Clusters: 132 \quad --- \quad \acoords: 20 \quad --- \quad \xcoords: 140 \hfill \\[1em]

\begin{tabular}{ | l | l | l |}
\multicolumn{1}{c}{Type} &  \multicolumn{1}{c}{Subpolytopes}  &  \multicolumn{1}{c}{Subalgebras} \\
\hline \(A_2\) & 120 & 56 \\ 
\hline \(A_1 \times A_1\) & 180 & 144 \\ \hline 
\hline \(A_3\) & 36 & 28 \\ 
\hline \(A_2 \times A_1\) & 72 & 72 \\ 
\hline \(A_1 \times A_1 \times A_1\) & 12 & 12 \\ \hline 
\hline \(D_4\) & 0 & 0 \\ 
\hline \(A_4\) & 8 & 8 \\ 
\hline \(A_3 \times A_1\) & 8 & 8 \\ 
\hline \(A_2 \times A_2\) & 4 & 4 \\ 
\hline \(A_2 \times A_1 \times A_1\) & 0 & 0 \\ 
\hline \(A_1 \times A_1 \times A_1 \times A_1\) & 0 & 0 \\ 
\hline
\end{tabular} 
\vspace{.5cm}

{ \qquad \qquad \Large{\(D_5\)} \hfill}

\vspace{.2cm}
\qquad Clusters: 182 \quad --- \quad \acoords: 25 \quad --- \quad \xcoords: 260 \hfill \\[1em]

\begin{tabular}{ | l | l | l |} 
\multicolumn{1}{c}{Type} &  \multicolumn{1}{c}{Subpolytopes}  &  \multicolumn{1}{c}{Subalgebras} \\
\hline \(A_2\) & 180 & 125 \\ 
\hline \(A_1 \times A_1\) & 230 & 145 \\ \hline 
\hline \(A_3\) & 70 & 65 \\ 
\hline \(A_2 \times A_1\) & 60 & 50 \\ 
\hline \(A_1 \times A_1 \times A_1\) & 30 & 30 \\ \hline 
\hline \(D_4\) & 5 & 5 \\ 
\hline \(A_4\) & 10 & 10 \\ 
\hline \(A_3 \times A_1\) & 5 & 5 \\ 
\hline \(A_2 \times A_2\) & 0 & 0 \\ 
\hline \(A_2 \times A_1 \times A_1\) & 5 & 5 \\ 
\hline \(A_1 \times A_1 \times A_1 \times A_1\) & 0 & 0 \\ 
\hline
\end{tabular}
\vspace{.5cm}

\newpage

{ \qquad \qquad \Large{\(E_6\)} \hfill}

\vspace{.2cm}
\qquad Clusters: 833 \quad --- \quad \acoords: 42 \quad --- \quad \xcoords: 770 \hfill \\[1em]

\begin{tabular}{ | l | l | l |}
\multicolumn{1}{c}{Type} &  \multicolumn{1}{c}{Subpolytopes}  &  \multicolumn{1}{c}{Subalgebras} \\
\hline \(A_2\) & 1071 & 504 \\ 
\hline \(A_1 \times A_1\) & 1785 & 833 \\ \hline 
\hline \(A_3\) & 476 & 364 \\ 
\hline \(A_2 \times A_1\) & 714 & 490 \\ 
\hline \(A_1 \times A_1 \times A_1\) & 357 & 357 \\ \hline 
\hline \(D_4\) & 35 & 35 \\ 
\hline \(A_4\) & 112 & 98 \\ 
\hline \(A_3 \times A_1\) & 112 & 112 \\ 
\hline \(A_2 \times A_2\) & 21 & 14 \\ 
\hline \(A_2 \times A_1 \times A_1\) & 119 & 119 \\ 
\hline \(A_1 \times A_1 \times A_1 \times A_1\) & 0 & 0 \\ \hline 
\hline \(D_5\) & 14 & 14 \\ 
\hline \(A_5\) & 7 & 7 \\ 
\hline \(D_4 \times A_1\) & 0 & 0 \\ 
\hline \(A_4 \times A_1\) & 14 & 14 \\ 
\hline \(A_3 \times A_2\) & 0 & 0 \\ 
\hline \(A_3 \times A_1 \times A_1\) & 0 & 0 \\ 
\hline \(A_2 \times A_2 \times A_1\) & 7 & 7 \\ 
\hline \(A_2 \times A_1 \times A_1 \times A_1\) & 0 & 0 \\ 
\hline \(A_1 \times A_1 \times A_1 \times A_1 \times A_1\) & 0 & 0 \\ 
\hline
\end{tabular}
\end{center}

\newpage

\section{Cobracket Spaces in Finite Cluster Algebras}\label{appendix:cobrackets}

\parbox{\textwidth}{In this appendix we tabulate the number of independent weight-four cluster polylogarithms that have a nonzero Lie cobracket on $\Gr(4,7) \simeq E_6$ and its suablagebras. In  \parfillskip=0pt}

\begin{table}[b!]
\begin{center}

\begin{tabular}{ | c | c | c | c | c |  c |}
\multicolumn{1}{c}{\multirow{2}{*}{Type}} & \multicolumn{1}{c}{\multirow{2}{*}{Nonclassical Cobrackets}} & \multicolumn{4}{c}{Automorphism Signature} \\
\multicolumn{1}{c}{} & \multicolumn{1}{c}{} & \multicolumn{1}{c}{$\sigma^+ \tau^+$} & \multicolumn{1}{c}{$\sigma^+ \tau^-$} & \multicolumn{1}{c}{$\sigma^- \tau^+$} & \multicolumn{1}{c}{$\sigma^- \tau^-$} \\
\hline \(A_2\) & 1 (0) & 0  & 1 (0) & 0 & 0 \\ 
\hline \(A_3\) & 6 (1) & 0 & 1 (0)  & 0 & 1 (1) \\ 
\hline \(A_4\) & 21 (6) & 0 & 3 (0) & 0  & 0 \\ 
\hline \(D_4\) & 34 (9) & \multicolumn{4}{r ;{3pt/2pt}}{\ \tikzmark{d4tablePosition2}} \\ 
\hline \(A_5\) & 56 (21) & 2 (1) & 5 (1) & 2 (0) & 5 (3) \\ 
\hline \(D_5\) & 116 (42) & \multicolumn{4}{r ;{3pt/2pt}}{\ \tikzmark{d5tablePosition2}} \\ 
\cline{1-2} \cdashline{3-6}[3pt/3pt] \(E_6\) & 448 (195) & \multicolumn{4}{r ;{3pt/2pt}}{\ \tikzmark{e6tablePosition2}}  \\ 
\cline{1-2} \cdashline{3-6}[3pt/3pt] 
\end{tabular} 

\vspace{1cm}
$D_4$ Automorphism Signatures 

\hspace{-1cm}\begin{tabular}{ c | c | c |}
\multicolumn{1}{c}{\tikzmark{d4topLeft2} } &\multicolumn{2}{c}{$\underline{\ \sigma_{\mathfrak{D}_4}^+ \tau_{\mathfrak{D}_4}^+\ }$} \\[-1em]
\multicolumn{1}{c}{} & \multicolumn{1}{c}{} & \multicolumn{1}{c}{} \\
\multicolumn{1}{c}{} & \multicolumn{1}{c}{$\sigma_{S_3}^+$} & \multicolumn{1}{c}{$\sigma_{S_3}^-$} \\[-1em]
\multicolumn{1}{c}{} & \multicolumn{1}{c}{} & \multicolumn{1}{c}{} \\
\cline{2-3} $\tau_{S_3}^+$ & 0 & 0 \\
\cline{2-3} $\tau_{S_3}^-$ & 1 (0)  & 0 \\
\cline{2-3}
\end{tabular} 
\hspace{.4cm}
\begin{tabular}{ c | c | c |}
\multicolumn{1}{c}{} &\multicolumn{2}{c}{$\ \underline{\sigma_{\mathfrak{D}_4}^+ \tau_{\mathfrak{D}_4}^-\ }$} \\[-1em]
\multicolumn{1}{c}{} & \multicolumn{1}{c}{} & \multicolumn{1}{c}{}\\
\multicolumn{1}{c}{} & \multicolumn{1}{c}{$\sigma_{S_3}^+$} & \multicolumn{1}{c}{$\sigma_{S_3}^-$} \\[-1em]
\multicolumn{1}{c}{} & \multicolumn{1}{c}{} & \multicolumn{1}{c}{} \\
\cline{2-3} $\tau_{S_3}^+$ & 2 (0) & 0 \\
\cline{2-3} $\tau_{S_3}^-$ & 0 & 0 \\
\cline{2-3}
\end{tabular}
\hspace{.4cm}
\begin{tabular}{ c | c | c |}
\multicolumn{1}{c}{} &\multicolumn{2}{c}{$\underline{\ \sigma_{\mathfrak{D}_4}^- \tau_{\mathfrak{D}_4}^+\ }$} \\[-1em]
\multicolumn{1}{c}{} & \multicolumn{1}{c}{} & \multicolumn{1}{c}{}\\
\multicolumn{1}{c}{} & \multicolumn{1}{c}{$\sigma_{S_3}^+$} & \multicolumn{1}{c}{$\sigma_{S_3}^-$} \\[-1em]
\multicolumn{1}{c}{} & \multicolumn{1}{c}{} & \multicolumn{1}{c}{} \\
\cline{2-3} $\tau_{S_3}^+$ & 1 (0) & 0 \\
\cline{2-3} $\tau_{S_3}^-$ & 1 (1) & 0 \\
\cline{2-3}
\end{tabular}
\hspace{.4cm}
\begin{tabular}{ c | c | c |}
\multicolumn{1}{c}{} &\multicolumn{2}{c}{$\underline{\ \sigma_{\mathfrak{D}_4}^- \tau_{\mathfrak{D}_4}^- \ }$} \\[-1em]
\multicolumn{1}{c}{} & \multicolumn{1}{c}{} & \multicolumn{1}{c}{}\\
\multicolumn{1}{c}{} & \multicolumn{1}{c}{$\sigma_{S_3}^+$} & \multicolumn{1}{c}{$\sigma_{S_3}^-$} \\[-1em]
\multicolumn{1}{c}{} & \multicolumn{1}{c}{} & \multicolumn{1}{c}{} \\
\cline{2-3} $\tau_{S_3}^+$ & 1 (0) & 0 \\
\cline{2-3} $\tau_{S_3}^-$ & 0 & 0 \tikzmark{d4bottomRight2}  \\
\cline{2-3}
\end{tabular}

\vspace{1cm}
$D_5$ Automorphism Signatures 

\begin{tabular}{| c | c |}
\multicolumn{2}{c}{\tikzmark{d5topLeft2}  $\underline{\ \sigma_{\mathfrak{D_5}}^+ \tau_{\mathfrak{D_5}}^+\ }$} \\[-1em]
\multicolumn{1}{c}{} & \multicolumn{1}{c}{} \\
\multicolumn{1}{c}{$\mathbb{Z}_2^+$} & \multicolumn{1}{c}{$\mathbb{Z}_2^-$} \\[-1em]
\multicolumn{1}{c}{} & \multicolumn{1}{c}{} \\
\hline
5 (2) & 0 \\
\hline
\end{tabular} 
\hspace{1.2cm}
\begin{tabular}{| c | c |}
\multicolumn{2}{c}{$\underline{\ \sigma_{\mathfrak{D_5}}^+ \tau_{\mathfrak{D_5}}^-\ }$} \\[-1em]
\multicolumn{1}{c}{} & \multicolumn{1}{c}{} \\
\multicolumn{1}{c}{$\mathbb{Z}_2^+$} & \multicolumn{1}{c}{$\mathbb{Z}_2^-$} \\[-1em]
\multicolumn{1}{c}{} & \multicolumn{1}{c}{} \\
\hline
9 (2) & 0 \\
\hline
\end{tabular} 
\hspace{1.2cm}
\begin{tabular}{| c | c |}
\multicolumn{2}{c}{$\underline{\ \sigma_{\mathfrak{D_5}}^- \tau_{\mathfrak{D_5}}^+\ }$} \\[-1em]
\multicolumn{1}{c}{} & \multicolumn{1}{c}{} \\
\multicolumn{1}{c}{$\mathbb{Z}_2^+$} & \multicolumn{1}{c}{$\mathbb{Z}_2^-$} \\[-1em]
\multicolumn{1}{c}{} & \multicolumn{1}{c}{} \\
\hline
0 & 3 (1) \\
\hline
\end{tabular} 
\hspace{1.2cm}
\begin{tabular}{| c | c |}
\multicolumn{2}{c}{$\underline{\ \sigma_{\mathfrak{D_5}}^- \tau_{\mathfrak{D_5}}^- \ }$} \\[-1em]
\multicolumn{1}{c}{} & \multicolumn{1}{c}{} \\
\multicolumn{1}{c}{$\mathbb{Z}_2^+$} & \multicolumn{1}{c}{$\mathbb{Z}_2^-$} \\[-1em]
\multicolumn{1}{c}{} & \multicolumn{1}{c}{} \\
\hline
0 & 7 (5) \tikzmark{d5bottomRight2}  \\
\hline
\end{tabular} 

\vspace{1cm}
$E_6$ Automorphism Signatures

\begin{tabular}{| c | c |}
\multicolumn{2}{c}{\tikzmark{e6topLeft2} $\underline{\ \sigma_{\mathfrak{D}_{14}}^+ \tau_{\mathfrak{D}_{14}}^+\ }$} \\[-1em]
\multicolumn{1}{c}{} & \multicolumn{1}{c}{} \\
\multicolumn{1}{c}{$\mathbb{Z}_2^+$} & \multicolumn{1}{c}{$\mathbb{Z}_2^-$} \\[-1em]
\multicolumn{1}{c}{} & \multicolumn{1}{c}{} \\
\hline
12 (6) & 14 (6) \\
\hline
\end{tabular} 
\hspace{1.2cm}
\begin{tabular}{| c | c |}
\multicolumn{2}{c}{$\underline{\ \sigma_{\mathfrak{D}_{14}}^+ \tau_{\mathfrak{D}_{14}}^-\ }$} \\[-1em]
\multicolumn{1}{c}{} & \multicolumn{1}{c}{} \\
\multicolumn{1}{c}{$\mathbb{Z}_2^+$} & \multicolumn{1}{c}{$\mathbb{Z}_2^-$} \\[-1em]
\multicolumn{1}{c}{} & \multicolumn{1}{c}{} \\
\hline
21 (6) & 17 (9) \\
\hline
\end{tabular} 
\hspace{1.2cm} 
\begin{tabular}{| c | c |}
\multicolumn{2}{c}{$\underline{\ \sigma_{\mathfrak{D}_{14}}^- \tau_{\mathfrak{D}_{14}}^+\ }$} \\[-1em]
\multicolumn{1}{c}{} & \multicolumn{1}{c}{} \\
\multicolumn{1}{c}{$\mathbb{Z}_2^+$} & \multicolumn{1}{c}{$\mathbb{Z}_2^-$} \\[-1em]
\multicolumn{1}{c}{} & \multicolumn{1}{c}{} \\
\hline
0 & 0 \\
\hline
\end{tabular} 
\hspace{1.2cm}
\begin{tabular}{| c | c |}
\multicolumn{2}{c}{$\underline{\ \sigma_{\mathfrak{D}_{14}}^- \tau_{\mathfrak{D}_{14}}^- \ }$} \\[-1em]
\multicolumn{1}{c}{} & \multicolumn{1}{c}{} \\
\multicolumn{1}{c}{$\mathbb{Z}_2^+$} & \multicolumn{1}{c}{$\mathbb{Z}_2^-$} \\[-1em]
\multicolumn{1}{c}{} & \multicolumn{1}{c}{} \\
\hline
0 & 0 \tikzmark{e6bottomRight2} \\
\hline
\end{tabular}
\end{center}
\begin{tikzpicture}[overlay, remember picture,decoration={markings,mark=at position .99 with {\arrow[scale=1.4,>=stealth]{>}}}]
    \draw [dashed,rounded corners=1cm] ([yshift=1cm,xshift=-.8cm]{pic cs:d4topLeft2}) rectangle ([yshift=-.6cm,xshift=1cm]{pic cs:d4bottomRight2}) ;
    \draw [dashed,rounded corners=1cm] ([yshift=1cm,xshift=-1.2cm]{pic cs:d5topLeft2}) rectangle ([yshift=-.6cm,xshift=1cm]{pic cs:d5bottomRight2}) ;
    \draw [dashed,rounded corners=1cm] ([yshift=1cm,xshift=-1.3cm]{pic cs:e6topLeft2}) rectangle ([yshift=-.6cm,xshift=1cm]{pic cs:e6bottomRight2}) ;
    \draw [dashed,shorten >=0.14cm,postaction={decorate}] ([yshift=1cm,xshift=1cm]{pic cs:d4bottomRight2}) [bend right] to[out=-60,in=-100] ([yshift=0.08cm,xshift=.22cm]{pic cs:d4tablePosition2}) ;
    \draw [dashed,shorten >=0.14cm,postaction={decorate}] ([yshift=1cm,xshift=1cm]{pic cs:d5bottomRight2}) [bend right] to[out=-80,in=-100] ([yshift=0.08cm,xshift=.16cm]{pic cs:d5tablePosition2}) ;
    \draw [dashed,shorten >=0.14cm,postaction={decorate}] ([yshift=1cm,xshift=1cm]{pic cs:e6bottomRight2}) [bend right] to[out=-100,in=-100] ([yshift=0.08cm,xshift=.16cm]{pic cs:e6tablePosition2}) ;
\end{tikzpicture}
\caption{The number of nonclassical weight-four cluster polylogarithms on various finite cluster algebras, prior to consideration of their automorphism group and after requiring specific automorphism signatures. The number of polylogarithms that can also be made to satisfy cobracket-level cluster adjacency is given in parentheses.}
\vspace{-1cm}
\label{table:nonclassical_dimensions}
\end{table}
 
\noindent\parbox{\textwidth}{table~\ref{table:nonclassical_dimensions} we first record the number of cluster polylogarithms that have a nonzero $\delta_{2,2}$ cobracket component, as considered in section~\ref{sec:sub-constructibility}. We tabulate the number of functions on each cluster algebra both before and after requiring a specific automorphism \parfillskip=0pt}

\begin{table}[t]
\begin{center}
\vspace{-1cm}
\begin{tabular}{ | c | c | c | c | c |  c |}
\multicolumn{1}{c}{\multirow{2}{*}{Type}} & \multicolumn{1}{c}{\multirow{2}{*}{Integrable Cobrackets}} & \multicolumn{4}{c}{Automorphism Signature} \\
\multicolumn{1}{c}{} & \multicolumn{1}{c}{} & \multicolumn{1}{c}{$\sigma^+ \tau^+$} & \multicolumn{1}{c}{$\sigma^+ \tau^-$} & \multicolumn{1}{c}{$\sigma^- \tau^+$} & \multicolumn{1}{c}{$\sigma^- \tau^-$} \\
\hline \(A_2\) & 6 (5) & 1 (1) & 1 (0) & 0 & 0 \\ 
\hline \(A_3\) & 21 (16)  & 2 (2)  & 2 (1)  & 0 & 3 (3) \\ 
\hline \(A_4\) & 56 (41) & 4 (4) & 4 (1)  & 0  & 0 \\ 
\hline \(D_4\) & 86 (61) & \multicolumn{4}{r ;{3pt/2pt}}{\ \tikzmark{d4tablePosition3}} \\ 
\hline \(A_5\) & 126 (91) & 8 (7) & 8 (4) & 5 (3) & 11 (9) \\ 
\hline \(D_5\) & 246 (172) & \multicolumn{4}{r ;{3pt/2pt}}{\ \tikzmark{d5tablePosition3}} \\ 
\cline{1-2} \cdashline{3-6}[3pt/3pt]  \(E_6\) & 833 (580) & \multicolumn{4}{r ;{3pt/2pt}}{\ \tikzmark{e6tablePosition3}}   \\ 
\cline{1-2} \cdashline{3-6}[3pt/3pt] 
\end{tabular} 

\vspace{1cm}
$D_4$ Automorphism Signatures 

\hspace{-1cm}\begin{tabular}{ c | c | c |}
\multicolumn{1}{c}{\tikzmark{d4topLeft3}} &\multicolumn{2}{c}{$\underline{\ \sigma_{\mathfrak{D}_4}^+ \tau_{\mathfrak{D}_4}^+\ }$} \\[-1em]
\multicolumn{1}{c}{} & \multicolumn{1}{c}{} & \multicolumn{1}{c}{} \\
\multicolumn{1}{c}{} & \multicolumn{1}{c}{$\sigma_{S_3}^+$} & \multicolumn{1}{c}{$\sigma_{S_3}^-$} \\[-1em]
\multicolumn{1}{c}{} & \multicolumn{1}{c}{} & \multicolumn{1}{c}{} \\
\cline{2-3} $\tau_{S_3}^+$ & 3 (3) & 0 \\
\cline{2-3} $\tau_{S_3}^-$ & 1 (0) & 0 \\
\cline{2-3}
\end{tabular} 
\hspace{.4cm}
\begin{tabular}{ c | c | c |}
\multicolumn{1}{c}{} &\multicolumn{2}{c}{$\underline{\ \sigma_{\mathfrak{D}_4}^+ \tau_{\mathfrak{D}_4}^-\ }$} \\[-1em]
\multicolumn{1}{c}{} & \multicolumn{1}{c}{} & \multicolumn{1}{c}{}\\
\multicolumn{1}{c}{} & \multicolumn{1}{c}{$\sigma_{S_3}^+$} & \multicolumn{1}{c}{$\sigma_{S_3}^-$} \\[-1em]
\multicolumn{1}{c}{} & \multicolumn{1}{c}{} & \multicolumn{1}{c}{} \\
\cline{2-3} $\tau_{S_3}^+$ & 3 (1) & 0 \\
\cline{2-3} $\tau_{S_3}^-$ & 1 (1) & 0 \\
\cline{2-3}
\end{tabular}
\hspace{.4cm}
\begin{tabular}{ c | c | c |}
\multicolumn{1}{c}{} &\multicolumn{2}{c}{$\underline{\ \sigma_{\mathfrak{D}_4}^- \tau_{\mathfrak{D}_4}^+\ }$} \\[-1em]
\multicolumn{1}{c}{} & \multicolumn{1}{c}{} & \multicolumn{1}{c}{}\\
\multicolumn{1}{c}{} & \multicolumn{1}{c}{$\sigma_{S_3}^+$} & \multicolumn{1}{c}{$\sigma_{S_3}^-$} \\[-1em]
\multicolumn{1}{c}{} & \multicolumn{1}{c}{} & \multicolumn{1}{c}{} \\
\cline{2-3} $\tau_{S_3}^+$ & 3 (2) & 0 \\
\cline{2-3} $\tau_{S_3}^-$ & 2 (2) & 0 \\
\cline{2-3}
\end{tabular}
\hspace{.4cm}
\begin{tabular}{ c | c | c |}
\multicolumn{1}{c}{} &\multicolumn{2}{c}{$\underline{\ \sigma_{\mathfrak{D}_4}^- \tau_{\mathfrak{D}_4}^- \ }$} \\[-1em]
\multicolumn{1}{c}{} & \multicolumn{1}{c}{} & \multicolumn{1}{c}{}\\
\multicolumn{1}{c}{} & \multicolumn{1}{c}{$\sigma_{S_3}^+$} & \multicolumn{1}{c}{$\sigma_{S_3}^-$} \\[-1em]
\multicolumn{1}{c}{} & \multicolumn{1}{c}{} & \multicolumn{1}{c}{} \\
\cline{2-3} $\tau_{S_3}^+$ & 3 (2) & 0 \\
\cline{2-3} $\tau_{S_3}^-$ & 0 & 0 \tikzmark{d4bottomRight3} \\
\cline{2-3}
\end{tabular}

\vspace{1cm}
$D_5$ Automorphism Signatures 

\begin{tabular}{| c | c |}
\multicolumn{2}{c}{\tikzmark{d5topLeft3} $\underline{\ \sigma_{\mathfrak{D_5}}^+ \tau_{\mathfrak{D_5}}^+ \ }$} \\[-1em]
\multicolumn{1}{c}{} & \multicolumn{1}{c}{} \\
\multicolumn{1}{c}{$\mathbb{Z}_2^+$} & \multicolumn{1}{c}{$\mathbb{Z}_2^-$} \\[-1em]
\multicolumn{1}{c}{} & \multicolumn{1}{c}{} \\
\hline
16 (13) & 0 \\
\hline
\end{tabular} 
\hspace{1.2cm}
\begin{tabular}{| c | c |}
\multicolumn{2}{c}{$\underline{\ \sigma_{\mathfrak{D_5}}^+ \tau_{\mathfrak{D_5}}^- \ }$} \\[-1em]
\multicolumn{1}{c}{} & \multicolumn{1}{c}{} \\
\multicolumn{1}{c}{$\mathbb{Z}_2^+$} & \multicolumn{1}{c}{$\mathbb{Z}_2^-$} \\[-1em]
\multicolumn{1}{c}{} & \multicolumn{1}{c}{} \\
\hline
16 (9) & 0 \\
\hline
\end{tabular} 
\hspace{1.2cm}
\begin{tabular}{| c | c |}
\multicolumn{2}{c}{$\underline{\ \sigma_{\mathfrak{D_5}}^- \tau_{\mathfrak{D_5}}^+ \ }$} \\[-1em]
\multicolumn{1}{c}{} & \multicolumn{1}{c}{} \\
\multicolumn{1}{c}{$\mathbb{Z}_2^+$} & \multicolumn{1}{c}{$\mathbb{Z}_2^-$} \\[-1em]
\multicolumn{1}{c}{} & \multicolumn{1}{c}{} \\
\hline
0 & 5 (3) \\
\hline
\end{tabular} 
\hspace{1.2cm}
\begin{tabular}{| c | c |}
\multicolumn{2}{c}{$\underline{\ \sigma_{\mathfrak{D_5}}^- \tau_{\mathfrak{D_5}}^-\ }$} \\[-1em]
\multicolumn{1}{c}{} & \multicolumn{1}{c}{} \\
\multicolumn{1}{c}{$\mathbb{Z}_2^+$} & \multicolumn{1}{c}{$\mathbb{Z}_2^-$} \\[-1em]
\multicolumn{1}{c}{} & \multicolumn{1}{c}{} \\
\hline
0 & 13 (11) \tikzmark{d5bottomRight3} \\
\hline
\end{tabular} 

\vspace{1cm}
$E_6$ Automorphism Signatures

\begin{tabular}{| c | c |}
\multicolumn{2}{c}{\tikzmark{e6topLeft3} $\underline{\ \sigma_{\mathfrak{D}_{14}}^+ \tau_{\mathfrak{D}_{14}}^+\ }$} \\[-1em]
\multicolumn{1}{c}{} & \multicolumn{1}{c}{} \\
\multicolumn{1}{c}{$\mathbb{Z}_2^+$} & \multicolumn{1}{c}{$\mathbb{Z}_2^-$} \\[-1em]
\multicolumn{1}{c}{} & \multicolumn{1}{c}{} \\
\hline
32 (26) & 25 (17) \\
\hline
\end{tabular} 
\hspace{1.2cm}
\begin{tabular}{| c | c |}
\multicolumn{2}{c}{$\underline{\ \sigma_{\mathfrak{D}_{14}}^+ \tau_{\mathfrak{D}_{14}}^-\ }$} \\[-1em]
\multicolumn{1}{c}{} & \multicolumn{1}{c}{} \\
\multicolumn{1}{c}{$\mathbb{Z}_2^+$} & \multicolumn{1}{c}{$\mathbb{Z}_2^-$} \\[-1em]
\multicolumn{1}{c}{} & \multicolumn{1}{c}{} \\
\hline
32 (17) & 30 (22) \\
\hline
\end{tabular} 
\hspace{1.2cm}
\begin{tabular}{| c | c |}
\multicolumn{2}{c}{$\underline{\ \sigma_{\mathfrak{D}_{14}}^- \tau_{\mathfrak{D}_{14}}^+\ }$} \\[-1em]
\multicolumn{1}{c}{} & \multicolumn{1}{c}{} \\
\multicolumn{1}{c}{$\mathbb{Z}_2^+$} & \multicolumn{1}{c}{$\mathbb{Z}_2^-$} \\[-1em]
\multicolumn{1}{c}{} & \multicolumn{1}{c}{} \\
\hline
0 & 0 \\
\hline
\end{tabular} 
\hspace{1.2cm}
\begin{tabular}{| c | c |}
\multicolumn{2}{c}{$\underline{\ \sigma_{\mathfrak{D}_{14}}^- \tau_{\mathfrak{D}_{14}}^- \ }$} \\[-1em]
\multicolumn{1}{c}{} & \multicolumn{1}{c}{} \\
\multicolumn{1}{c}{$\mathbb{Z}_2^+$} & \multicolumn{1}{c}{$\mathbb{Z}_2^-$} \\[-1em]
\multicolumn{1}{c}{} & \multicolumn{1}{c}{} \\
\hline
0 & 0 \tikzmark{e6bottomRight3} \\
\hline
\end{tabular}
\end{center}
\begin{tikzpicture}[overlay, remember picture,decoration={markings,mark=at position .99 with {\arrow[scale=1.4,>=stealth]{>}}}]
    \draw [dashed,rounded corners=1cm] ([yshift=1cm,xshift=-.8cm]{pic cs:d4topLeft3}) rectangle ([yshift=-.6cm,xshift=1cm]{pic cs:d4bottomRight3}) ;
    \draw [dashed,rounded corners=1cm] ([yshift=1cm,xshift=-1.4cm]{pic cs:d5topLeft3}) rectangle ([yshift=-.6cm,xshift=1cm]{pic cs:d5bottomRight3}) ;
    \draw [dashed,rounded corners=1cm] ([yshift=1cm,xshift=-1.5cm]{pic cs:e6topLeft3}) rectangle ([yshift=-.6cm,xshift=1cm]{pic cs:e6bottomRight3}) ;
    \draw [dashed,shorten >=0.14cm,postaction={decorate}] ([yshift=1cm,xshift=1cm]{pic cs:d4bottomRight3}) [bend right] to[out=-60,in=-100] ([yshift=0.08cm,xshift=.22cm]{pic cs:d4tablePosition3}) ;
    \draw [dashed,shorten >=0.14cm,postaction={decorate}] ([yshift=1cm,xshift=1cm]{pic cs:d5bottomRight3}) [bend right] to[out=-80,in=-100] ([yshift=0.08cm,xshift=.16cm]{pic cs:d5tablePosition3}) ;
    \draw [dashed,shorten >=0.14cm,postaction={decorate}] ([yshift=1cm,xshift=1cm]{pic cs:e6bottomRight3}) [bend right] to[out=-100,in=-100] ([yshift=0.08cm,xshift=.16cm]{pic cs:e6tablePosition3}) ;
\end{tikzpicture}

\caption{The number of weight-four cluster polylogarithms on various finite cluster algebras that have nonzero cobrackets, prior to consideration of their automorphism group and after requiring specific automorphism signatures. The number of polylogarithms that can also be made to satisfy cobracket-level cluster adjacency is given in parentheses.}
\label{table:total_dimensions}
\end{table}
\vspace{1cm}

\noindent signature, and include the number of functions that also respect cobracket-level cluster adjacency in parentheses. (Some of these numbers can also be found in~\cite{Harrington:2015bdt}.) In table~\ref{table:total_dimensions} we record the same information, but for all weight-four cluster polylogarithms that have any nonzero cobracket component.

\newpage

\bibliographystyle{JHEP}
\bibliography{subalgebras}

\end{document}